\documentclass{acm_proc_article-sp}

\newtheorem{theorem}{Theorem}
\usepackage{amsmath,url,cite}
\usepackage{amsfonts}
\usepackage{graphicx}
\usepackage{subfigure}

\begin{document}

\title{b-Bit Minwise Hashing}

\numberofauthors{2}
\author{
\alignauthor
Ping Li\\\affaddr{Department of Statistical Science}\\
       \affaddr{Cornell University}\\
              \affaddr{Ithaca, NY 14853}\\
       \email{pingli@cornell.edu}
\alignauthor
Arnd Christian K\"{o}nig\\
       \affaddr{Microsoft Research}\\
       \affaddr{Microsoft Corporation}\\
              \affaddr{Redmond, WA 98052}\\
       \email{chrisko@microsoft.com}
}

\maketitle
\begin{abstract}

This\footnote{This version slightly modified the first draft written in August, 2009.} paper establishes the theoretical framework of \textbf{\em $b$-bit minwise hashing}. The original  {\em minwise hashing} method\cite{Proc:Broder} has become a standard technique for estimating set similarity (e.g., {\em resemblance}) with applications in information retrieval, data management, social networks and computational advertising.

By only storing the lowest $b$ bits of each (minwise) hashed value (e.g., $b=1$ or 2), one can gain substantial advantages  in terms of computational efficiency and storage space. We prove the basic theoretical results and provide an unbiased estimator of the resemblance for any $b$. We demonstrate that, even in the least favorable scenario, using $b=1$  may reduce the storage space at least by a factor of 21.3 (or 10.7) compared to using $b=64$ (or $b=32$), if one is interested in resemblance $\geq 0.5$.

\end{abstract}



\section{Introduction}

Computing the size of set intersections is a fundamental problem in information retrieval, databases, and  machine learning. Given two sets, $S_1$ and $S_2$, where
\begin{align}\notag
S_1,  S_2 \subseteq \Omega = \{0, 1, 2, ..., D-1\},
 \end{align}
 a basic task is to compute the joint size $a = |S_1\cap S_2|$, which measures the (un-normalized) similarity between $S_1$ and $S_2$. The so-called \textbf{\em resemblance}, denoted by $R$, provides a normalized similarity measure:
\begin{align}\notag
R = \frac{|S_1 \cap S_2|}{S_1 \cup S_2|} = \frac{a}{f_1 + f_2 - a}, \hspace{0.15in} \text{where} \  f_1 = |S_1|, \  f_2 = |S_2|.
\end{align}
It is known that $1-R$, the {\em resemblance distance}, is a metric, i.e., satisfying the triangle inequality\cite{Proc:Broder,Proc:Charikar}.  

In large datasets  encountered in information retrieval and databases, efficiently computing the joint sizes is often highly challenging\cite{Proc:Brin_Sigmod95,Article:Henzinger04}. Detecting (nearly) duplicate web pages is a classical example\cite{Proc:Broder,Proc:Broder_WWW97}.

Typically, each Web document can be processed as ``a bag of shingles,''  where a shingle consists of $w$ contiguous words in a document. Here $w$ is a tuning parameter and was set to be $w=5$ in several studies\cite{Proc:Broder,Proc:Broder_WWW97,Proc:Fetterly_WWW03}.

Clearly, the total number of possible shingles is huge.  Considering merely $10^5$ unique English words, the total number of possible $5$-shingles should be $D = (10^5)^5=O(10^{25})$.  Prior studies used $D = 2^{64}$ \cite{Proc:Fetterly_WWW03} and $D = 2^{40}$\cite{Proc:Broder,Proc:Broder_WWW97}.

\subsection{Minwise Hashing}

In their seminal work,  Broder and his colleagues developed  {\em minwise hashing} and successfully applied the technique to the task of duplicate document removal at the Web scale\cite{Proc:Broder,Proc:Broder_WWW97}. Since then, there have been considerable theoretical and methodological developments\cite{Article:Indyk2001,Proc:Charikar,Proc:Itoh_STOC03,Article:Broder_min-wise,Article:Kaplan09,Article:Li_Church_CL07,Proc:Li_Church_Hastie_NIPS08}.

As a general technique for estimating set similarity, {\em minwise hashing}  has been applied to a  wide range of applications, for example, content matching for online advertising\cite{Proc:Pandey_WWW09}, detection of large-scale redundancy in enterprise file systems\cite{Article:Forman09}, syntactic similarity algorithms for enterprise information management\cite{Proc:Cherkasova_KDD09}, compressing social networks\cite{Proc:Chierichetti_KDD09},
advertising diversification\cite{Proc:Gollapudi_WWW09}, community extraction and classification in the Web graph\cite{Article:Dourisboure09}, graph sampling\cite{Proc:Najork_WSDM09}, wireless sensor networks\cite{Article:Kalpakis08}, Web spam\cite{Article:Urvoy08,Proc:Nitin_WSDM08}, Web graph compression
\cite{Proc:Buehrer_WSDM08}, text reuse in the Web\cite{Proc:Bendersky_WSDM09}, and many more.


Here, we give a brief introduction to this algorithm. Suppose a random permutation $\pi$ is performed on $\Omega$, i.e.,
\begin{align}\notag
\pi: \ \Omega \longrightarrow \Omega, \hspace{0.3in} \text{where} \ \ \ \Omega=\{0, 1, ..., D-1\}
\end{align}
An elementary probability argument can show
\begin{align}
\mathbf{Pr}\left(\text{min}({\pi}(S_1)) = \text{min}({\pi}(S_2)) \right) = \frac{|S_1
  \cap S_2|}{|S_1 \cup S_2|}=R.
\end{align}

After $k$ minwise independent permutations, denoted by $\pi_1$,
$\pi_2$, ..., $\pi_k$,  one can  estimate $R$ without bias,  as a binomial probability, i.e.,
\begin{align}
&\hat{R}_{M} = \frac{1}{k}\sum_{j=1}^{k}1\{{\min}({\pi_j}(S_1)) =
  {\min}({\pi_j}(S_2))\}, \\\label{eqn_Var_M}
&\text{Var}\left(\hat{R}_{M}\right) = \frac{1}{k}R(1-R).
\end{align}

Throughout the paper, we will frequently use the term ``sample,'' corresponding to the term ``sample size'' (denoted by $k$). In {\em minwise hashing}, a sample is a hashed value, e.g., ${\min}({\pi_j}(S_i))$, which may require
 e.g., 64 bits to store\cite{Proc:Fetterly_WWW03}, depending on the universal size $D$. The total storage for each set would be $bk$ bits, where $b=64$ is possible.

After the samples have been collected, the storage and computational   cost is proportional to $b$. Therefore, reducing the number of bits for each hashed value would be useful, not only for  saving significant storage space but also for considerably improving the computational efficiency.

\subsection{Our Main Contributions}

In this paper, we establish a unified theoretical framework for \textbf{\em b-bit minwise hashing}. Instead of using $b=64$ bits\cite{Proc:Fetterly_WWW03}  or $40$ bits\cite{Proc:Broder,Proc:Broder_WWW97}, our theoretical results suggest using as few as $b=1$ or $b=2$ bits can yield significant improvements.\\

In $b$-bit minwise hashing, a ``sample'' consists of $b$ bits only, as opposed to e.g., 64 bits in the original minwise hashing.

Intuitively, using fewer bits per sample will increase the estimation variance, compared to (\ref{eqn_Var_M}), at the same ``sample size'' $k$. Thus, we will have to increase $k$ to maintain the same accuracy. Interestingly, our theoretical results will demonstrate that, when resemblance is not too small (e.g., $R\geq 0.5$, the threshold used in\cite{Proc:Broder,Proc:Broder_WWW97}), we do not have to increase $k$ much. This means that, compared to the earlier approach, the b-bit minwise hashing can be used to improve estimation accuracy and significantly reduce storage requirements at the same time.

For example, when $b=1$ and $R=0.5$, the estimation variance will increase at most by a factor of 3 (even in the least favorable scenario). This means, in order not to lose accuracy, we have to increase the sample size by a factor of 3. If we originally stored each hashed value using 64 bits\cite{Proc:Fetterly_WWW03}, the improvement by using $b=1$ will be $64/3=21.3$.

\section{The Fundamental Results}

Consider two sets, $S_1$ and $S_2$,
\begin{align}\notag
&S_1, S_2 \subseteq \Omega = \{0,1, 2, ..., D-1\},\\\notag
&f_1=|S_1|,\ f_2 = |S_2|, \ a = |S_1\cap S_2|
\end{align}
Apply a random permutation $\pi$ on $S_1$ and $S_2$: $\pi: \Omega \longrightarrow \Omega$.  Define the minimum values under  $\pi$ to be $z_1$ and $z_2$:
\begin{align}\notag
z_1 = \min\left(\pi\left(S_1\right)\right), \hspace{0.2in} z_2 = \min\left(\pi\left(S_2\right)\right).
\end{align}

Define $e_{1,i} = i$th lowest bit of $z_1$, and $e_{2,i} = i$th lowest bit of $z_2$.  Theorem \ref{The_basic} derives the analytical expression for $E_b$:
 \begin{align}\label{eqn_basic}
E_b = \mathbf{Pr}\left(\prod_{i=1}^b 1\{e_{1,i}=e_{2,i}\} =1\right).
 \end{align}

\begin{theorem}\label{The_basic}
Assume $D$ is large.
\begin{align}
\mathbf{Pr}\left(\prod_{i=1}^b1\left\{e_{1,i} = e_{2,i}\right\} = 1\right) = C_{1,b} + \left(1-C_{2,b}\right) R
\end{align}
where
\begin{align}
&r_1 = \frac{f_1}{D}, \hspace{0.2in} r_2 = \frac{f_2}{D},\\
&C_{1,b} = A_{1,b} \frac{r_2}{r_1+r_2} + A_{2,b}\frac{r_1}{r_1+r_2},\\
&C_{2,b} = A_{1,b} \frac{r_1}{r_1+r_2} + A_{2,b}\frac{r_2}{r_1+r_2},\\
&A_{1,b} = \frac{r_1\left[1-r_1\right]^{2^b-1}}{1-\left[1-r_1\right]^{2^b}},\\
&A_{2,b} = \frac{r_2\left[1-r_2\right]^{2^b-1}}{1-\left[1-r_2\right]^{2^b}}.
\end{align}

For a fixed $r_j$ (where $j\in\{1,2\}$),  $A_{j,b}$ is a monotonically decreasing function of $b=1,2,3,...$.

For a fixed $b$, $A_{j,b}$ is a monotonically decreasing function of $r_j\in[0,1]$, with the limit to be
\begin{align}
\lim_{r_{j}\rightarrow 0}A_{j,b} = \frac{1}{2^b}.
\end{align}
\textbf{Proof}: See Appendix \ref{app_proof_basic}.$\Box$
\end{theorem}

Theorem \ref{The_basic} says that, for a given $b$, the desired probability (\ref{eqn_basic}) is determined by $R$ and the ratios, $r_1 = \frac{f_1}{D}$ and $r_2 = \frac{f_2}{D}$. The only assumption needed in the proof of Theorem \ref{The_basic} is that $D$ should be large, which is always satisfied in practice.

$A_{j,b}$ ($j\in \{1,2\}$) is a  decreasing function of $r_j$ and $A_{j,b}\leq\frac{1}{2^b}$. As $b$ increases, $A_{j,b}$ converges to zero very quickly. In fact, when $b\geq 32$, one can essentially view $A_{j,b}=0$.

%

\subsection{The Unbiased Estimator}\label{sec_estimator}
Theorem \ref{The_basic} naturally suggests an unbiased estimator of $R$, denoted by $\hat{R}_b$:
\begin{align}\label{eqn_R_b}
&\hat{R}_b = \frac{\hat{E}_b - C_{1,b}}{1-C_{2,b}},\\
&\hat{E}_{b} = \frac{1}{k}\sum_{j=1}^{k}\left\{ \prod_{i=1}^b1\{e_{1,i,\pi_j} = e_{2,i,\pi_j}\}=1\right\},
\end{align}
where $e_{1,i,\pi_j}$ ($e_{2,i,\pi_j}$) denotes the $i$th lowest bit of $z_1$ ($z_2$), under the permutation $\pi_j$.

Following property of binomial distribution, we obtain
\begin{align}\notag
&\text{Var}\left(\hat{R}_b\right) = \frac{\text{Var}\left(\hat{E}_b\right)}{\left[1-C_{2,b}\right]^2} = \frac{1}{k}\frac{E_b(1-E_b)}{\left[1-C_{2,b}\right]^2}\\\label{eqn_Var_b}
=&\frac{1}{k}\frac{\left[C_{1,b}+(1-C_{2,b})R\right]\left[1-C_{1,b}-(1-C_{2,b})R\right]}{\left[1-C_{2,b}\right]^2}
\end{align}

For large $b$ (i.e., $A_{1,b}, A_{2,b}\rightarrow 0$ and $C_{1,b}, C_{2,b}\rightarrow 0$), $\text{Var}\left(\hat{R}_b\right)$ converges to the variance of $\hat{R}_{M}$, the estimator for the original minwise hashing:
\begin{align}\notag
\lim_{b\rightarrow \infty} \text{Var}\left(\hat{R}_b\right) = \frac{R(1-R)}{k} =\text{Var}\left(\hat{R}_M\right)
\end{align}


\subsection{The Variance-Space Trade-off}

 As we decrease $b$, the space needed for storing each ``sample'' will be smaller; the estimation variance (\ref{eqn_Var_b}) at the same sample size $k$, however, will increase.

%

 This variance-space trade-off can be precisely quantified by the \textbf{\em storage factor} $B(b;R,r_1, r_2)$:
\begin{align}\notag
&B(b;R,r_1,r_2) = b\times \text{Var}\left(\hat{R}_b\right)\times k \\\label{eqn_B(b)} =&\frac{b\left[C_{1,b}+(1-C_{2,b})R\right]\left[1-C_{1,b}-(1-C_{2,b})R\right]}{\left[1-C_{2,b}\right]^2}.
\end{align}
Lower $B(b)$ values are more desirable.

\begin{figure}[h]
\begin{center}
\mbox{
{\includegraphics[width = 1.8  in]{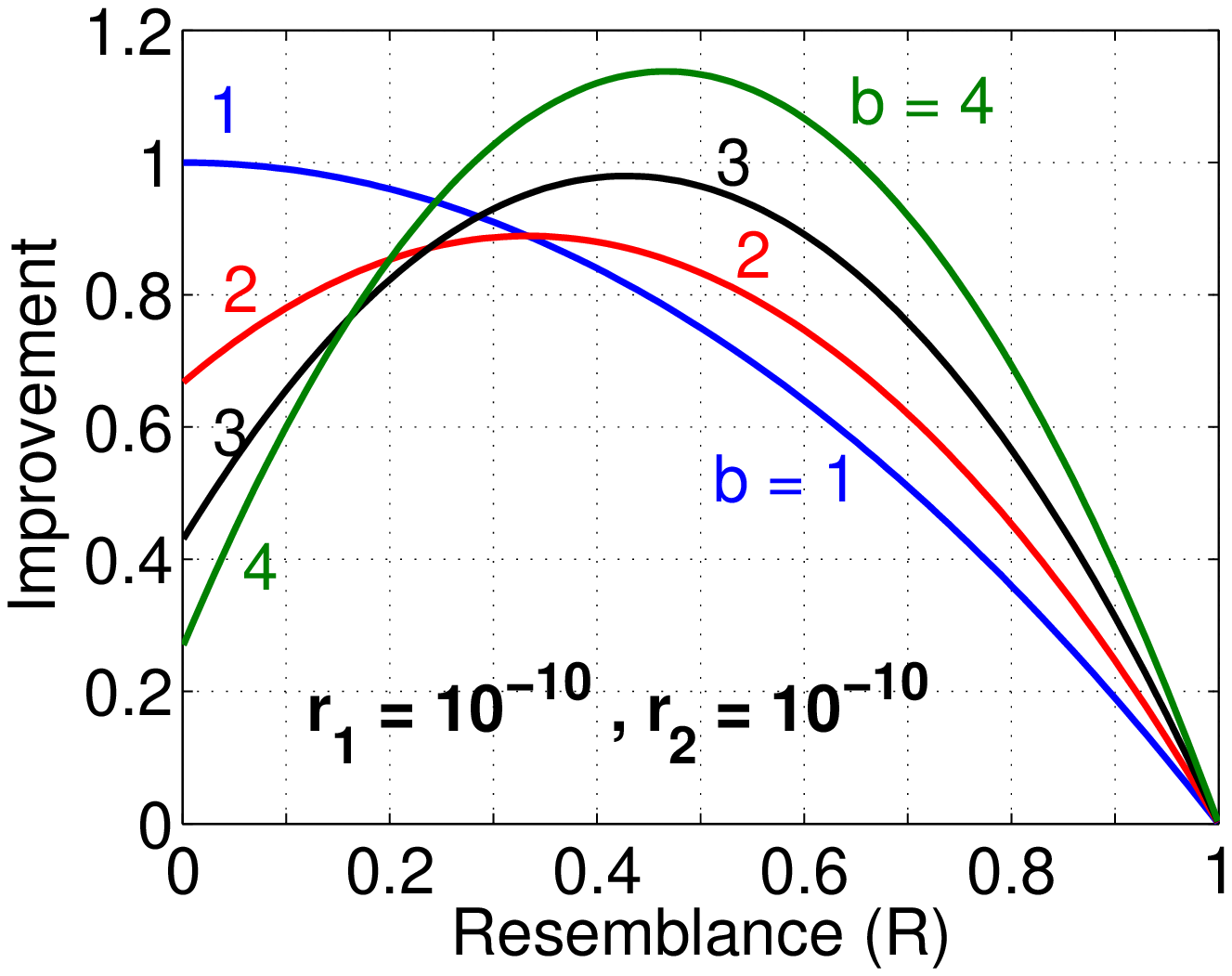}}\hspace{-0.13in}
{\includegraphics[width = 1.8  in]{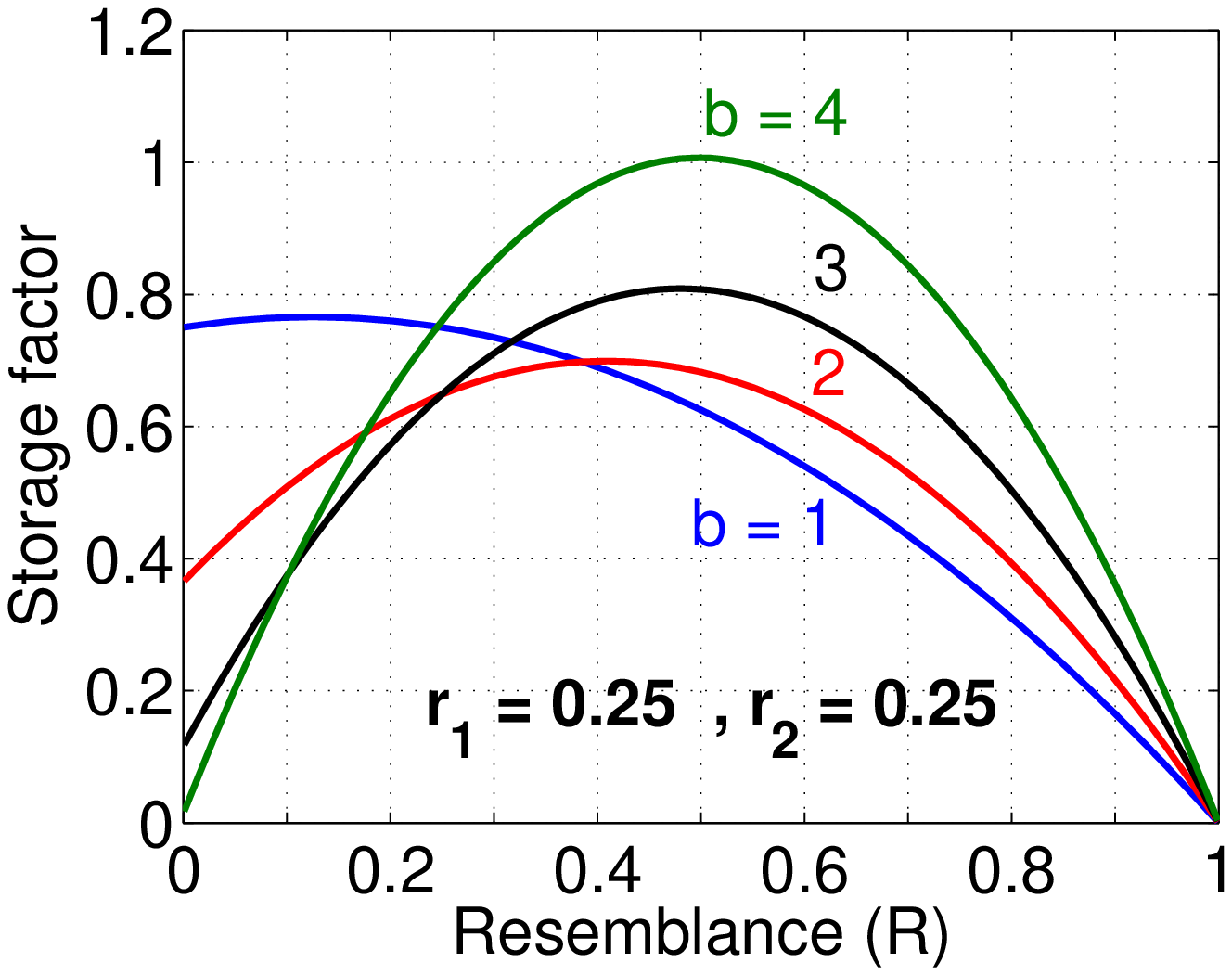}}}

\mbox{
{\includegraphics[width = 1.8  in]{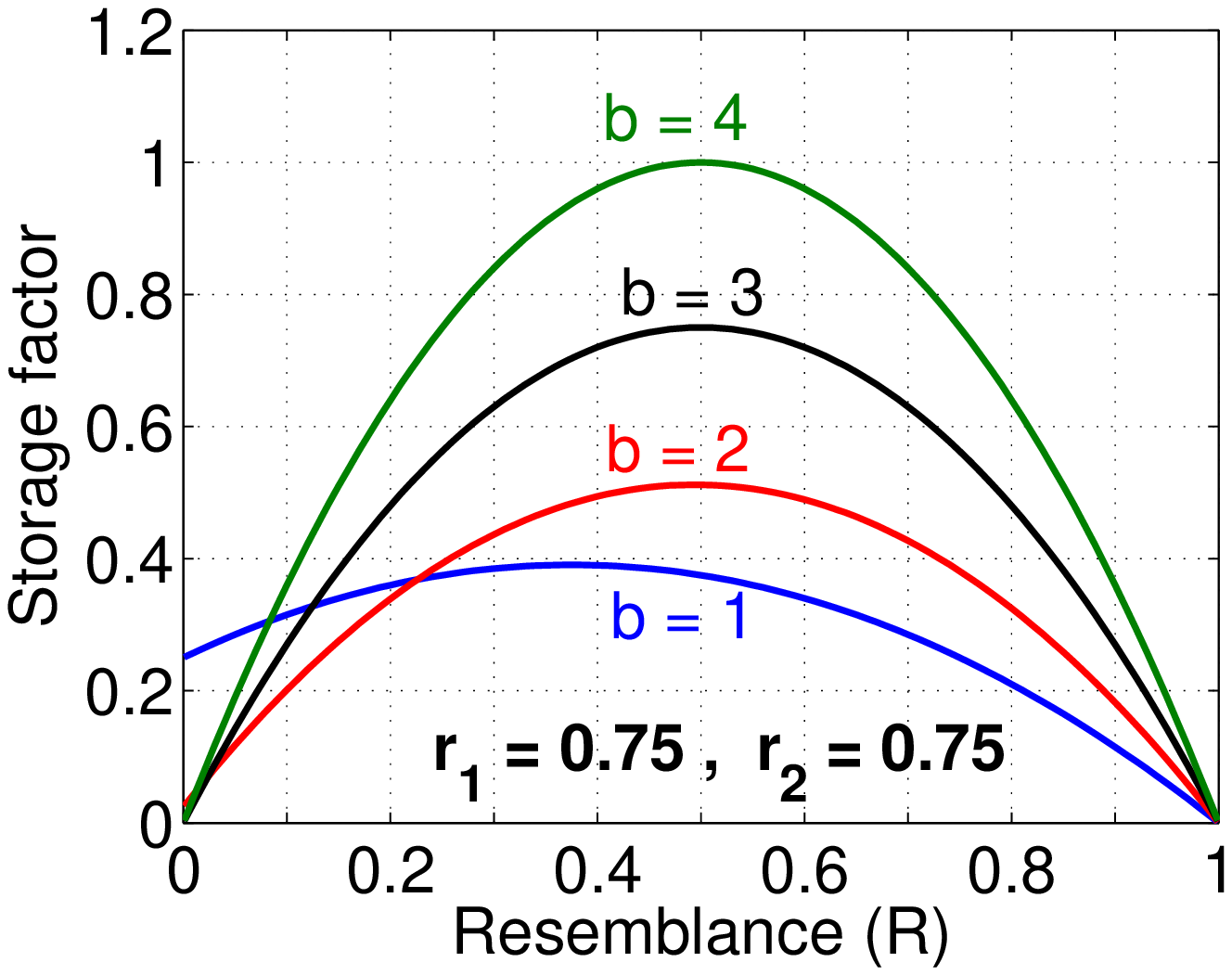}}\hspace{-0.13in}
{\includegraphics[width = 1.8  in]{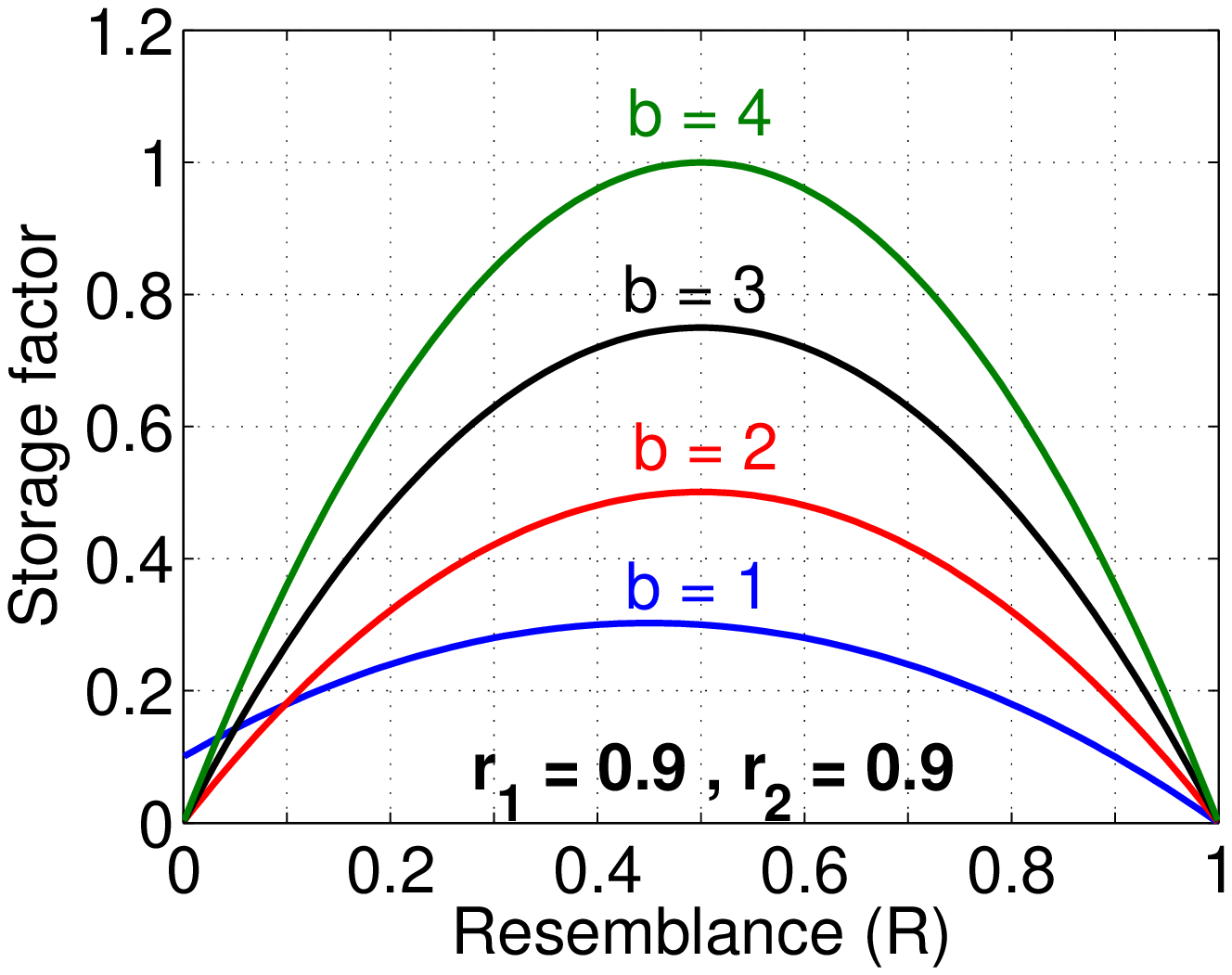}}}
\end{center}
\caption{$B(b;R,r_1,r_2)$ in (\ref{eqn_B(b)}). The lower the better. }\label{fig_B(b)}
\end{figure}

Figure \ref{fig_B(b)} plots $B(b)$ for the whole range of $R\in(0,1)$ and four selected $r_1=r_2$ values (from $10^{-10}$ to 0.9). Figure \ref{fig_B(b)} shows that when the ratios, $r_1$ and $r_2$, are close to 1, it is  always desirable to use $b=1$, almost for the whole range of $R$. However, when $r_1$ and $r_2$ are close to 0, using $b=1$ has the advantage when about $R\geq0.4$. For small $R$ and $r_1$, $r_2$, it may be more advantageous to user lager $b$, e.g., $b\geq2$.

The ratio of storage factors, $\frac{B(b_1;R,r_1,r_2)}{B(b_2;R,r_1,r_2)}$, directly measures how much improvement using $b=b_2$ (e.g., $b_2=1$) can have over using $b=b_1$ (e.g., $b_1=64$ or 32).

Some algebraic manipulation yields the following Theorem.
\begin{theorem}\label{The_Br}
If $r_1 = r_2$ and $b_1>b_2$, then
\begin{align}
\frac{B(b_1;R,r_1,r_2)}{B(b_2;R,r_1,r_2)} = \frac{b_1}{b_2}\ \frac{A_{1,b_1}(1-R)+R}{A_{1,b_2}(1-R)+R} \ \frac{1-A_{1,b_2}}{1-A_{1,b_1}},
\end{align}
is a monotonically increasing function of $R\in[0,1]$.

If $R\rightarrow 1$ (which implies $r_1\rightarrow r_2$), then
\begin{align}
&\frac{B(b_1;R,r_1,r_2)}{B(b_2;R,r_1,r_2)}\rightarrow \frac{b_1}{b_2}\frac{1-A_{1,b_2}}{1-A_{1,b_1}}.
\end{align}

If $r_1=r_2$, $b_2=1$, $b_1\geq32$ (hence we treat $A_{1,b}=0$), then
\begin{align}\label{eqn_Bb21}
\frac{B(b_1;R,r_1,r_2)}{B(1;R,r_1,r_2)} = b_1 \frac{R}{R+1-r_1}
\end{align}
\textbf{Proof:}\ \ We omit the proof due to its simplicity.$\Box$
\end{theorem}

Suppose the original minwise hashing used $b=64$ bits to store each sample, then the maximum improvement of the $b$-bit minwise hashing would be 64-fold, attained when $r_1=r_2=1$ and $R=1$, according to (\ref{eqn_Bb21}). In the least favorable situation, i.e., $r_1, r_2\rightarrow 0$, the improvement will still be $\frac{64R}{R+1}$-fold, which is $\frac{64}{3}=21.3$-fold when $R=0.5$.

\begin{figure}[h]
\begin{center}
\mbox{
{\includegraphics[width = 1.8  in]{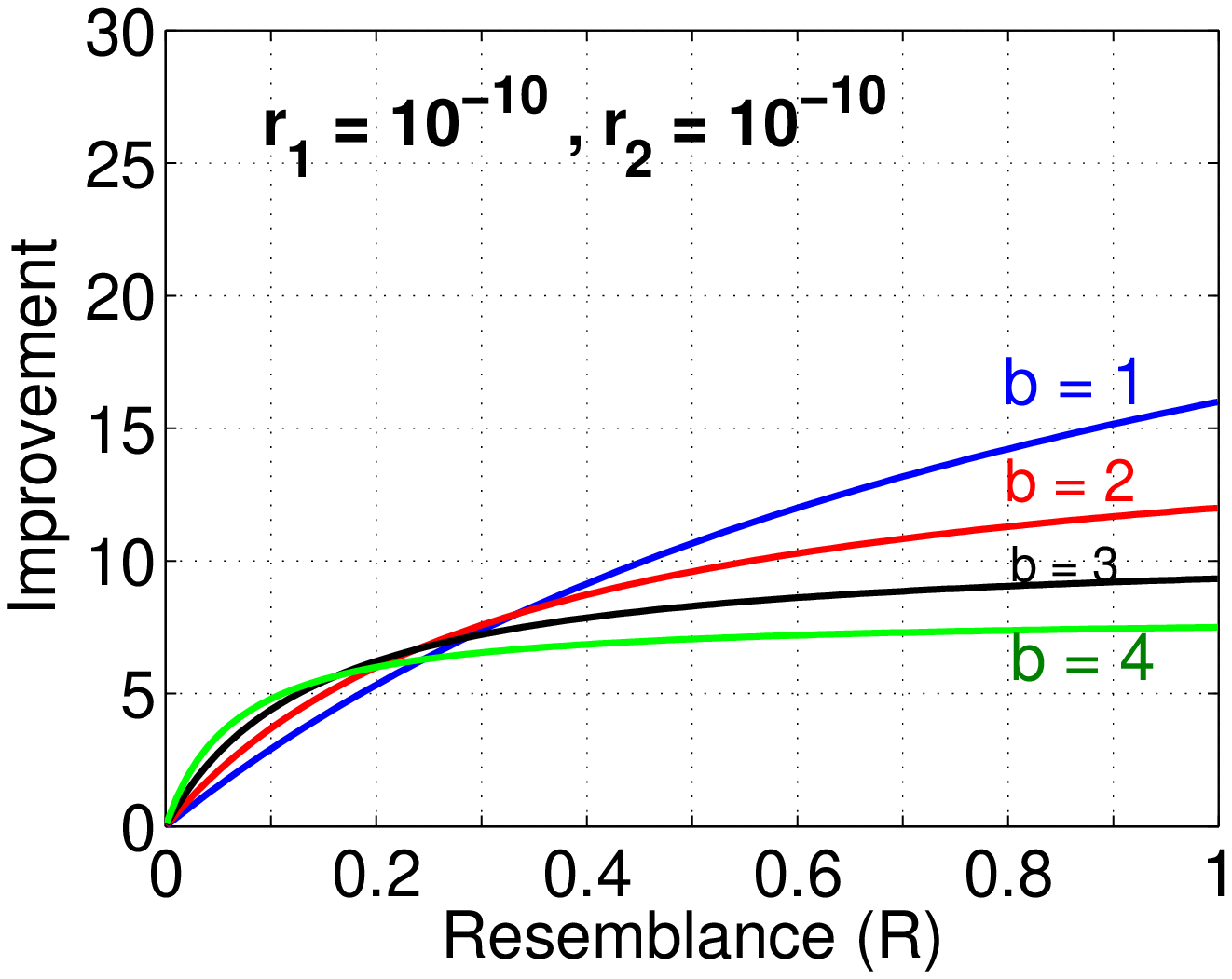}}\hspace{-0.13in}
{\includegraphics[width = 1.8  in]{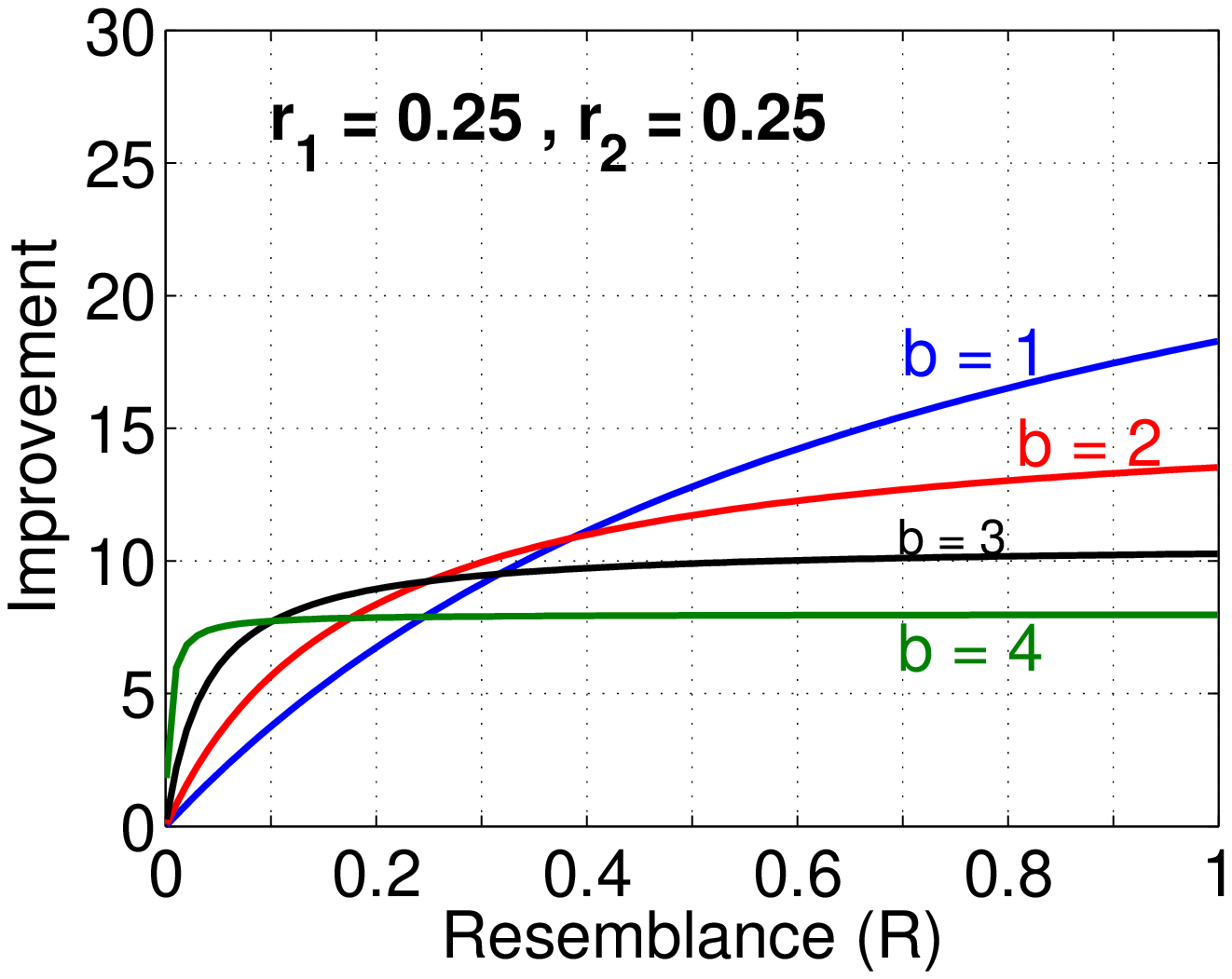}}}

\mbox{
{\includegraphics[width = 1.8  in]{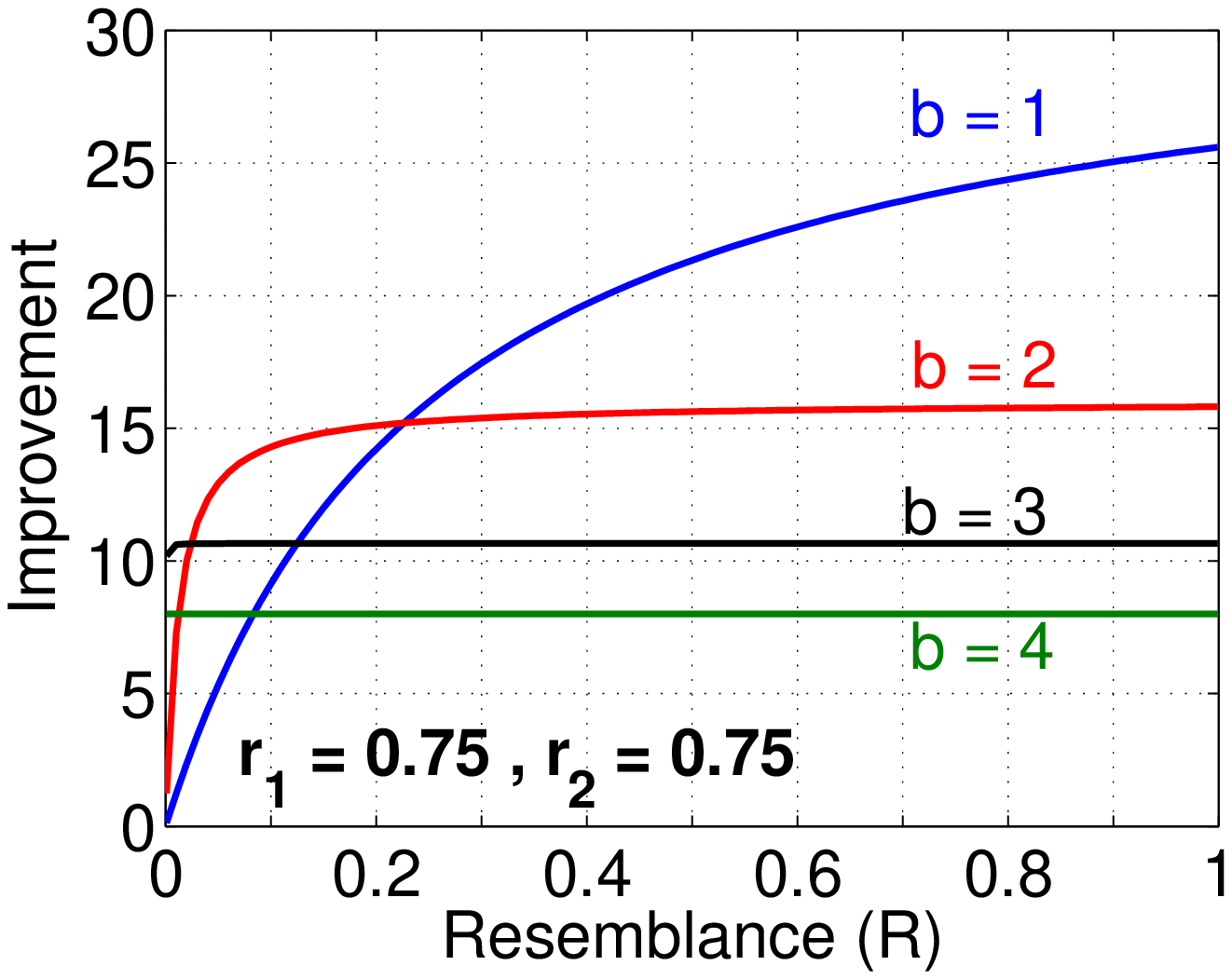}}\hspace{-0.13in}
{\includegraphics[width = 1.8  in]{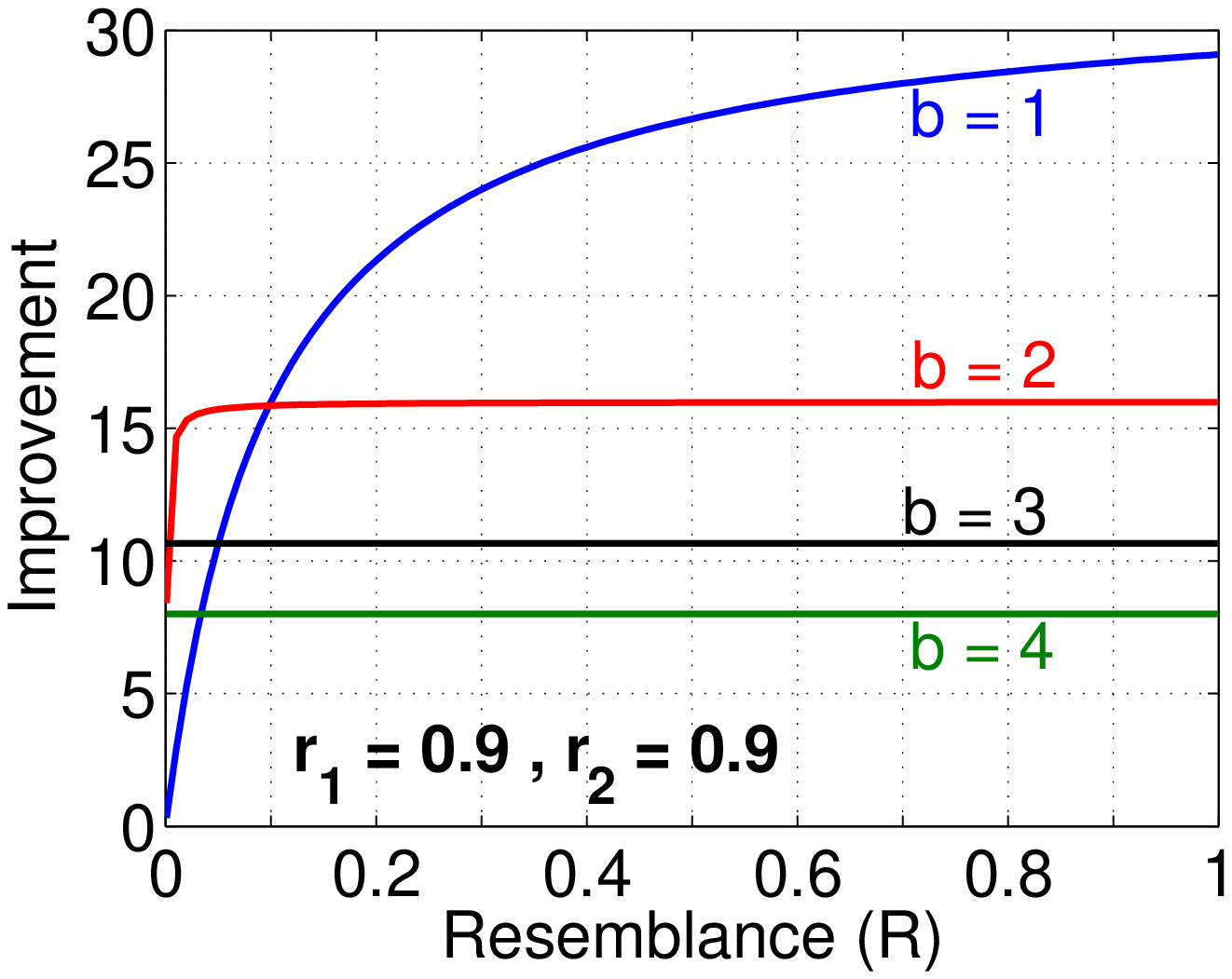}}}
\end{center}
\caption{$\frac{B(32)}{B(b)}$, the relative storage improvement of using $b=1,2,3,4$ bits, compared to using $32$ bits. }\label{fig_B(32)/B(b)}\vspace{-0.in}
\end{figure}

Figure \ref{fig_B(32)/B(b)} plots $\frac{B(32)}{B(b)}$, to directly visualize the relative improvement. The plots are, of course, consistent with what Theorem \ref{The_Br} would predict.

\section{Experiments}

We conducted three experiments. The first two experiments were based on a set of 2633 words, extracted from a chuck of MSN Web pages. Our third experiment used a set of 10000 news articles crawled from the Web.

Our first experiment is a sanity check, to verify the correctness of the theory. That is, our proposed estimator $\hat{R}_b$, (\ref{eqn_R_b}), is  unbiased and its variance (the same as the mean square error (MSE)) follows the prediction by our formula in (\ref{eqn_Var_b}).

\subsection{Experiment 1}

 For our first experiment, we selected 10 pairs of words to validate the theoretical estimator $\hat{R}_{M}$ and the variance formula $\text{Var}\left(\hat{R}_M\right)$, derived in in Sec. \ref{sec_estimator}.

Table \ref{tab_10pairs} summarizes the data and also provides the theoretical improvements $\frac{B(32)}{B(1)}$ and $\frac{B(64)}{B(1)}$. For each word, the data consist of the document IDs in which that word occurs. The words were selected to include highly frequent word pairs (e.g., ``OF-AND''), highly rare word pairs (e.g., ``GAMBIA-KIRIBATI''), highly unbalanced pairs (e.g., ''A-Test''), highly similar pairs (e.g, ``KONG-HONG''), as well as word pairs that are not quite similar (e.g., ``LOW-PAY'').

\begin{table}[h]
\caption{\small Ten pairs of words used in the experiments for validating the estimator and theoretical variance (\ref{eqn_Var_b}). Since the variance is determined by $r_1$, $r_2$, and $R$, words were selected to ensure a good coverage of scenarios.
 }
\begin{center}{\tiny
\begin{tabular}{l l l l l l l }
\hline \hline
Word 1 & Word 2 &$r_1$  &$r_2$  &$R$ &$\frac{B(32)}{B(1)}$ &$\frac{B(64)}{B(1)}$ \\\hline
KONG & HONG &0.0145 &0.0143 &0.925 &15.5 &31.0\\
RIGHTS & RESERVED  &0.187 &0.172 &0.877 &16.6&32.2\\
OF & AND &0.570 &0.554 &0.771 &20.4 &40.8\\
GAMBIA &KIRIBATI &0.0031   &0.0028  &0.712 &13.3 &26.6\\
UNITED &STATES &0.062 &0.061 &0.591 &12.4 &24.8\\
SAN &FRANCISCO &0.049     &0.025 &0.476 &10.7 &21.4\\
CREDIT & CARD &0.046 &0.041 &0.285 &7.3 &14.6\\
TIME & JOB &0.189        &0.05 &0.128 &4.3 &8.6\\
LOW  & PAY  &0.045        &0.043 &0.112 &3.4 &6.8\\
A  & TEST &0.596        &0.035 &0.052 &3.1 &6.2\\
\hline\hline
\end{tabular}
}
\end{center}
\label{tab_10pairs}
\end{table}

We estimate the resemblance using the original minwise hashing estimator $\hat{R}_M$ and the $b$-bit version for $b=1, 2, 3$.

\begin{figure}[h!]
\begin{center}
\mbox{
\includegraphics[width = 1.8in]{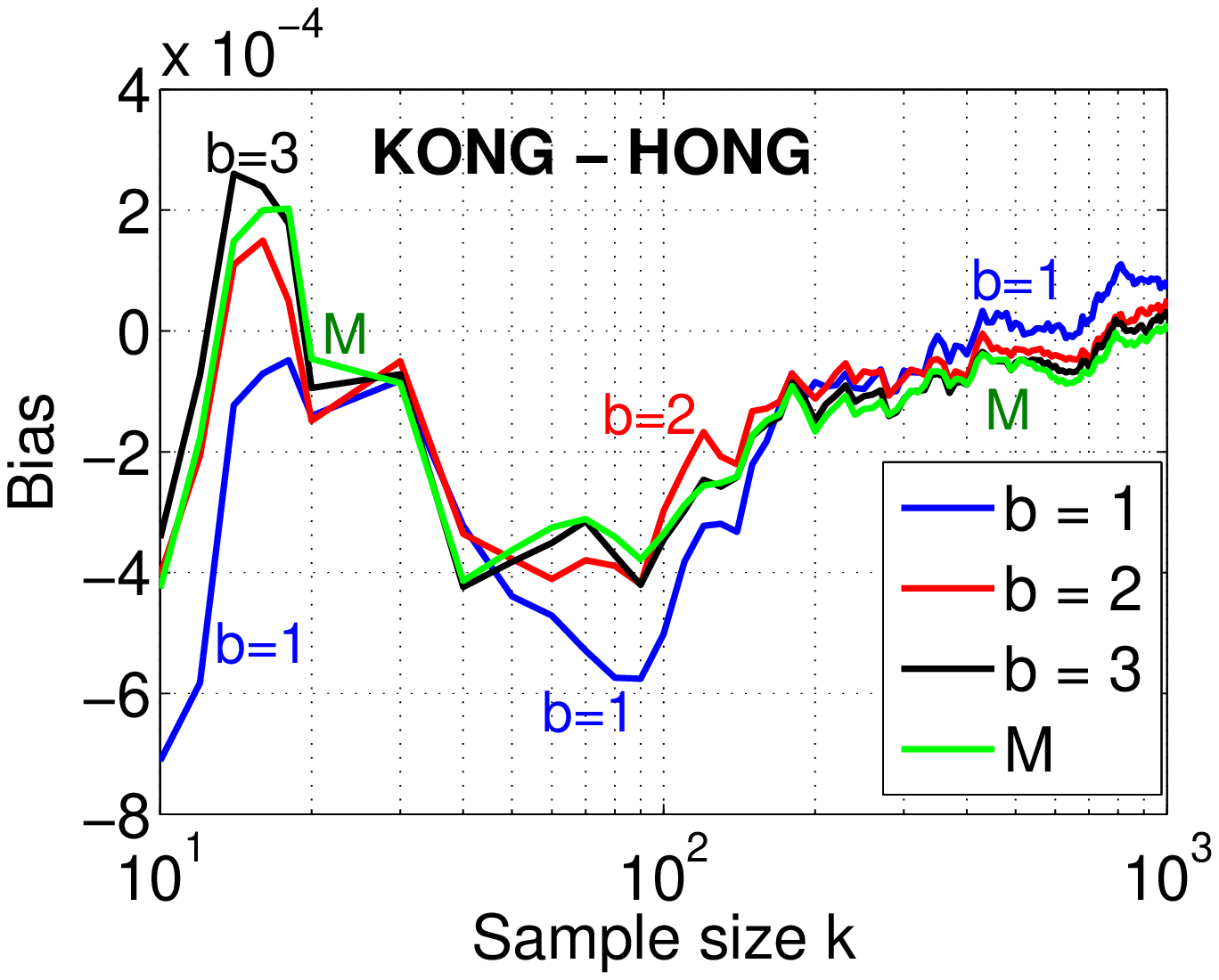}\hspace{-0.13in}
\includegraphics[width = 1.8in]{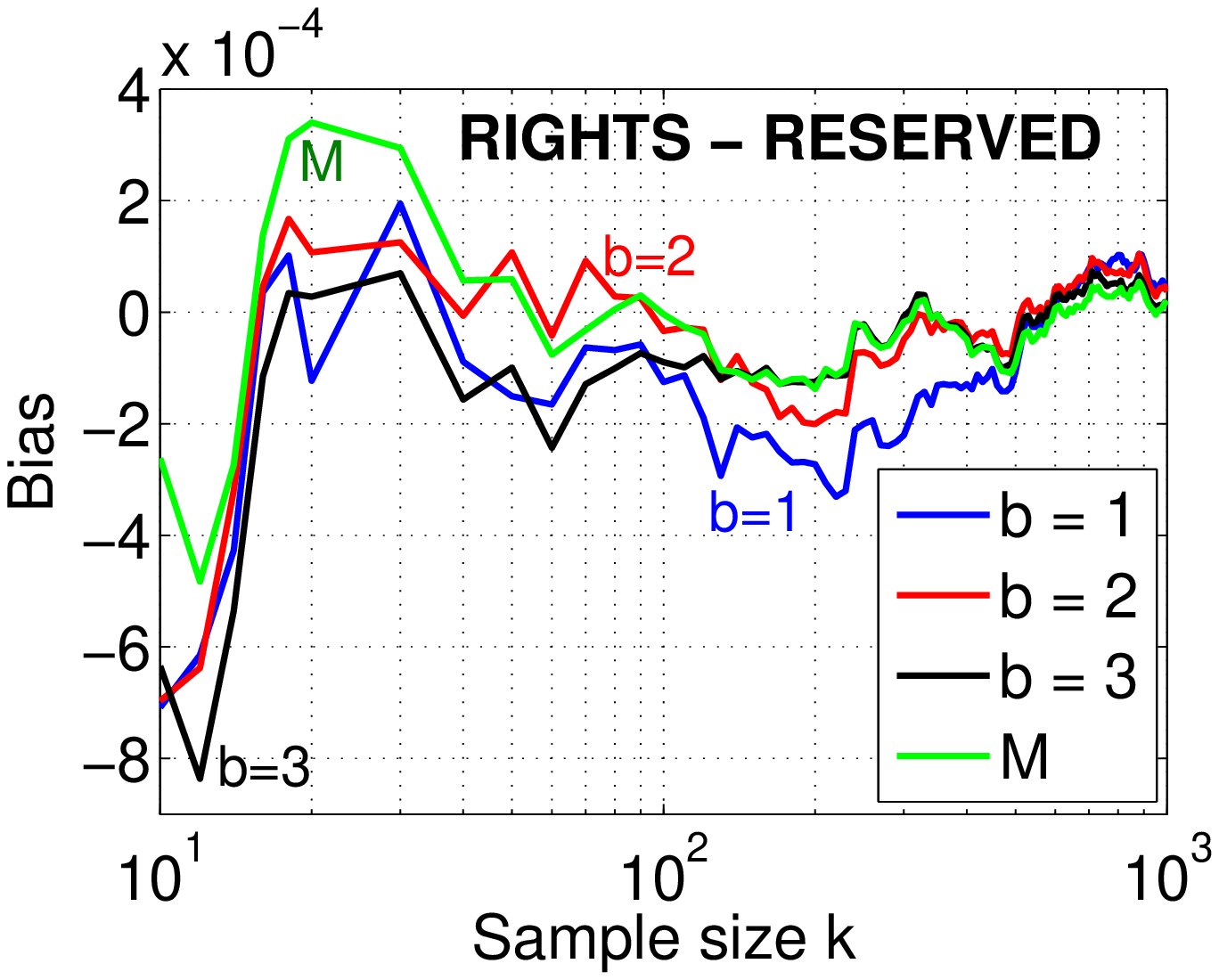}}\vspace{-0.in}
\mbox{
\includegraphics[width = 1.8in]{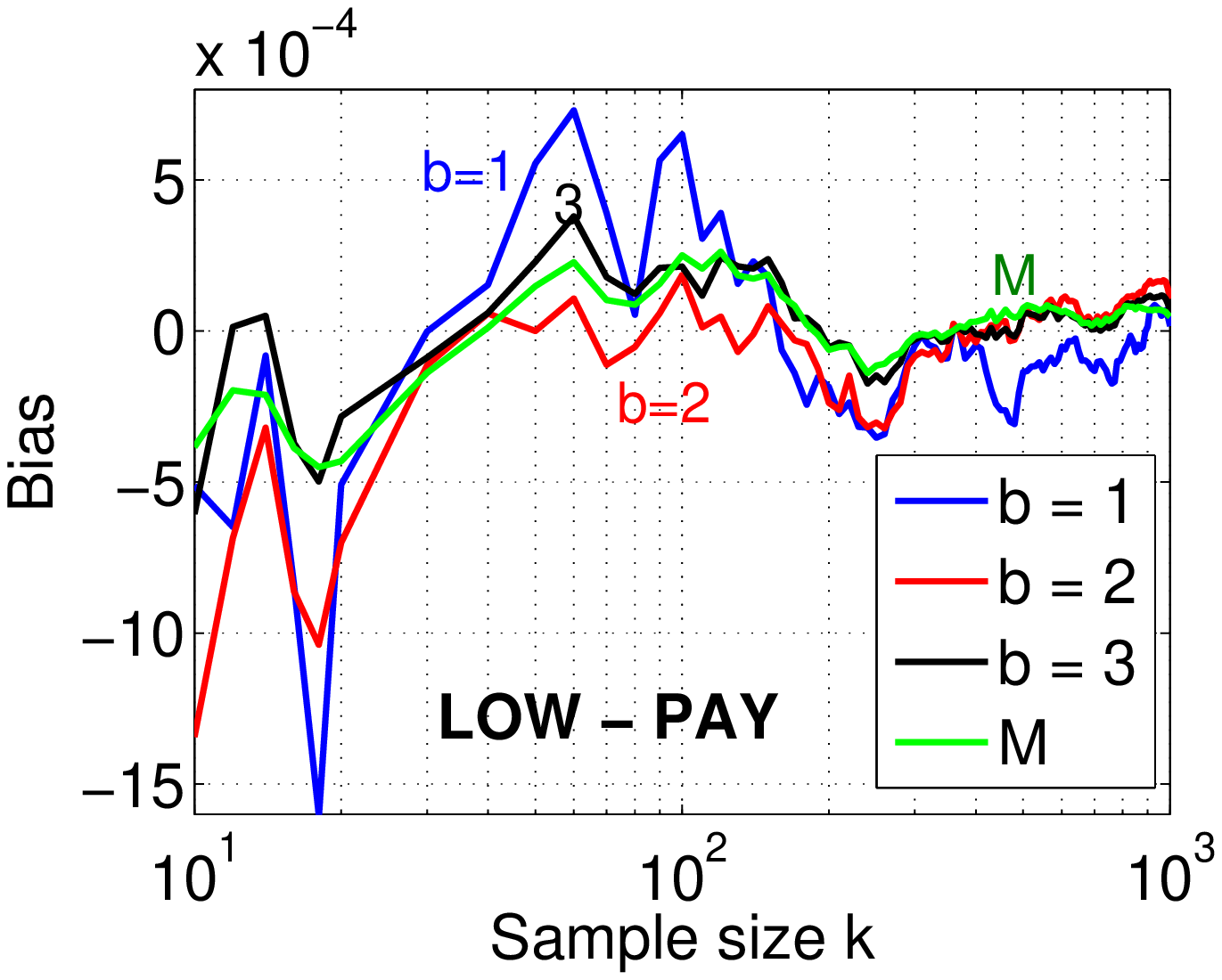}\hspace{-0.13in}
\includegraphics[width = 1.8in]{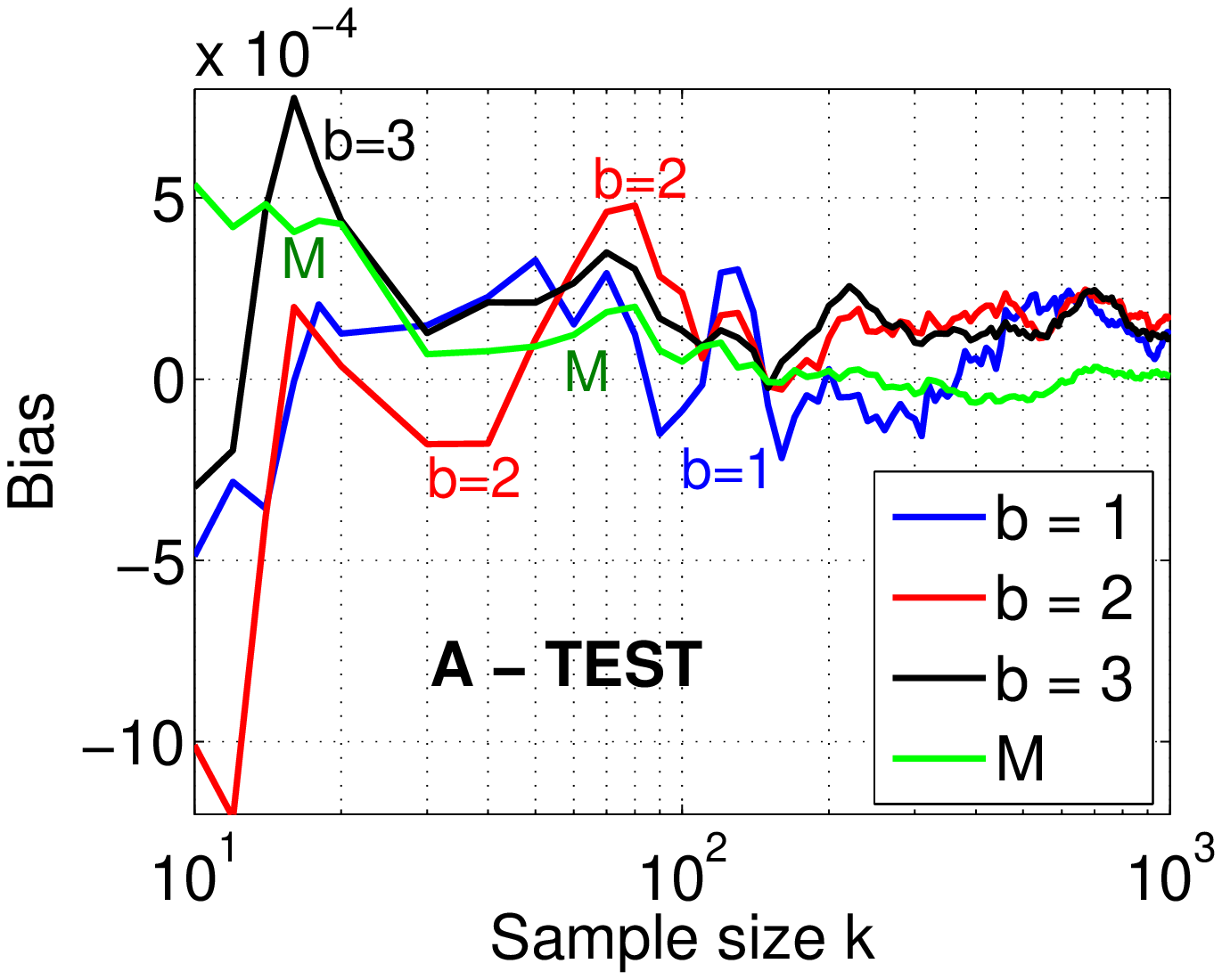}}\vspace{-0.25in}
\end{center}
\caption{\textbf{Biases}, empirically estimated from 25000 simulations  at each sample size $k$. ``M'' denotes the original minwise hashing.  }\label{fig_bias}\vspace{0.2in}
\end{figure}

Figure \ref{fig_bias} presents the estimation biases  for selected 4 word pairs. Theoretically, the estimator $\hat{R}_b$ is unbiased. Figure \ref{fig_bias} verifies this fact as the empirical biases are  all very small and no systematic biases can be observed.

\begin{figure}[h!]
\begin{center}
\mbox{
\includegraphics[width = 1.8in]{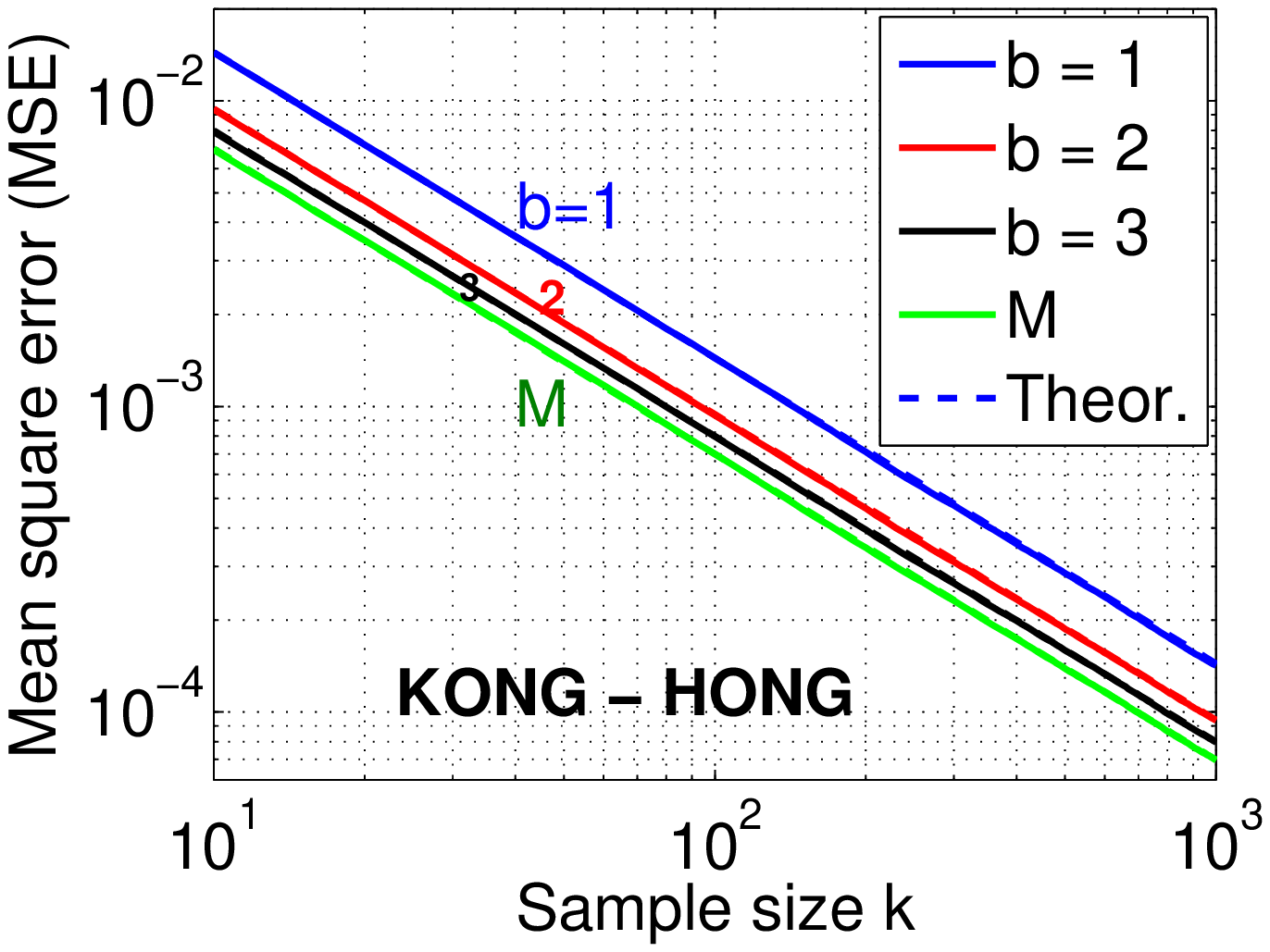}\hspace{-0.13in}
\includegraphics[width = 1.8in]{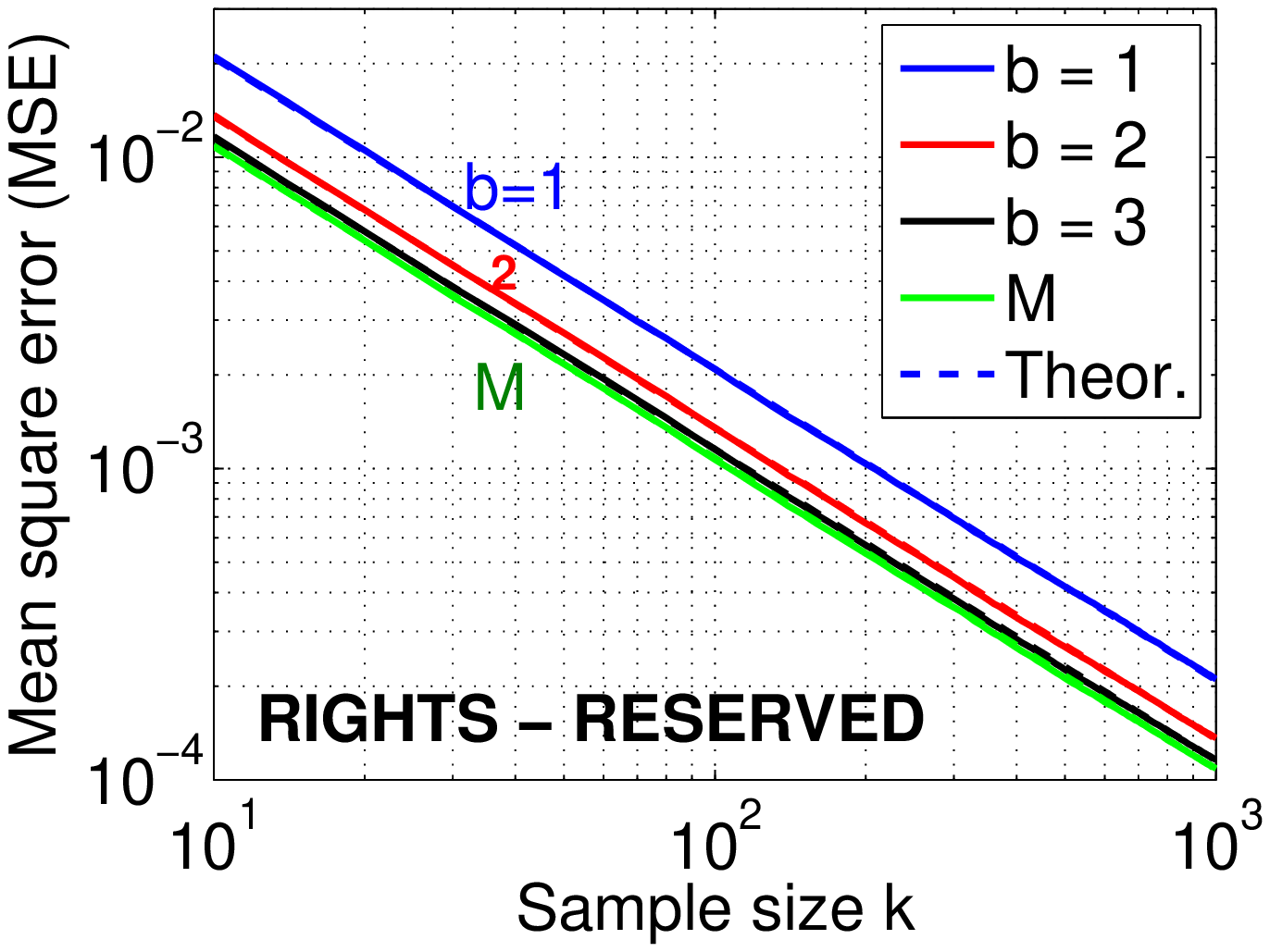}}\vspace{-0.0in}
\mbox{
\includegraphics[width = 1.8in]{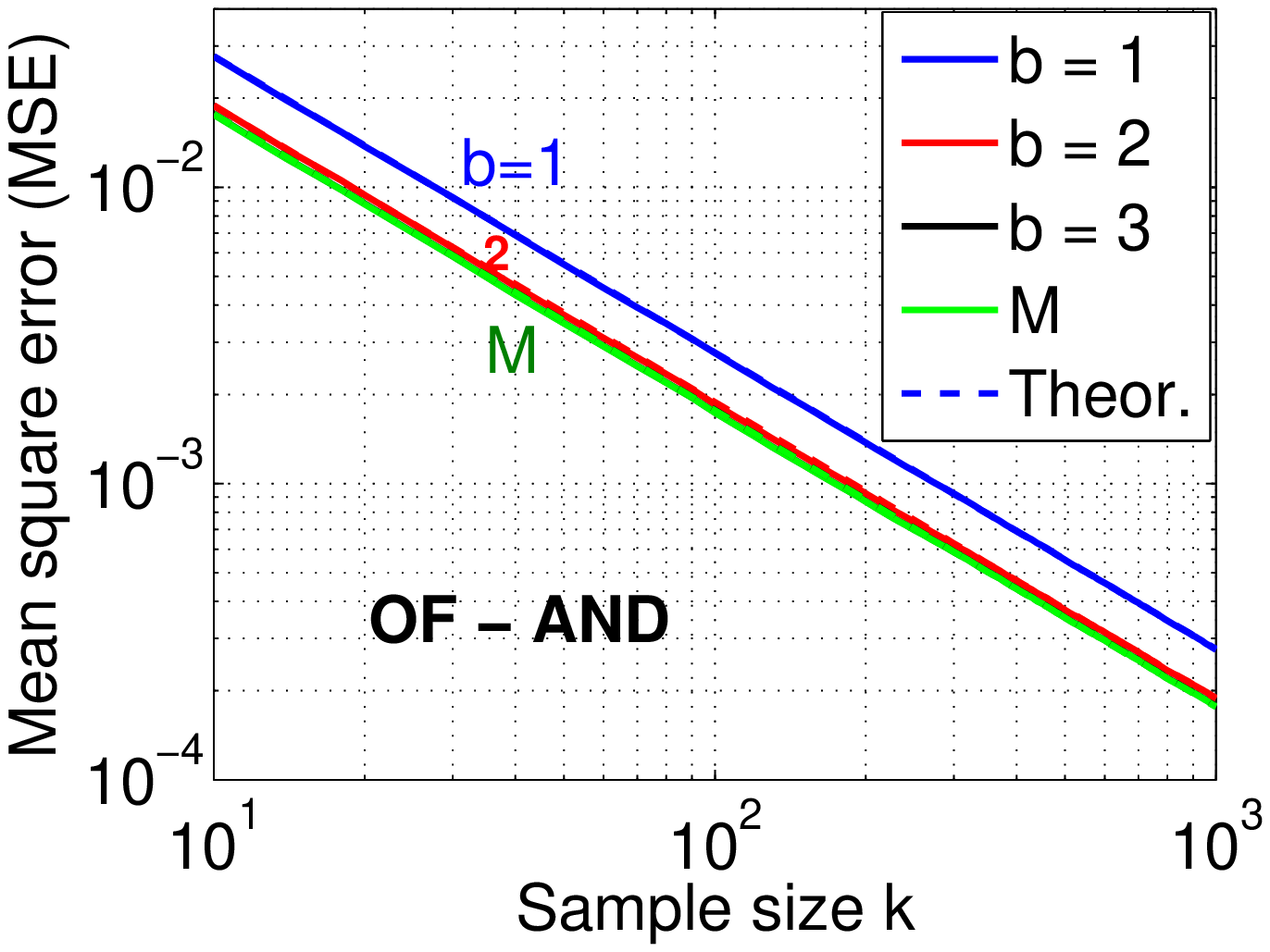}\hspace{-0.13in}
\includegraphics[width = 1.8in]{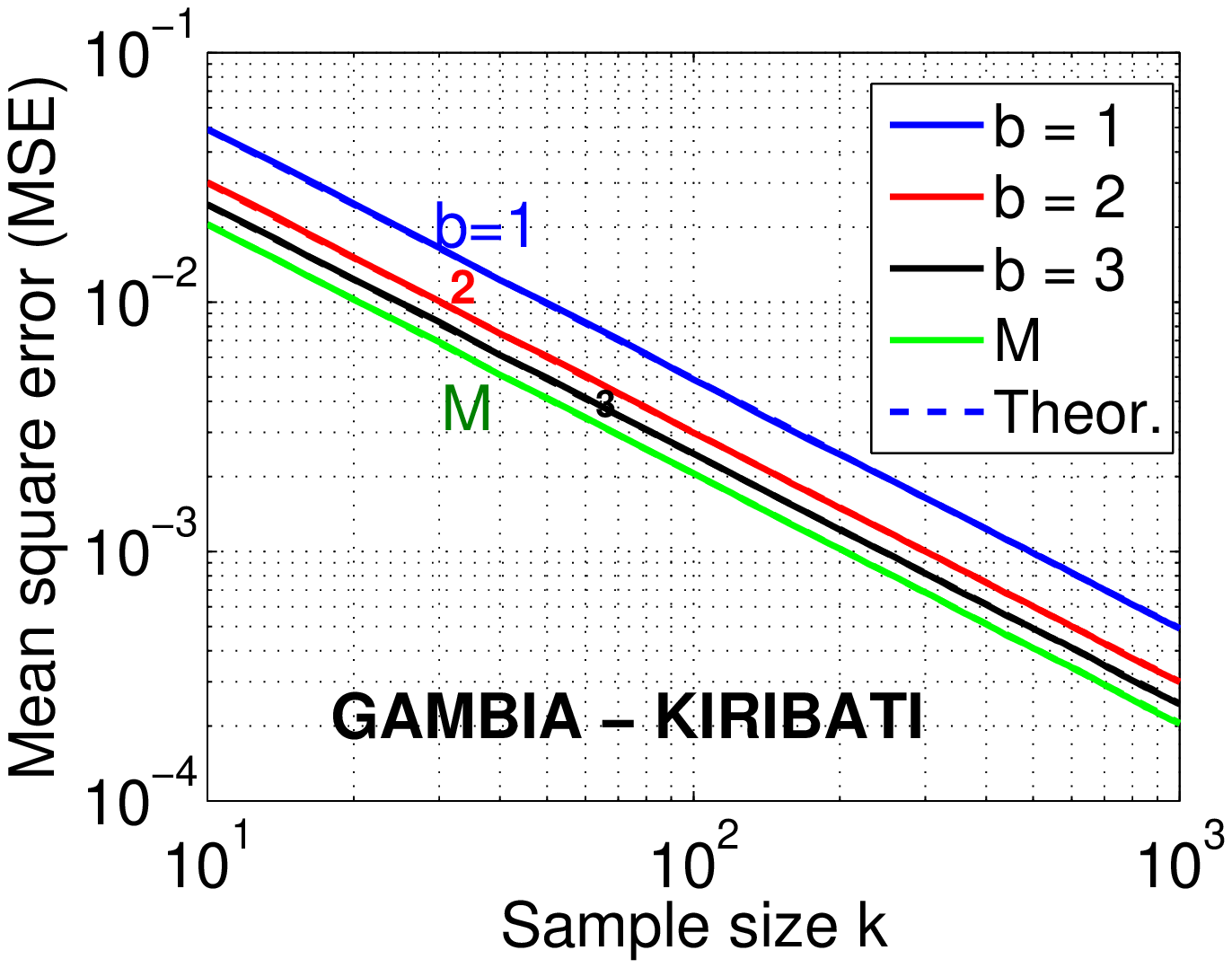}}\vspace{-0.in}
\mbox{
\includegraphics[width = 1.8in]{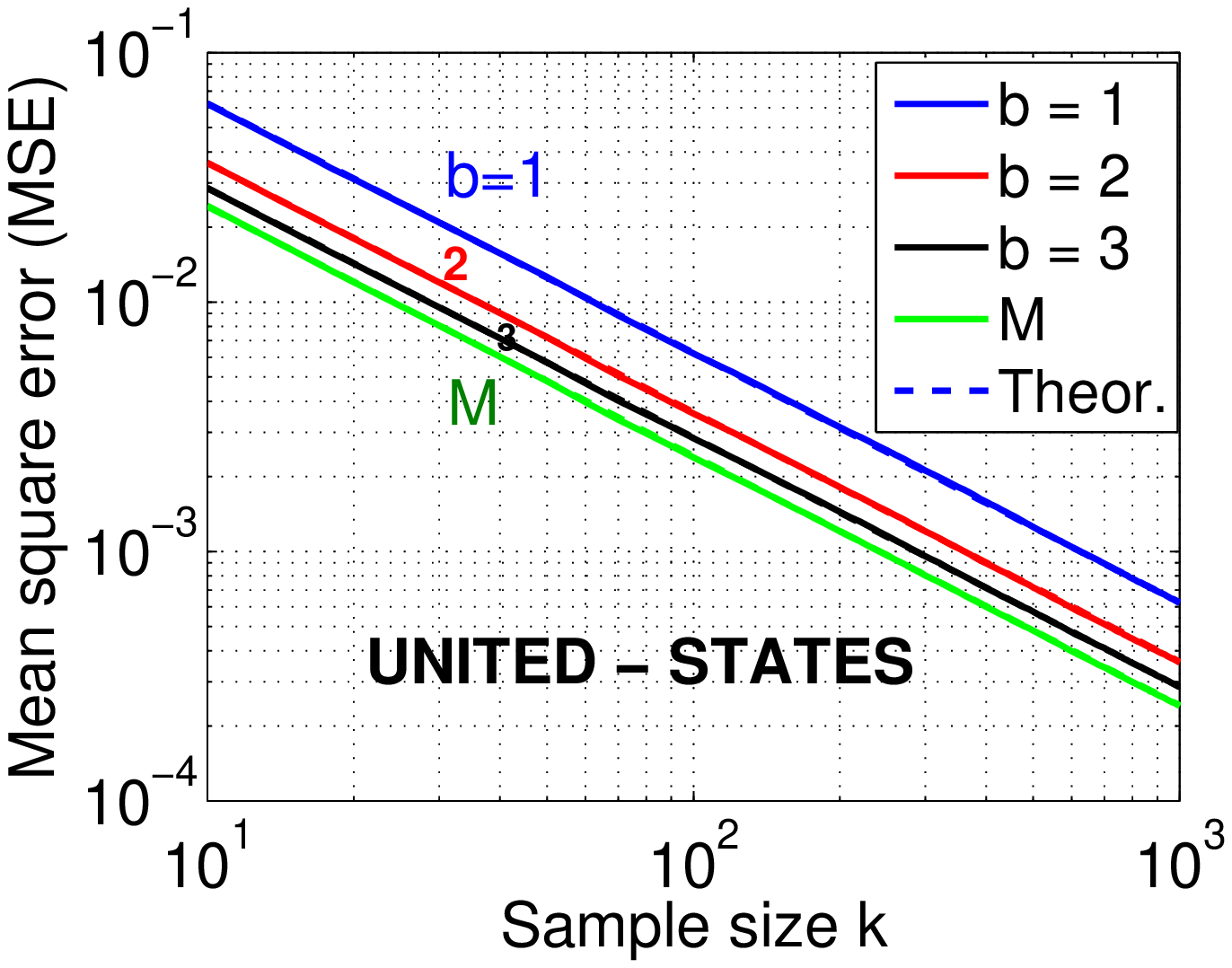}\hspace{-0.13in}
\includegraphics[width = 1.8in]{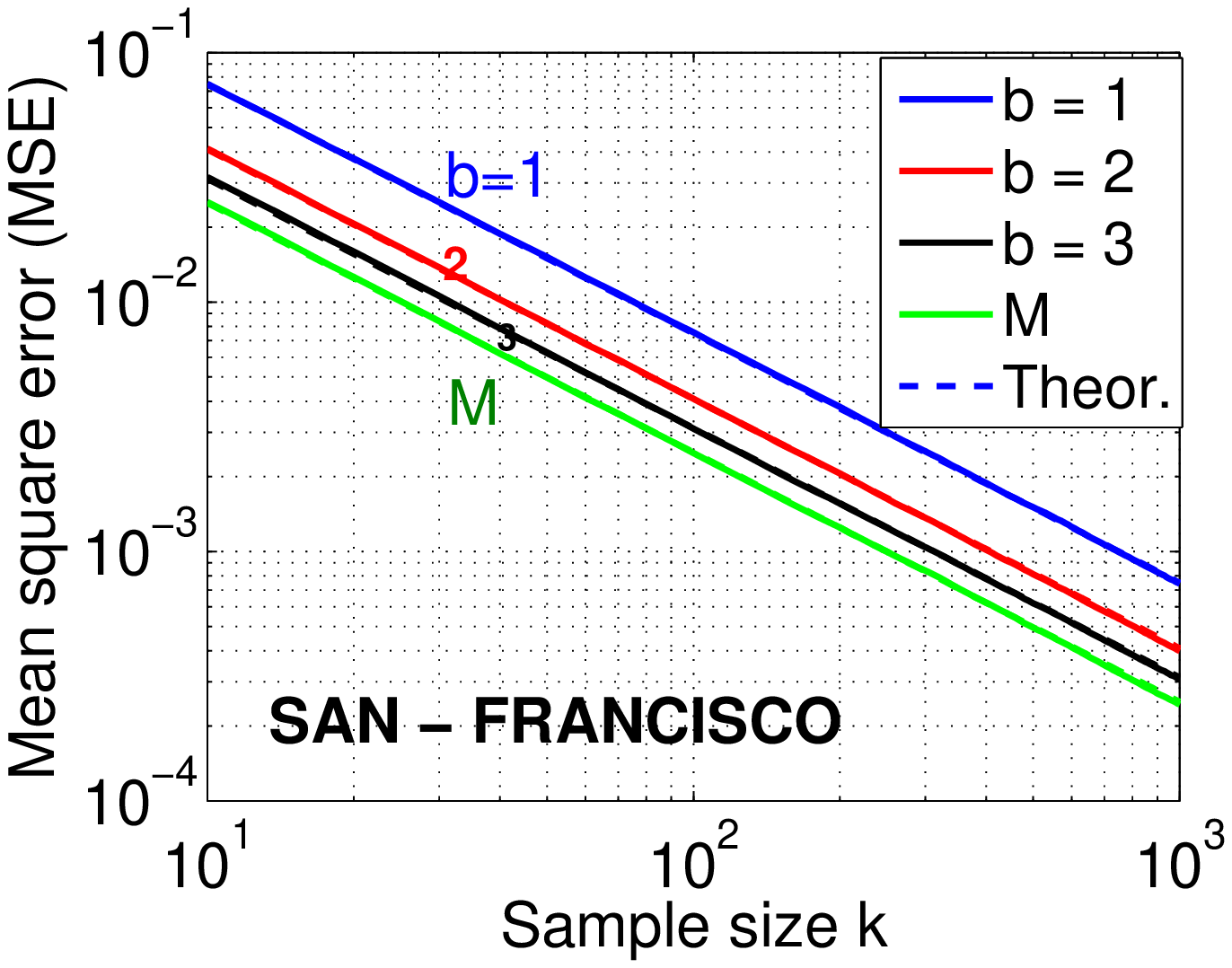}}\vspace{-0.in}
\mbox{
\includegraphics[width = 1.8in]{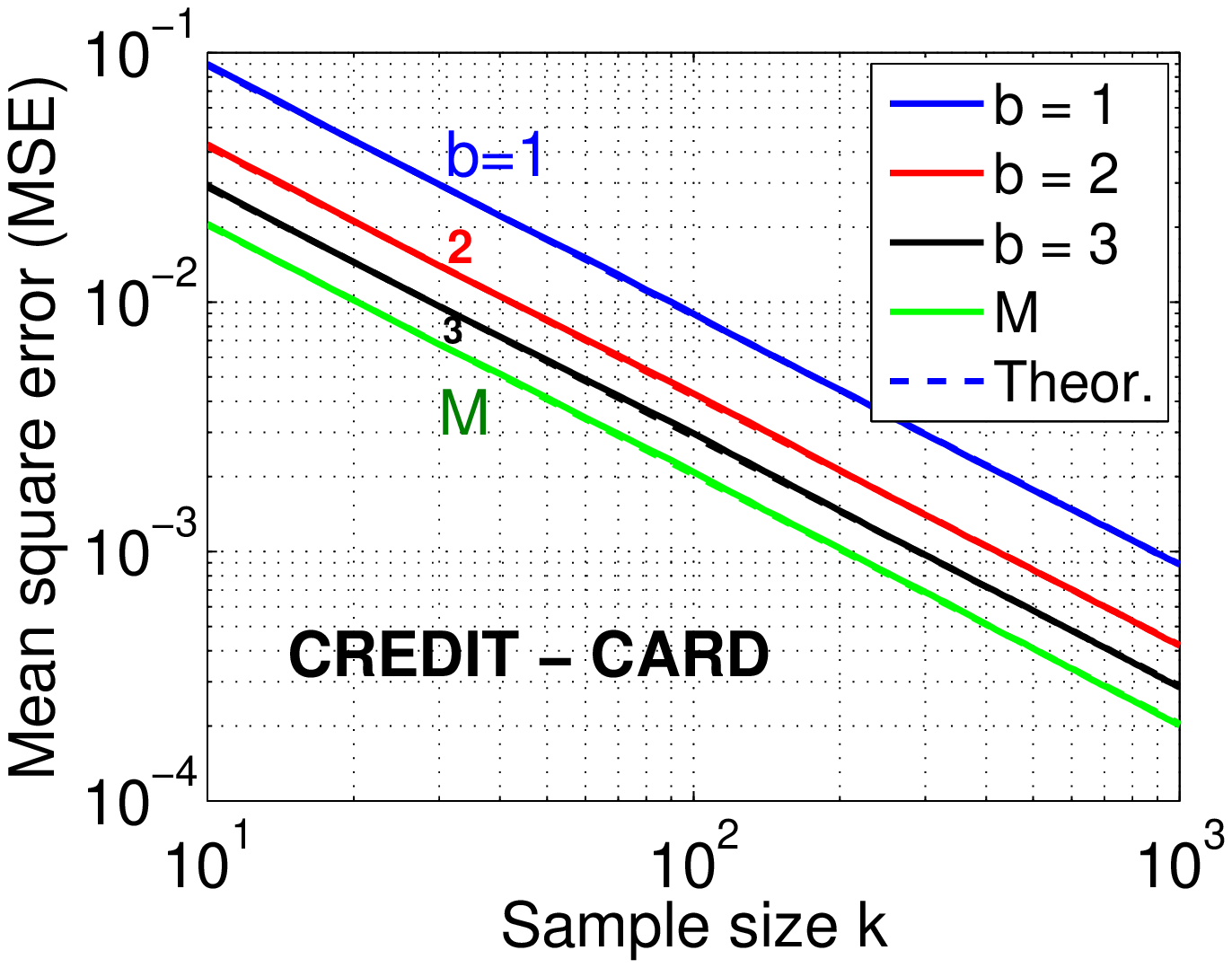}\hspace{-0.13in}
\includegraphics[width = 1.8in]{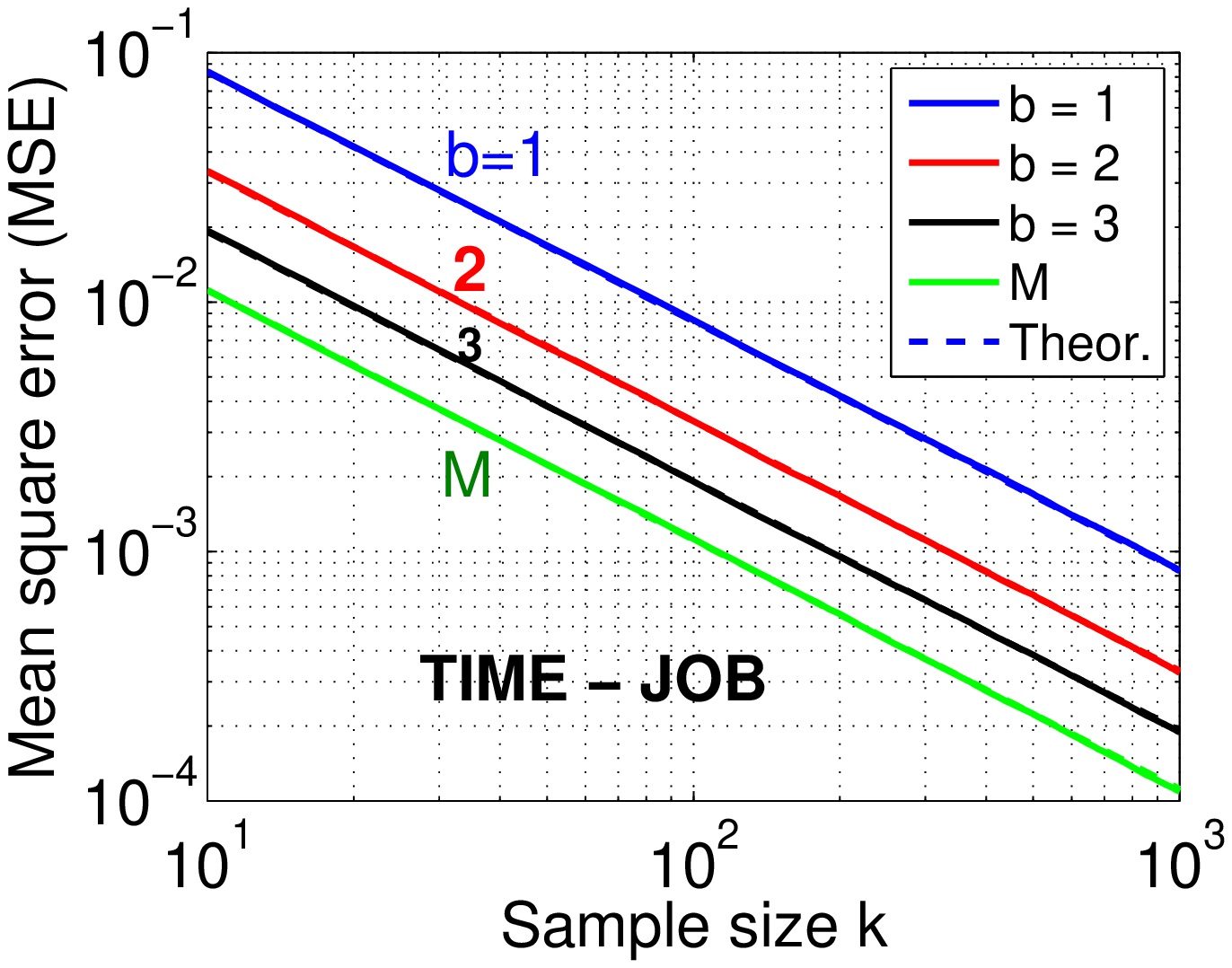}}\vspace{-0.in}
\mbox{
\includegraphics[width = 1.8in]{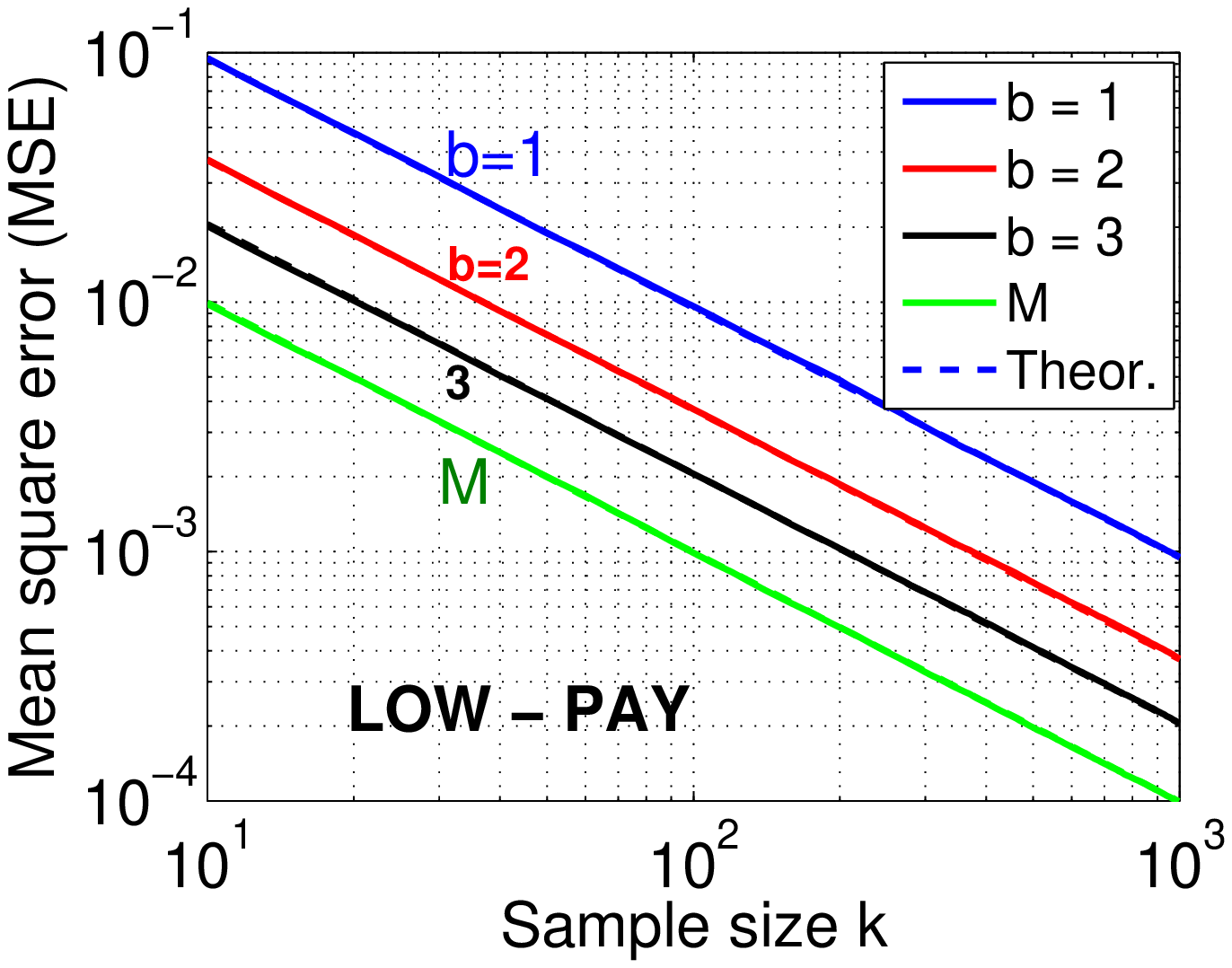}\hspace{-0.13in}
\includegraphics[width = 1.8in]{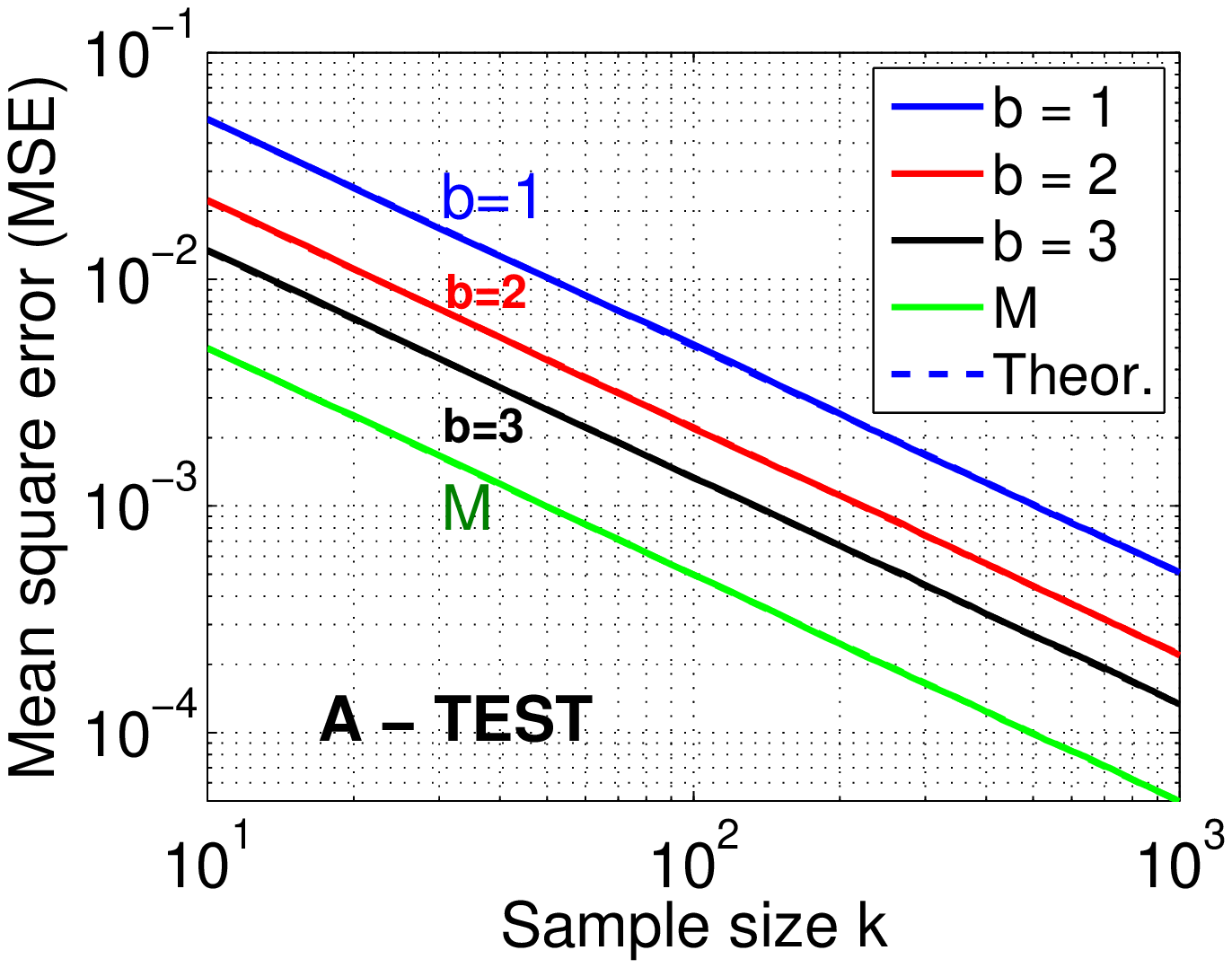}}\vspace{-0.25in}
\end{center}
\caption{\textbf{MSEs}, empirically estimated from simulations. ``M'' denotes the original minwise hashing. ``Theor.'' denotes the theoretical variances of $Var(\hat{R}_b)$ and $Var(\hat{R}_M)$. Those dashed curves, however, are invisible because the empirical results overlapped the theoretical predictions. At the same sample size $k$, we always have $Var(\hat{R}_1)>Var(\hat{R}_2)> Var(\hat{R}_3)>Var(\hat{R}_M)$. However, $\hat{R}_1$ only requires 1 bit per sample while $\hat{R}_2$ requires 2 bits, etc.  }\label{fig_mse}
\end{figure}

Figure \ref{fig_mse} plots the empirical mean square errors (MSE = variance + bias$^2$) and the theoretical variances (in dashed lines), for all 10 word pairs. However, all dashed lines overlapped with the corresponding solid curves. This figure satisfactorily illustrates  that the variance formula (\ref{eqn_Var_b}) is  accurate and $\hat{R}_b$ is indeed unbiased (because MSE=variance).

\clearpage
\subsection{Experiment 2}

This section presents an experiment for finding pairs whose resemblance values $\geq R_0$.  This experiment is in the same spirit as \cite{Proc:Broder,Proc:Broder_WWW97}. We use all 2633 words (i.e., 3465028 pairs) as described in Experiment 1. We use both $\hat{R}_M$ and $\hat{R}_b$ ($b=1, 2, 3, 4$) and then present the precision and recall curves, at different values of thresholds $R_0$ and sample sizes $k$.

\begin{figure}[h!]
\begin{center}
\mbox{
\includegraphics[width = 1.78in]{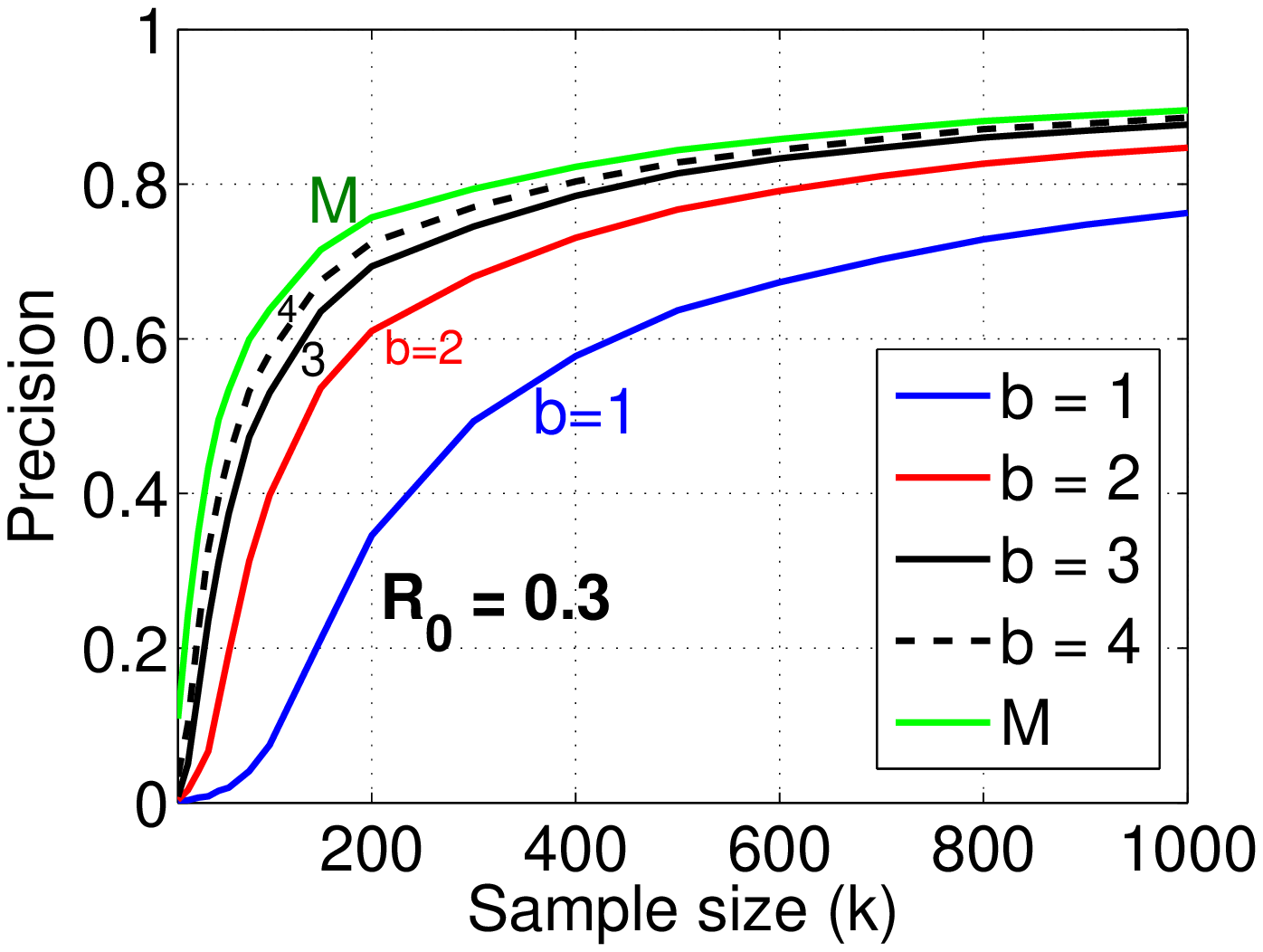}\hspace{-0.12in}
\includegraphics[width = 1.78in]{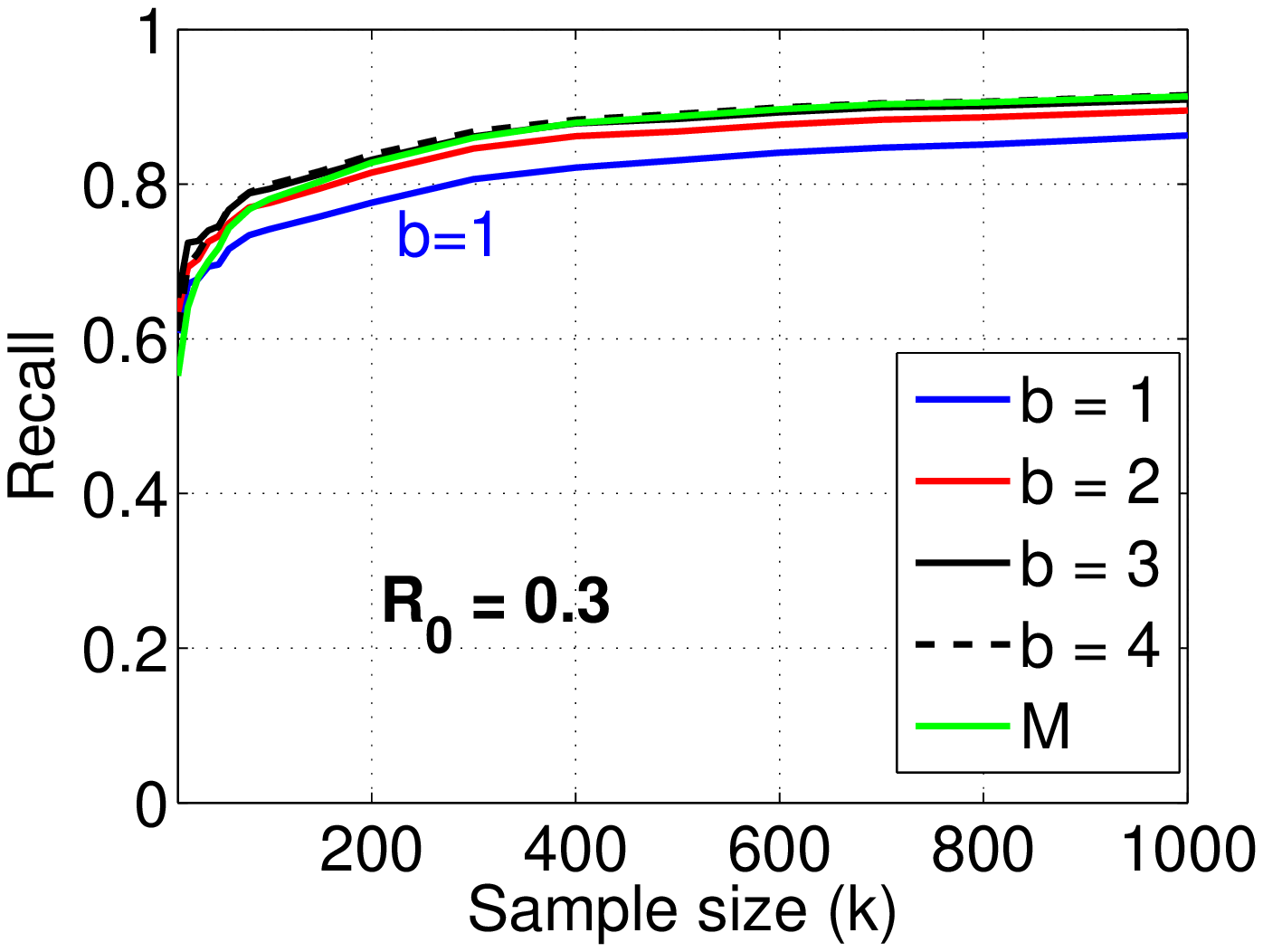}}\vspace{-0.093in}
\mbox{
\includegraphics[width = 1.78in]{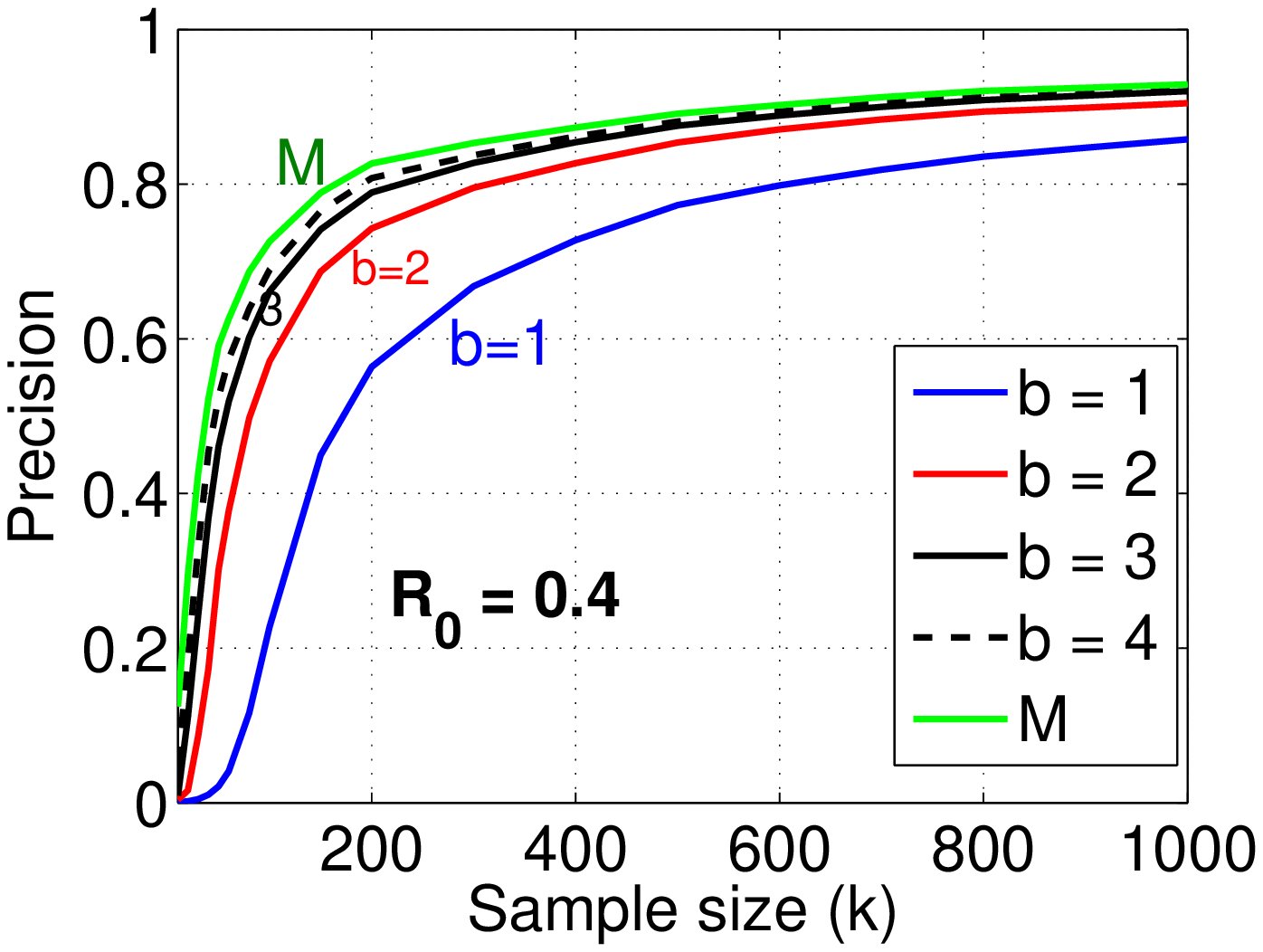}\hspace{-0.12in}
\includegraphics[width = 1.78in]{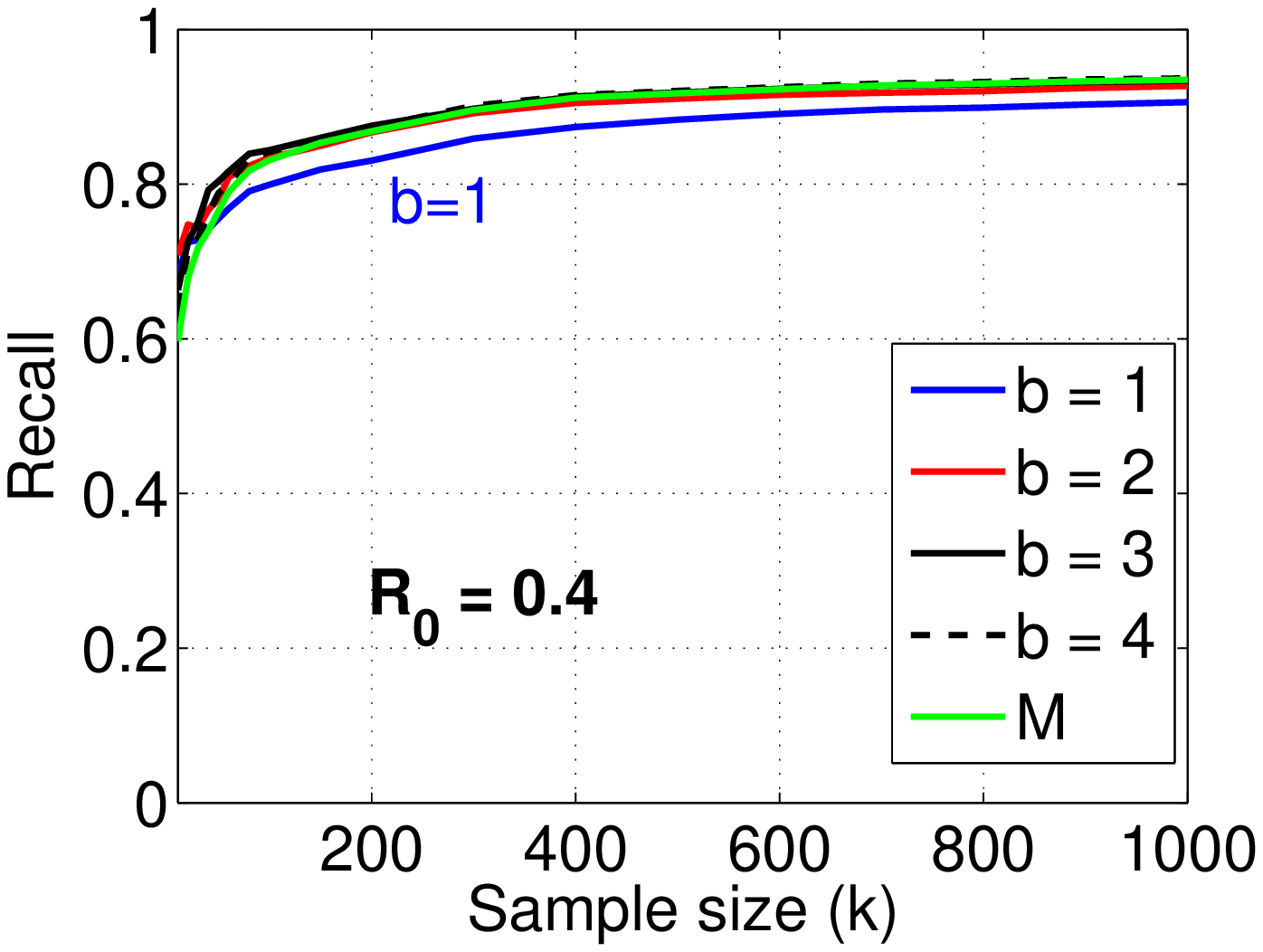}}\vspace{-0.093in}
\mbox{
\includegraphics[width = 1.78in]{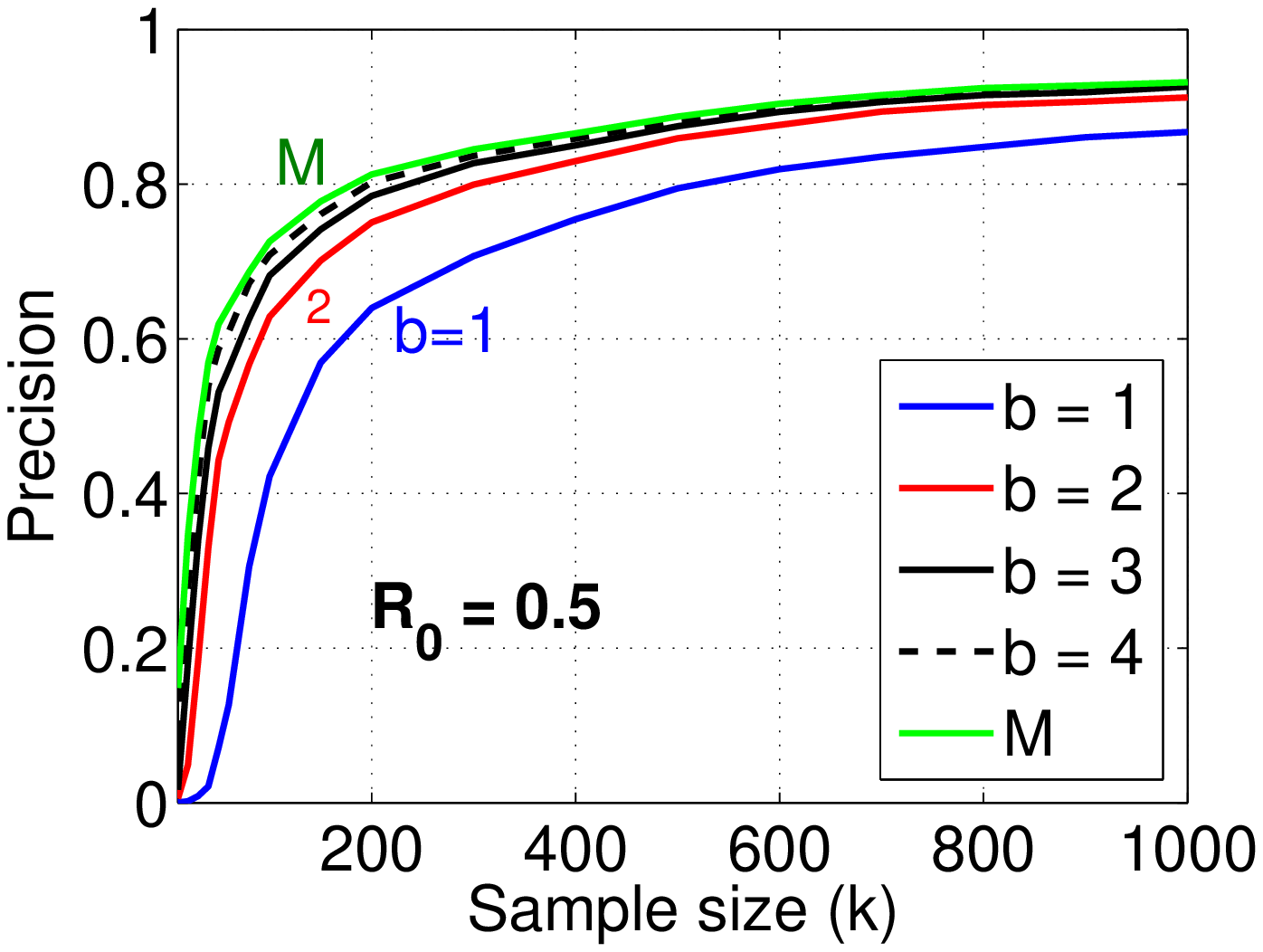}\hspace{-0.12in}
\includegraphics[width = 1.78in]{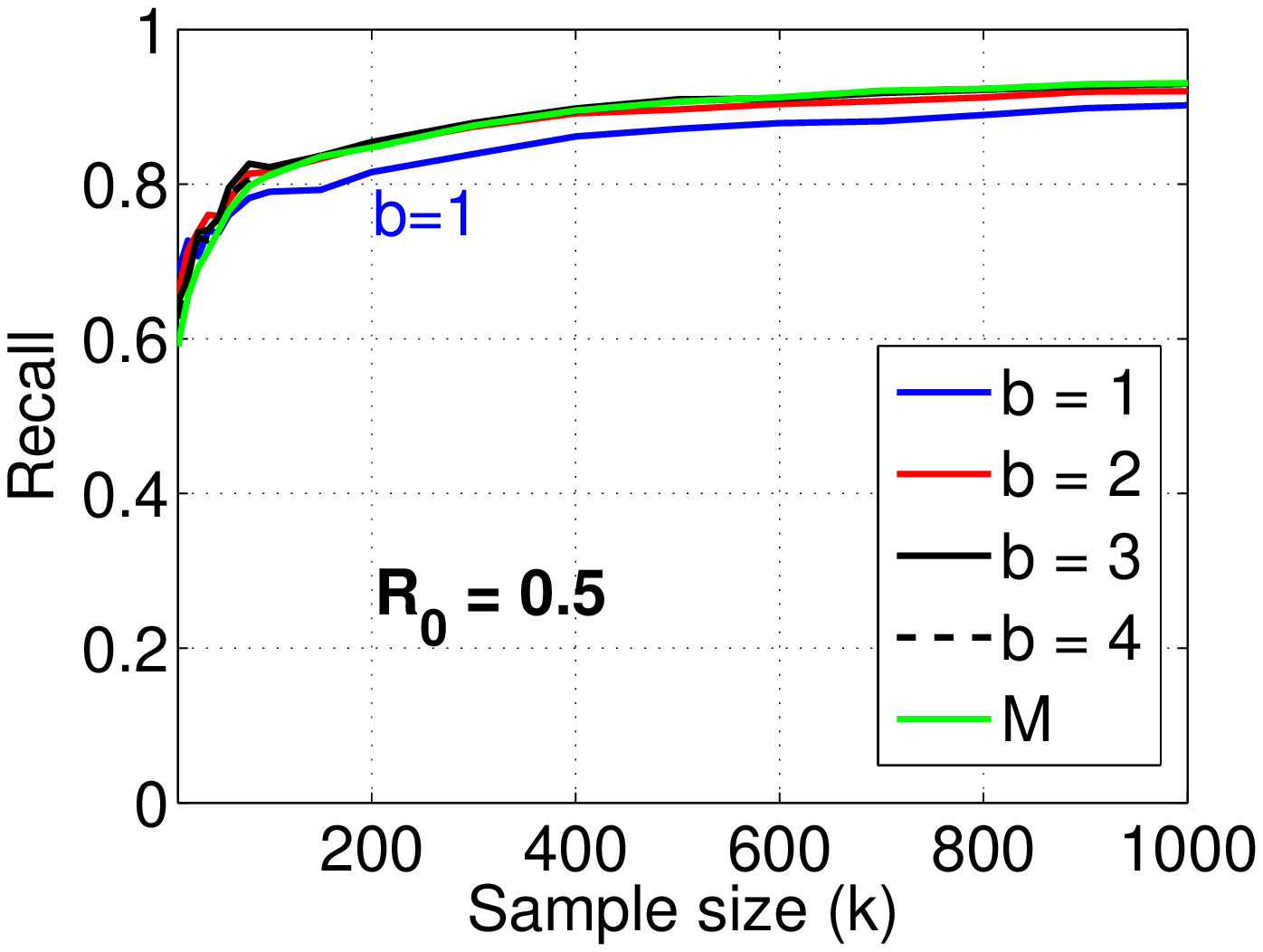}}\vspace{-0.093in}
\mbox{
\includegraphics[width = 1.78in]{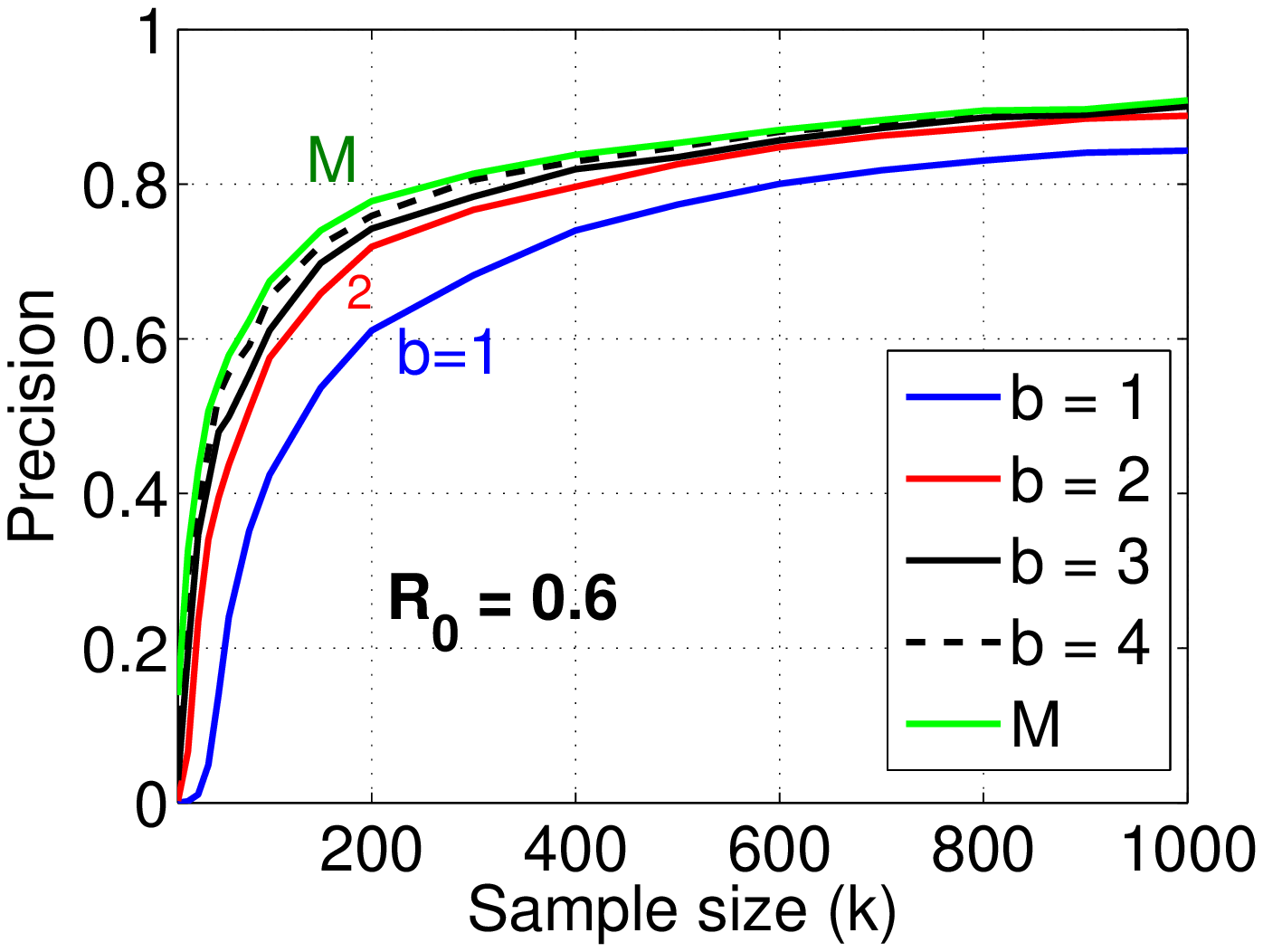}\hspace{-0.12in}
\includegraphics[width = 1.78in]{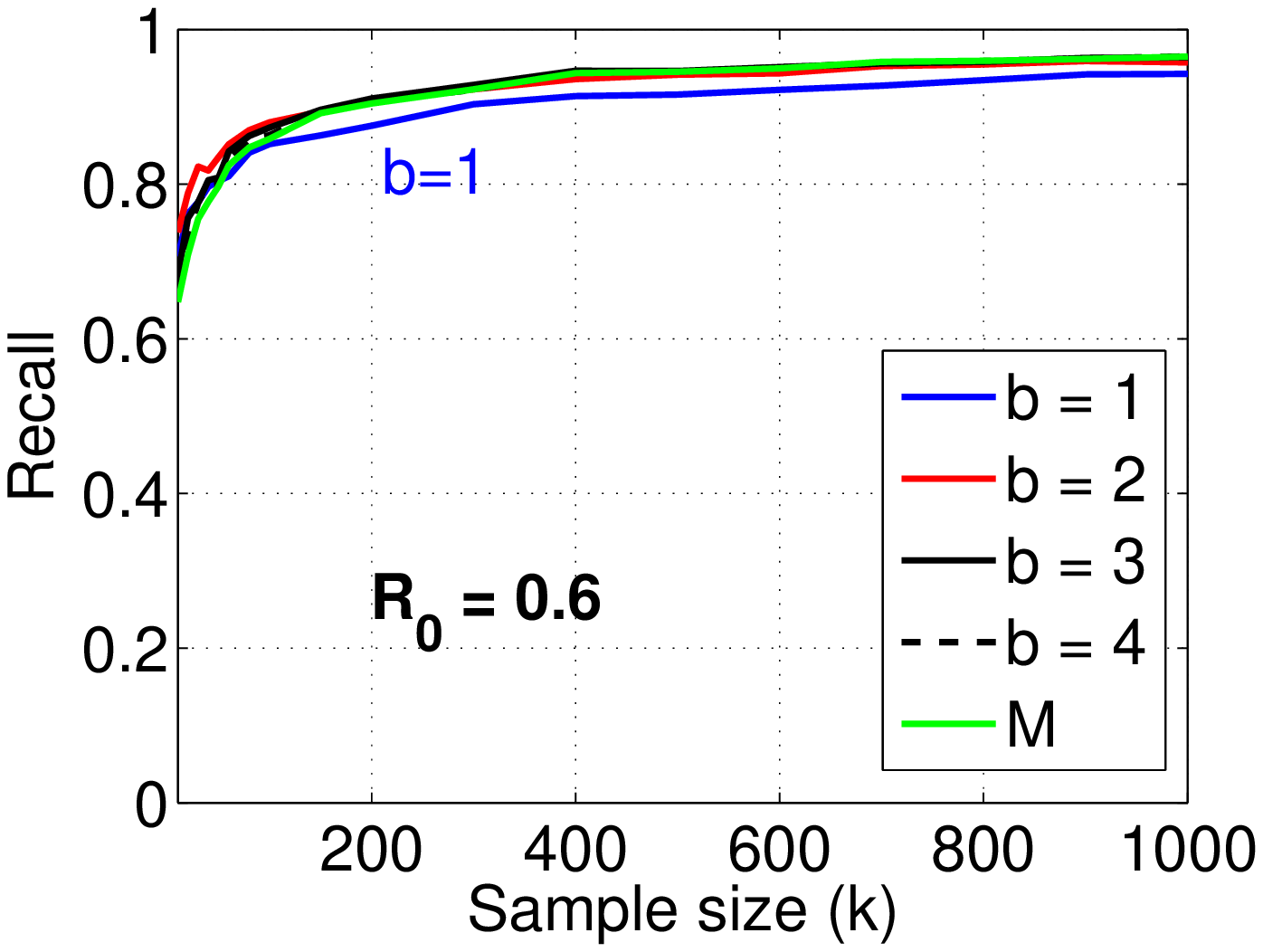}}\vspace{-0.093in}
\mbox{
\includegraphics[width = 1.78in]{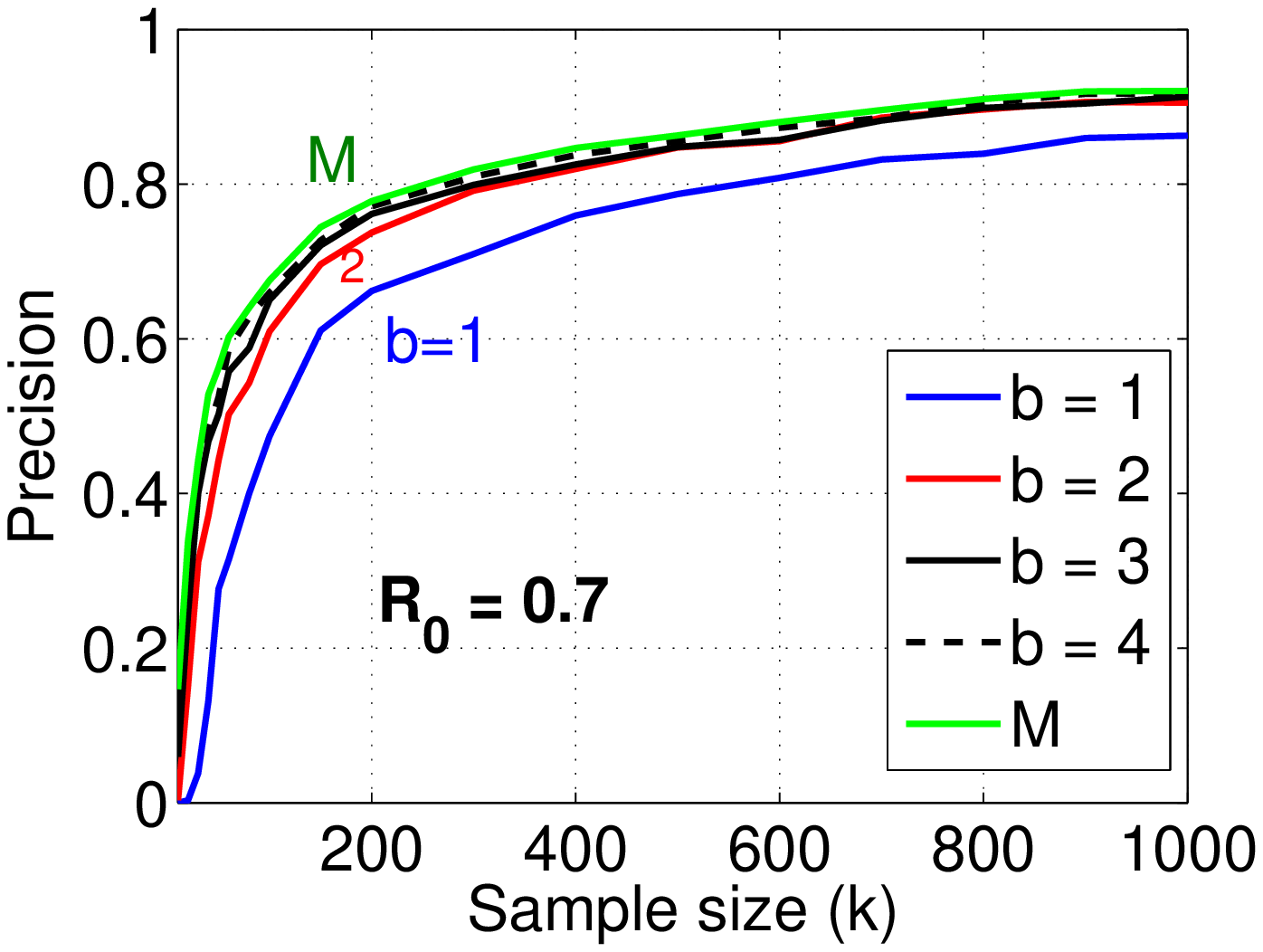}\hspace{-0.12in}
\includegraphics[width = 1.78in]{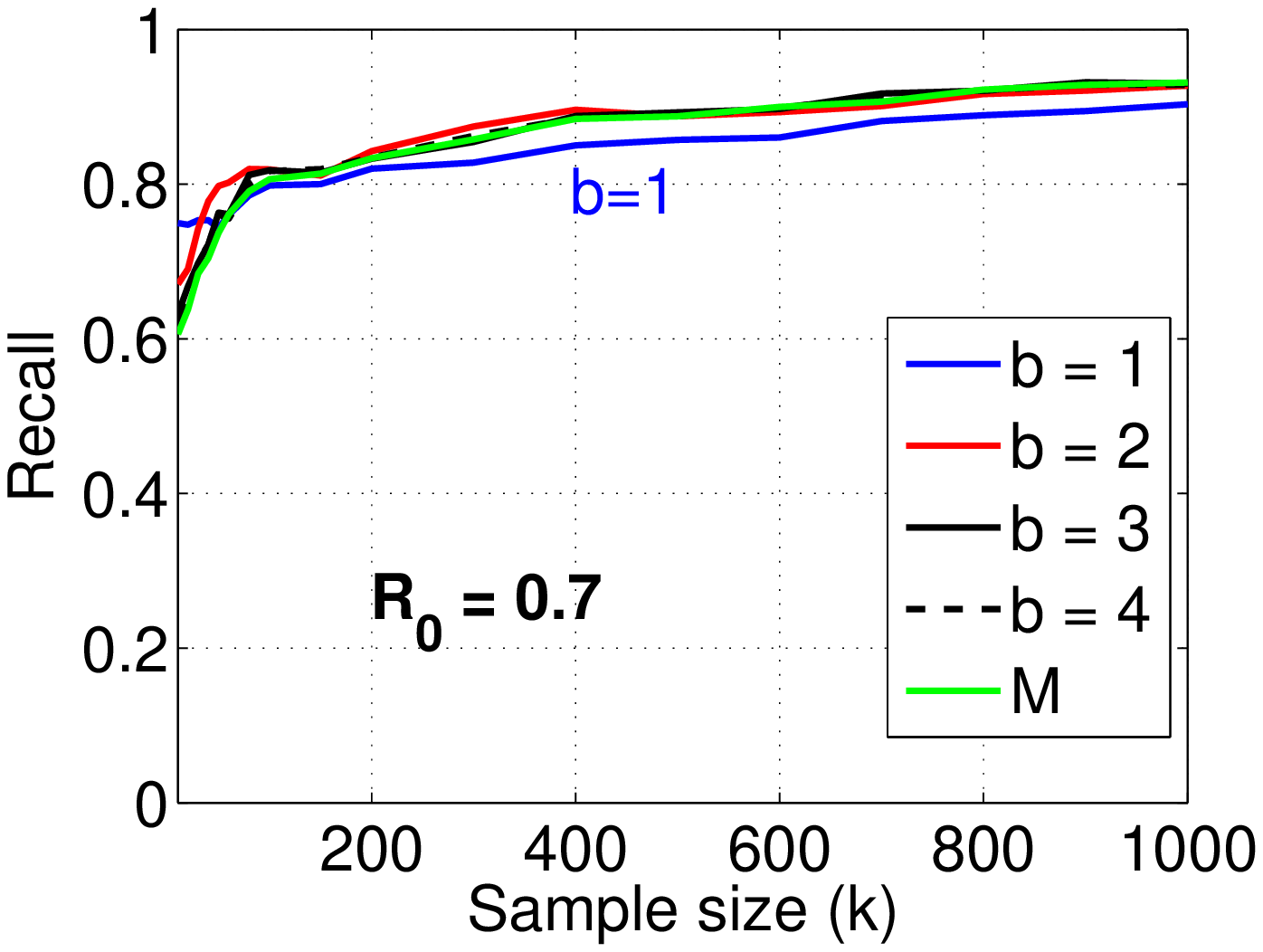}}\vspace{-0.093in}
\mbox{
\includegraphics[width = 1.78in]{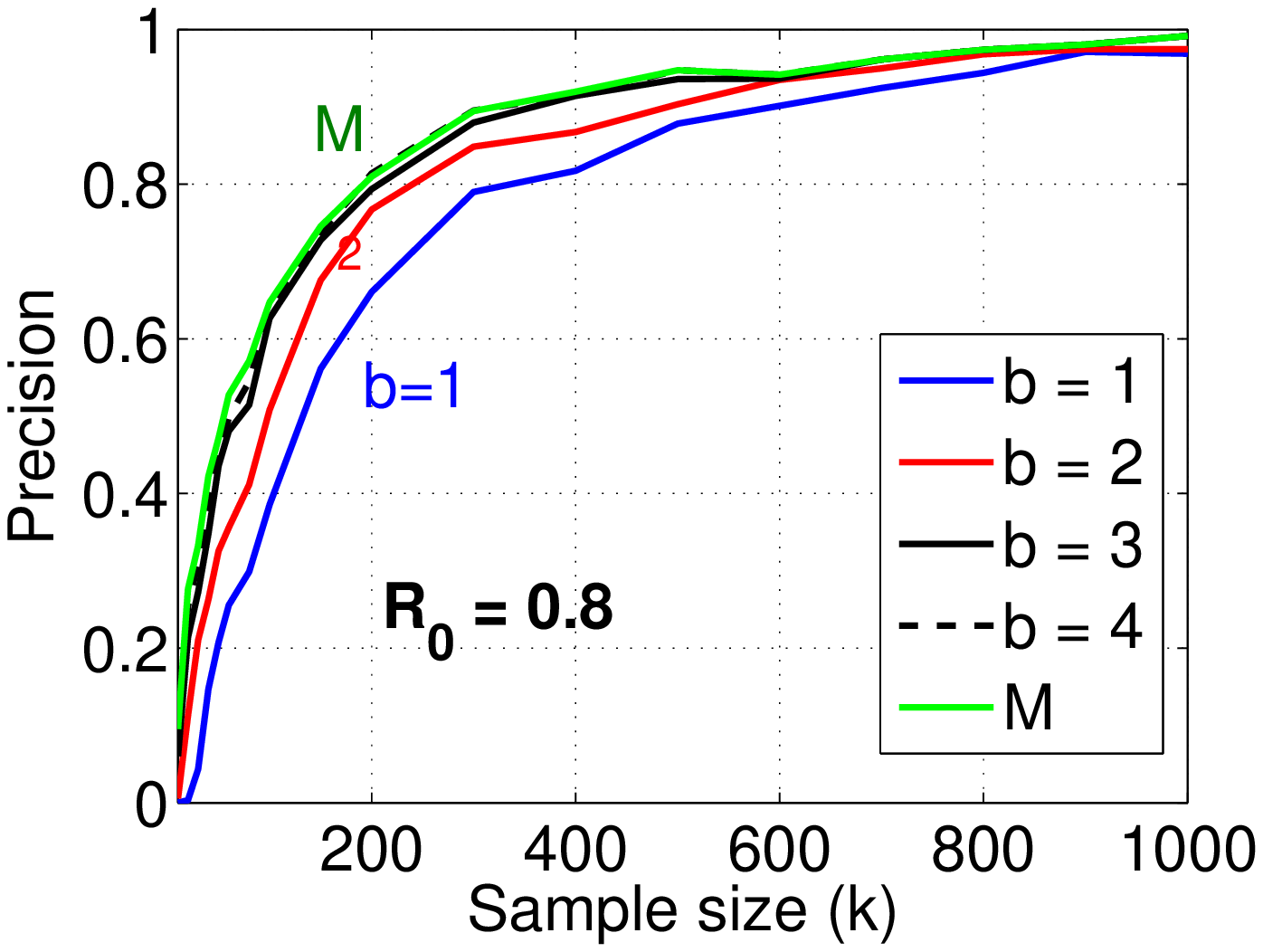}\hspace{-0.12in}
\includegraphics[width = 1.78in]{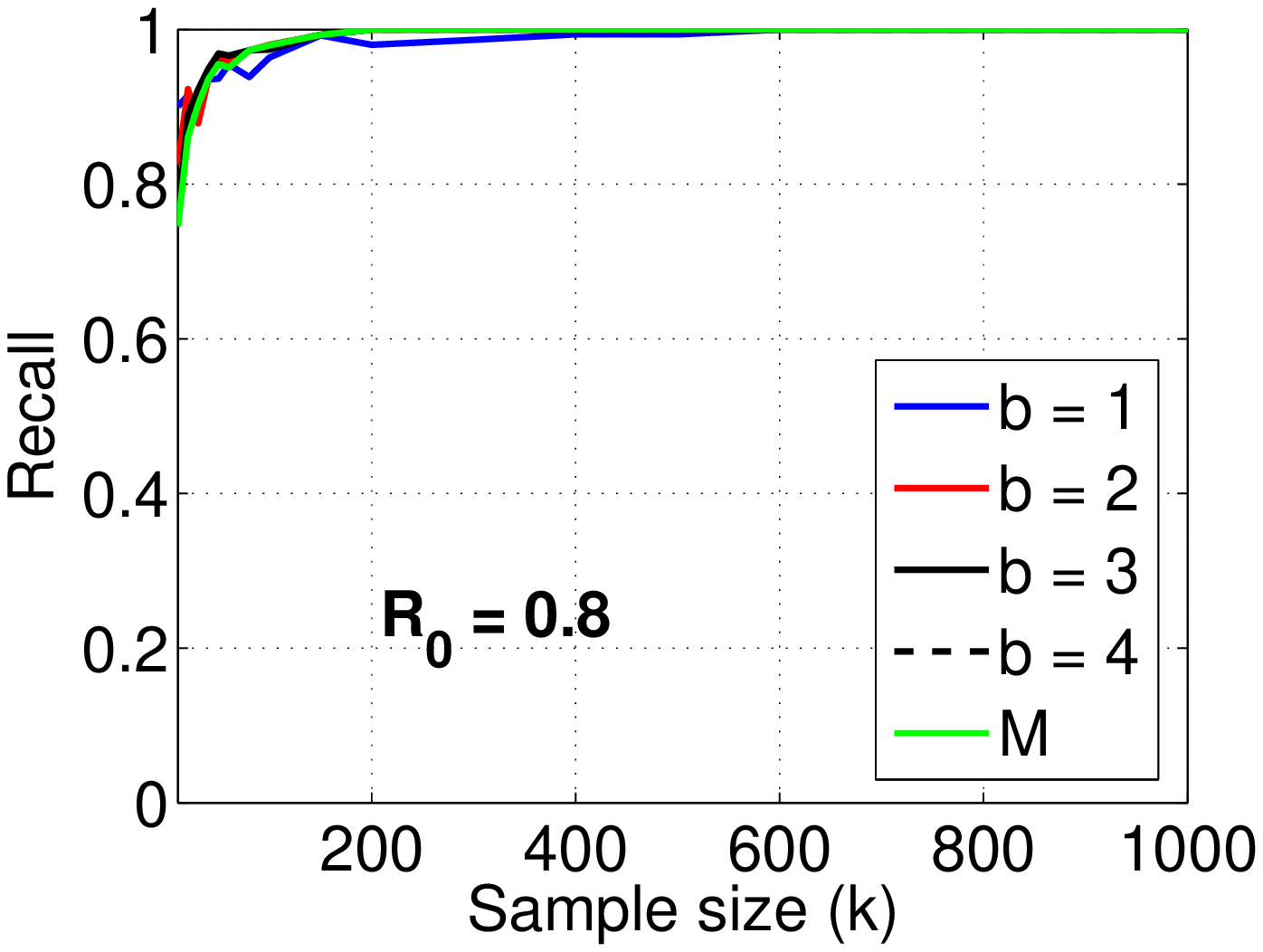}}\vspace{-0.25in}
\end{center}
\caption{\textbf{Precision \& Recall}, averaged over 700 repetitions. The task is to retrieve pairs with $R\geq R_0$.      }\label{fig_PR}
\end{figure}

Figure \ref{fig_PR} presents the precision and recall curves.  The recall curves (right panels) do not well differentiate $\hat{R}_M$ from $\hat{R}_b$ (unless $b=1$). The precision curves (left panel) show more clearly that using $b=1$ may result in lower precision values than $\hat{R}_M$ (especially when $R_0$ is small), at the same sample size $k$. When $b\geq3$, $\hat{R}_b$ performs very similarly to $\hat{R}_M$.

We stored each sample for $\hat{R}_M$ using 32 bits, although in real applications, 64 bits may be needed\cite{Proc:Broder,Proc:Broder_WWW97,Proc:Fetterly_WWW03}. Table \ref{tab_PR} summarizes the relative improvements of $\hat{R}_b$ over $\hat{R}_M$, in terms of bits, for each threshold $R_0$. Not surprisingly, the results are quite consistent with Figure \ref{fig_B(32)/B(b)}.

\begin{table}[h]
\caption{\small  Relative improvement (in space) of $\hat{R}_b$ (using $b$ bits per sample) over $\hat{R}_M$ (32 bits per sample). For precision = 0.7, 0.8, we find the required sample sizes (from Figure \ref{fig_PR}) for $\hat{R}_M$ and $\hat{R}_b$ and use them to estimate the required storage in bits. The values in the table are the ratios of the storage bits. For $b=1$ and threshold $R_0\geq0.5$, the improvements are roughly $10\sim18$-fold, consistent with the theoretical predictions in Figure \ref{fig_B(32)/B(b)}.
 }
\begin{center}{\small
\begin{tabular}{l l l}
\hline \hline
$R_0$ &Precision = $0.70$ &Precision = $0.80$\\&$b=1$\ \ \ 2\ \ \ \ \ 3\ \ \ \ \ 4&$b=1$\ \ \ 2\ \ \ \ \ 3\ \ \ \ \ 4\\\hline
    0.3    &6.49\ \   6.61\ \    7.04\ \  6.40  &  ---- \hspace{0.05in}  7.96\ \   7.58\ \     6.58\\
    0.4    &7.86\  \   8.65\ \    7.48\ \    6.50  &   8.63\ \    8.36\ \    7.71\  \   6.91\\
    0.5    &9.52\ \    9.23\  \   7.98\ \   7.24  & 11.1\ \    9.85\ \    8.22\ \    7.39\\
    0.6   &11.5\  \   10.3\  \  8.35\  \   7.10  & 13.9\ \   10.2\ \    8.07\ \    7.28\\
    0.7   &13.4\  \   12.2\  \  9.24\  \   7.20  & 14.5\ \   12.3\ \    8.91\ \    7.39\\
    0.8   &17.6\  \   12.5\  \  9.96\  \    7.76  & 18.2\ \   12.8\ \    9.90\  \    8.08\\
\hline\hline
\end{tabular}
}
\end{center}
\label{tab_PR}
\end{table}

\subsection{Experiment 3}

To illustrate the improvements  by the use of b-bit minwise hashing on a real-life application, we conducted a duplicate detection experiment using a corpus of 10000 news documents (49995000 pairs).  The dataset was crawled as part of the BLEWS project at Microsoft\cite{Proc:Gamom_AAAI08}. In the news domain, duplicate detection is an important problem as (e.g.) search engines must not serve up the same story multiple times and news stories (especially AP stories) are commonly copied with slight alterations/changes in bylines only.

In the experiments we computed the pairwise resemblances for all documents in the set; we present the data for retrieving document pairs with resemblance $R\geq R_0$ below.

\begin{figure}[h!]
\begin{center}
\mbox{
\includegraphics[width = 1.8in]{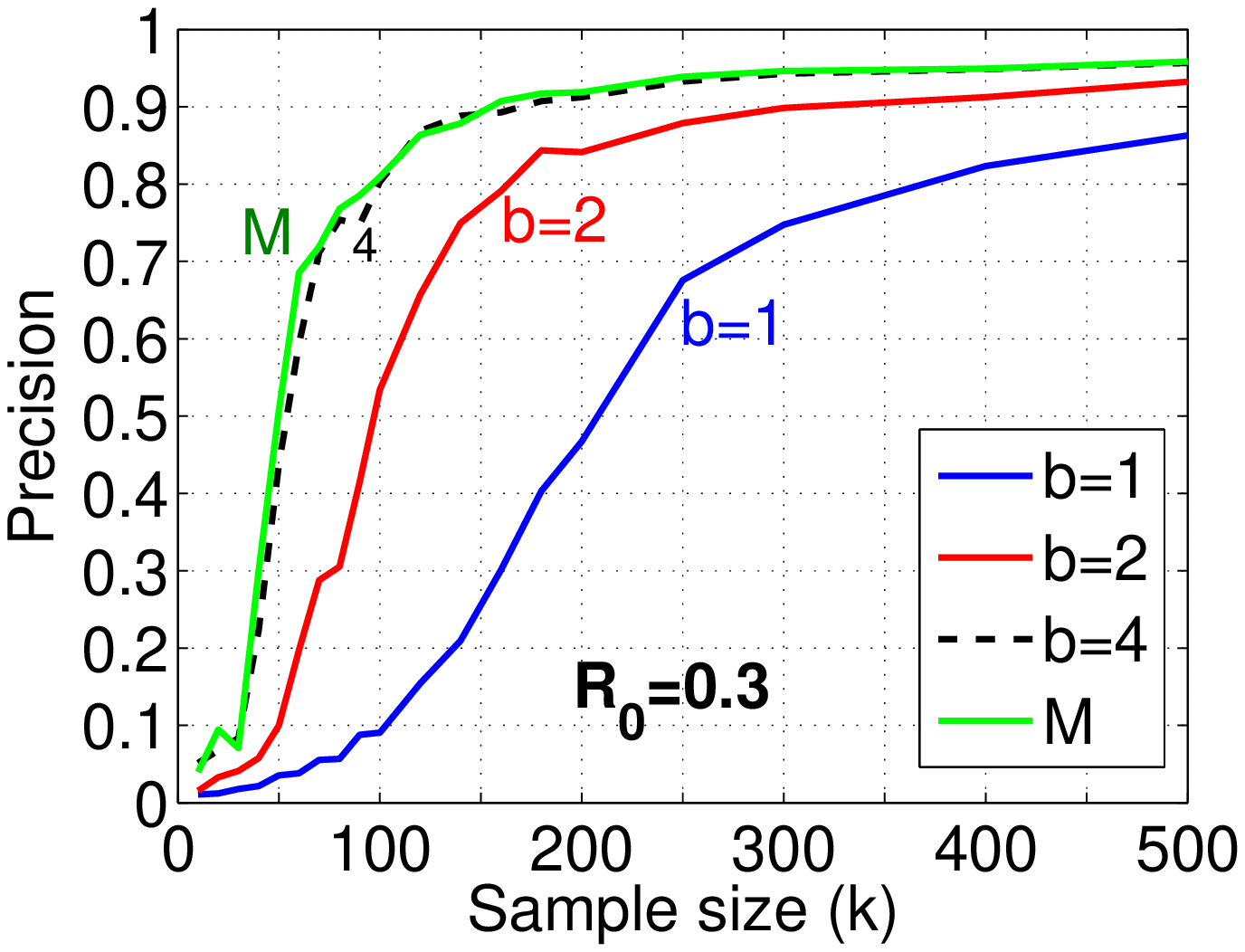}\hspace{-0.13in}
\includegraphics[width = 1.8in]{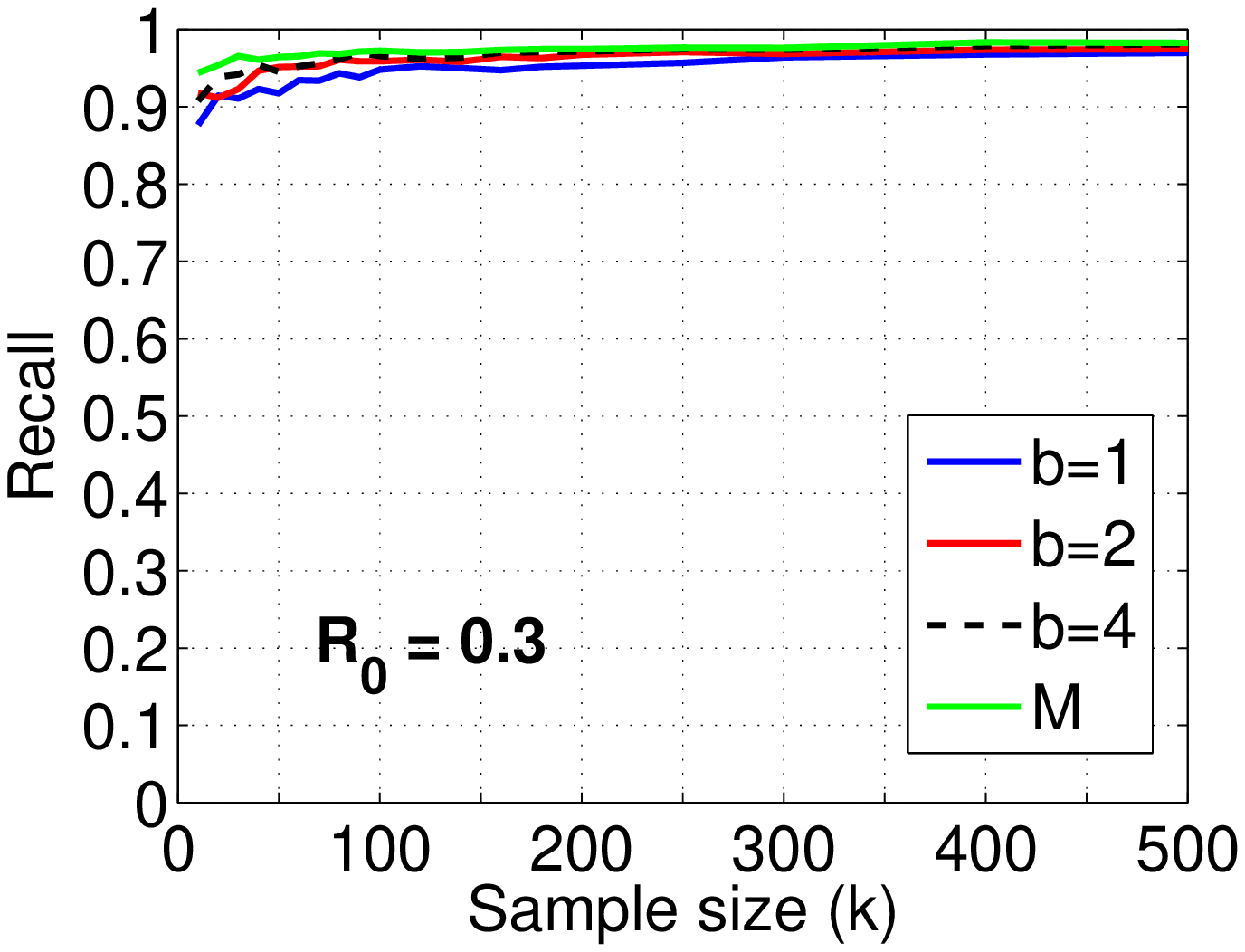}}

\mbox{
\includegraphics[width = 1.8in]{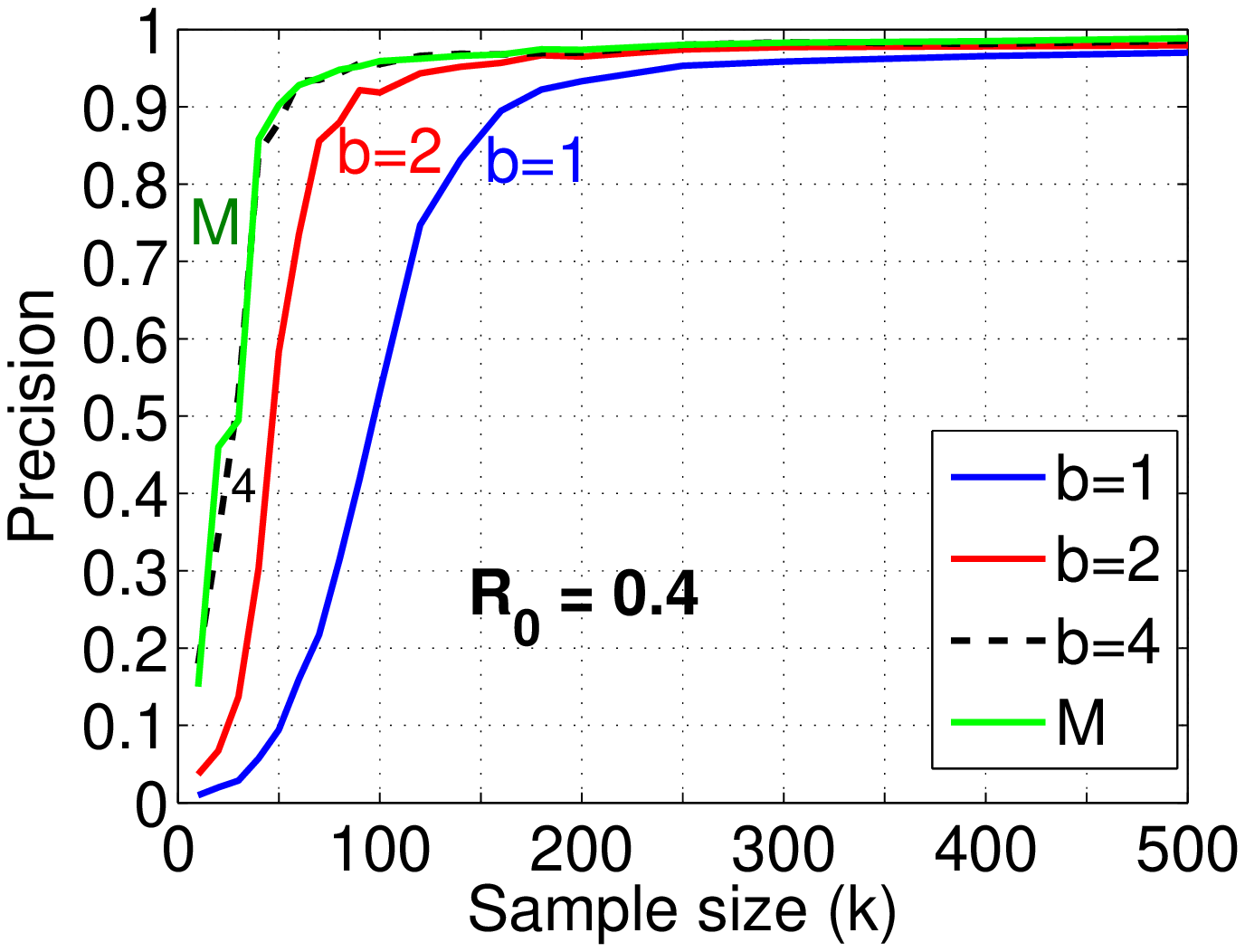}\hspace{-0.13in}
\includegraphics[width = 1.8in]{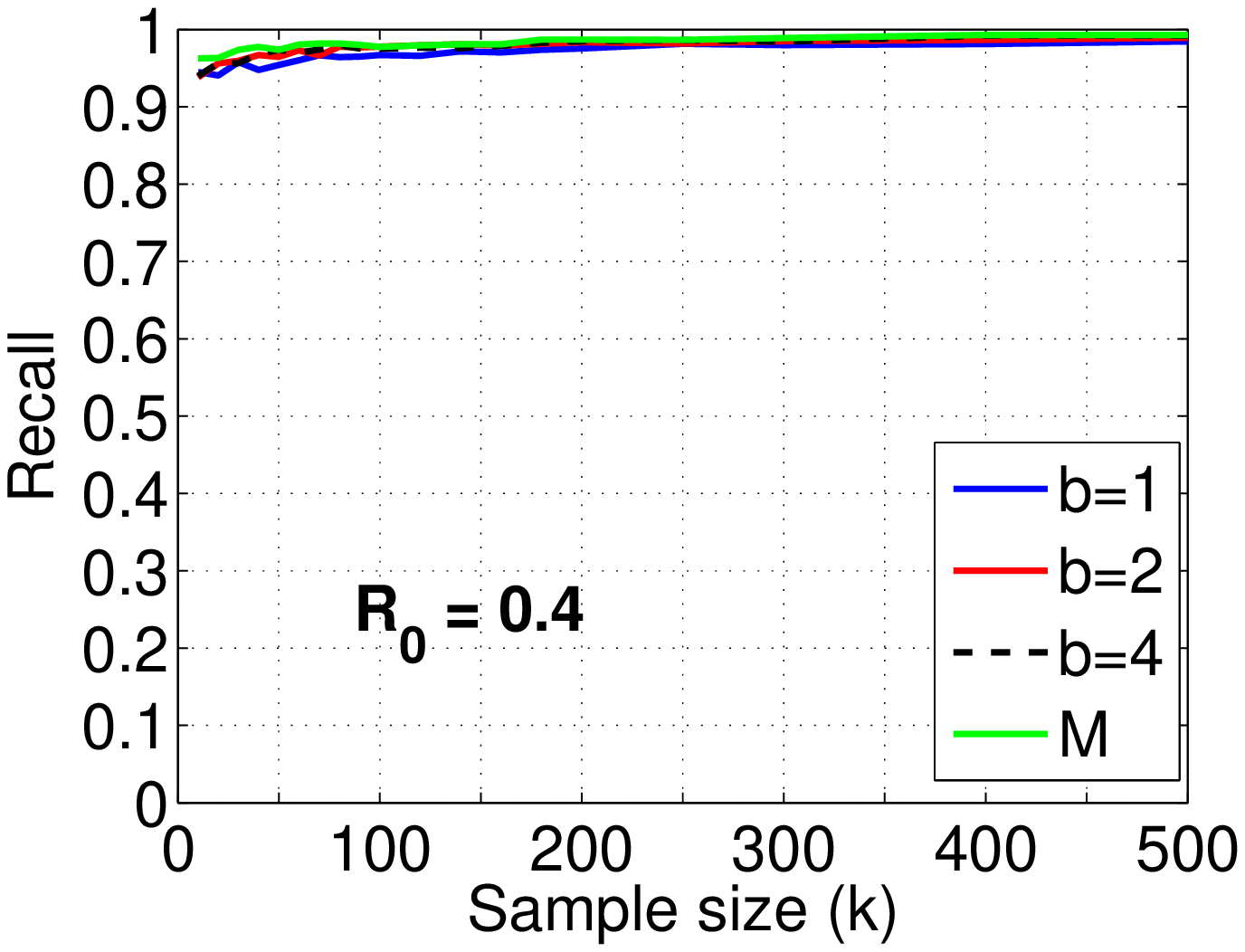}}

\mbox{
\includegraphics[width = 1.8in]{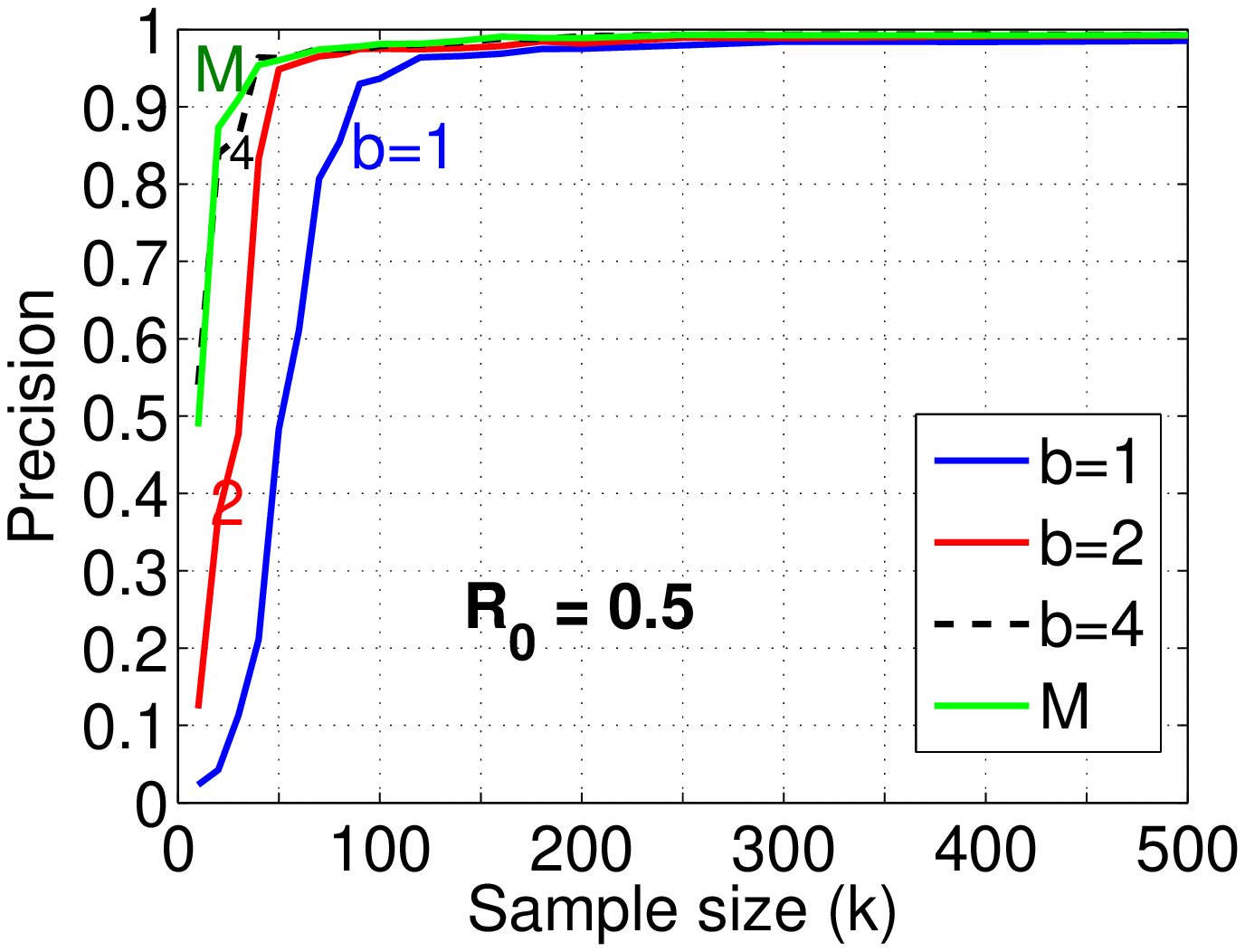}\hspace{-0.13in}
\includegraphics[width = 1.8in]{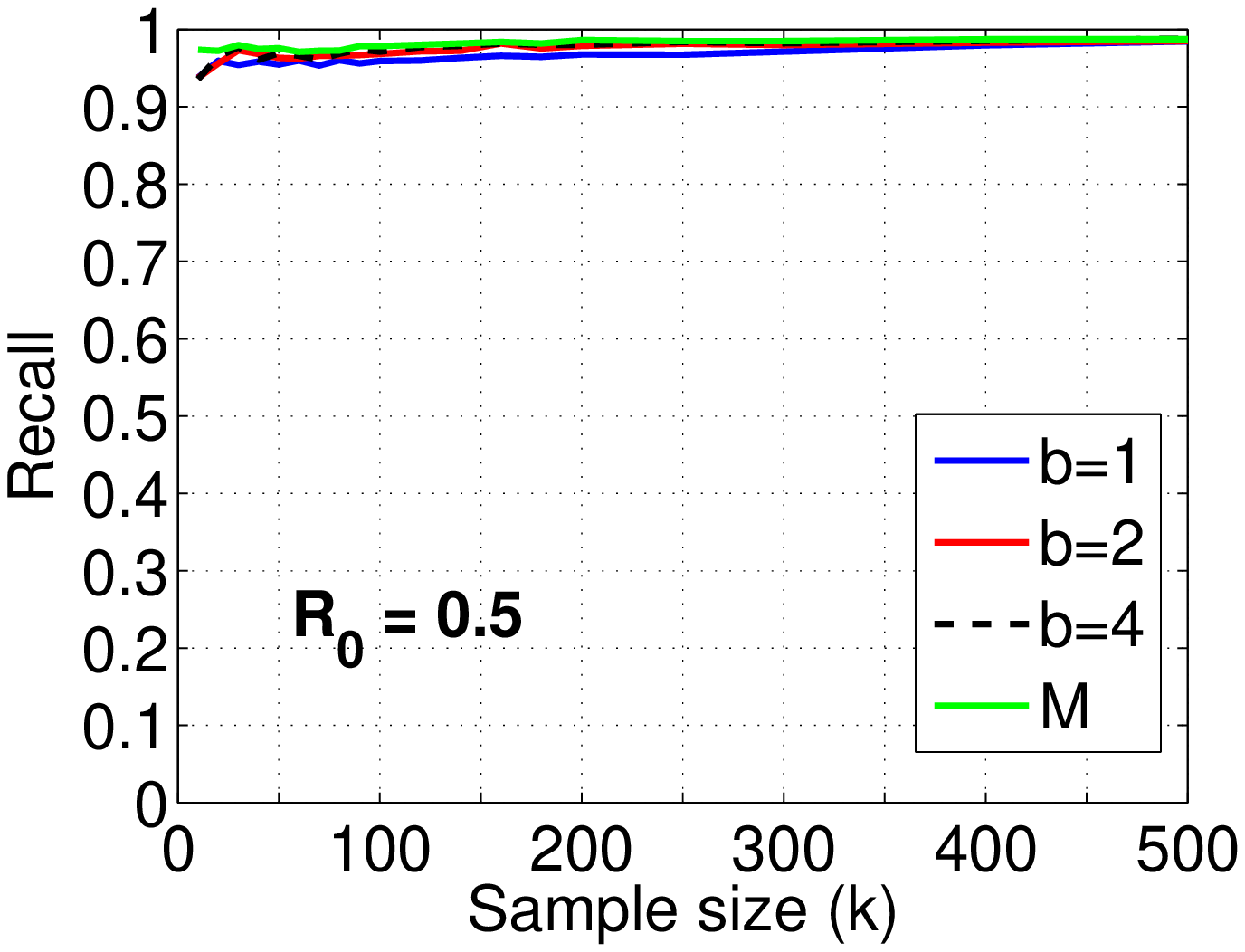}}

\mbox{
\includegraphics[width = 1.8in]{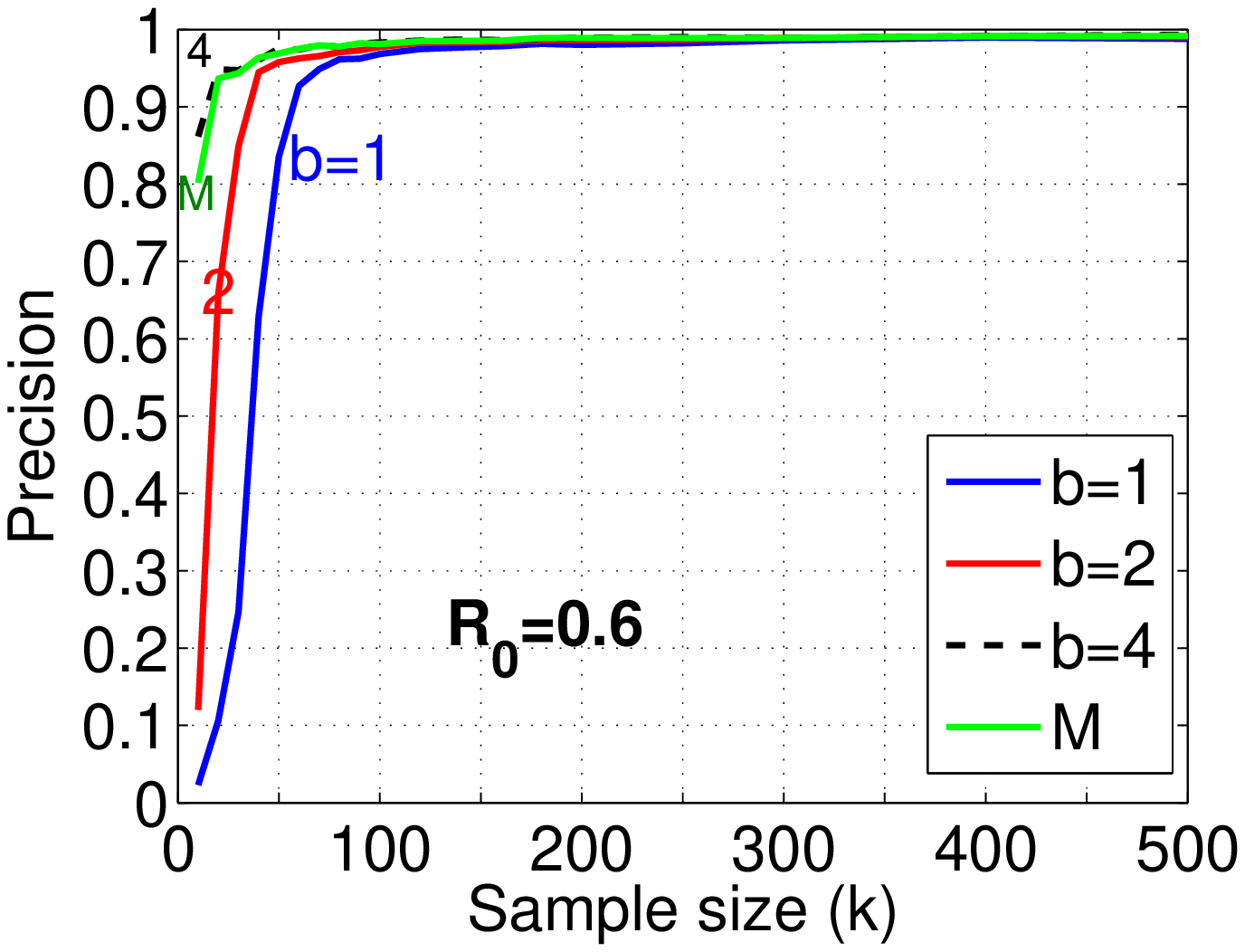}\hspace{-0.13in}
\includegraphics[width = 1.8in]{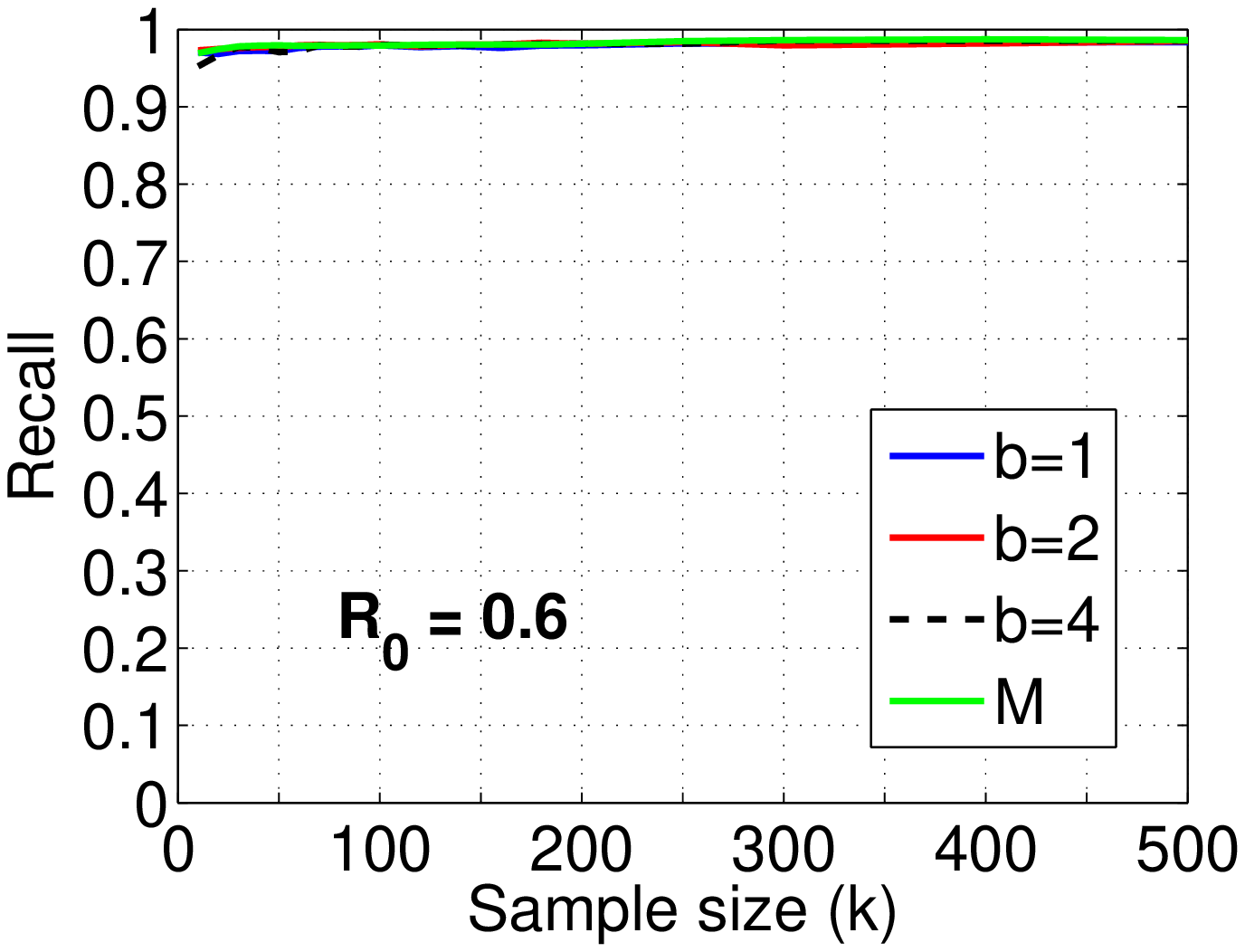}}
\end{center}
\caption{\textbf{News Precision \& Recall}. The task is to retrieve news article pairs with $R\geq R_0$.      }\label{fig_News_PR}
\end{figure}

We estimate the resemblances using $\hat{R}_b$ with $b=1$, 2, 4 bits, and the original minwise hashing (using 32 bits).  Figure \ref{fig_News_PR} presents the precision \& recall curves. The recall values are all very  high (mostly $>0.95$) and do not well differentiate various estimators.

The precision curves for $\hat{R}_4$ (using 4 bits per sample) and $\hat{R}_{M}$ (using 32 bits per sample) are almost indistinguishable, suggesting a 8-fold improvement in space using $b=4$.

When using $b=1$ or 2, the space improvements are normally around 10-fold to 15-fold, compared to $\hat{R}_M$, especially for achieving high precisions (e.g., $\geq 0.9$). This experiment again confirms the significant improvement of the $b$-bit minwise hashing using $b=1$ (or 2).

\newpage

Note that in the context of (Web) document duplicate detection, in addition to shingling, a number of specialized hash-signatures have been proposed, which leverage properties of natural-language text (such as the placement of stopwords\cite{Proc:Theobald_SIGIR08}). However, our approach is not aimed at any specific datesets, but is a general, domain-independent technique. Also, to the extent that other approaches rely on minwise hashing for signature computation, these may be combined with our techniques.

\section{Discussion: Combining Bits for Enhancing Performance}

Figure \ref{fig_B(b)} and Figure \ref{fig_B(32)/B(b)} have shown that, for about $R\geq0.4$, using $b=1$ always outperforms using $b>1$, even in the least favorable situation. This naturally leads to the conjecture that one may be able to further improve the performance using ``$b<1$'', when $R$ is close to 1.

One simple approach to implement ``$b<1$'' is to combine two bits from two permutations.

Recall $e_{1,1,\pi}$ denotes the lowest bit of the hashed value under $\pi$.
Theorem \ref{The_basic} has proved that
\begin{align}\notag
E_1=\mathbf{Pr}\left(e_{1,1,\pi} = e_{2,1,\pi}\right) = C_{1,1} + \left(1-C_{2,1}\right) R
\end{align}

Consider two permutations $\pi_1$ and $\pi_2$. We store
\begin{align}\notag
x_1 = \text{XOR}(e_{1,1,\pi_1},\  e_{1,1,\pi_2}), \hspace{0.2in} x_2 = \text{XOR}(e_{2,1,\pi_1},\  e_{2,1,\pi_2})
\end{align}
Then $x_1=x_2$ either when $e_{1,1,\pi_1}=e_{2,1,\pi_1}$ and $e_{1,1,\pi_2}=e_{2,1,\pi_2}$, or, when $e_{1,1,\pi_1}\neq e_{2,1,\pi_1}$ and $e_{1,1,\pi_2}\neq e_{2,1,\pi_2}$. Thus
\begin{align}
T = \mathbf{Pr}\left(x_1 = x_2\right) = E_1^2 + (1-E_1)^2,
\end{align}
which is a quadratic equation with solution
\begin{align}
R = \frac{\sqrt{2T-1} + 1-2C_{1,1}}{2-2C_{2,1}}.
\end{align}
We can estimate $T$ without bias as  a binomial. However, the resultant estimator for $R$ will be biased, at small sample size $k$, due to the nonlinearity. We will recommend the following estimator
\begin{align}
\hat{R}_{1/2} = \frac{\sqrt{\max\{2\hat{T}-1,0\}} + 1-2C_{1,1}}{2-2C_{2,1}}.
\end{align}
The truncation $\max\{, 0\}$ will introduce further bias; but it is necessary and is usually a good bias-variance trade-off.

We use $\hat{R}_{1/2}$ to indicate that two bits are combined into one. The asymptotic variance of $\hat{R}_{1/2}$ can be derived using the ``delta method'' in statistics:
\begin{align}\label{eqn_Var_1/2}
\text{Var}\left(\hat{R}_{1/2}\right) = \frac{1}{k}\frac{T(1-T)}{4(1-C_{2,1})^2(2T-1)}+O\left(\frac{1}{k^2}\right).
\end{align}

One should keep in mind that, in order to generate $k$ samples for $\hat{R}_{1/2}$, we have to conduct $2\times k$ permutations. Of course, each sample is still stored using 1 bit, despite that we  use ``$b=1/2$'' to denote this estimator.

Interestingly, as $R\rightarrow 1$, $\hat{R}_{1/2}$ does twice as well as $\hat{R}_1$:
\begin{align}
\lim_{R\rightarrow 1} \frac{\text{Var}\left(\hat{R}_1\right)}{\text{Var}\left(\hat{R}_{1/2}\right)} =
\lim_{R\rightarrow 1} \frac{2(1-2E_1)^2}{(1-E_1)^2+E_1^2} = 2.
\end{align}
Recall, if $R=1$, then $r_1=r_2$, $C_{1,1}=C_{2,1}$, and $E_1 = C_{1,1}+1-C_{2,1}=1$.

On the other hand, $\hat{R}_{1/2}$ may not be an ideal estimator when $R$ is not too large. For example, one can numerically show  that (as $k\rightarrow \infty$)
\begin{align}\notag
\text{Var}\left(\hat{R}_1\right) < \text{Var}\left(\hat{R}_{1/2}\right), \hspace{0.2in} \text{if} \ \ R<0.5774,\ r_1, \ r_2\rightarrow 0
\end{align}

%

\begin{figure}[h]
\begin{center}
\mbox{
\includegraphics[width = 1.8in]{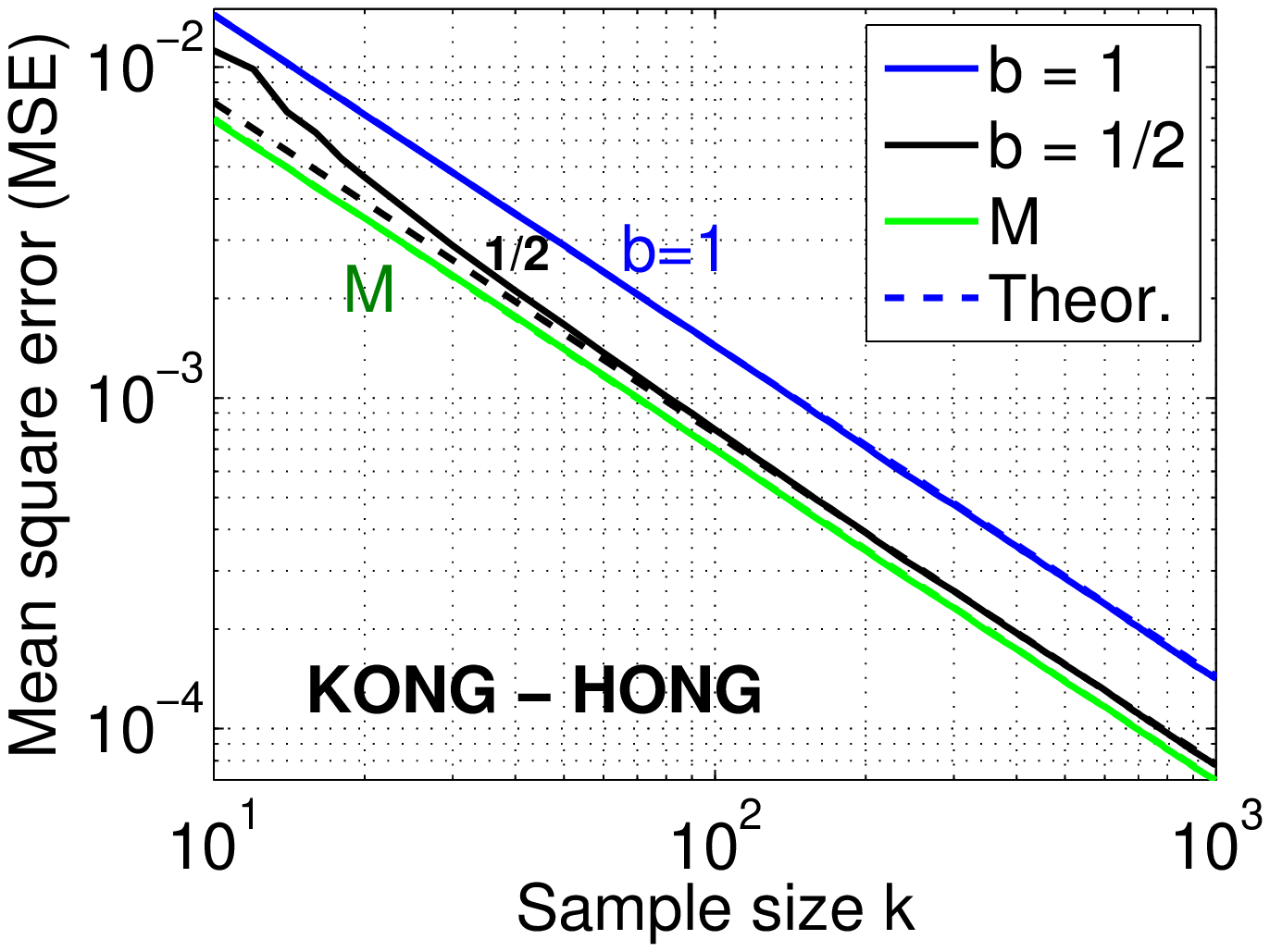}\vspace{-0.1in}
\includegraphics[width = 1.8in]{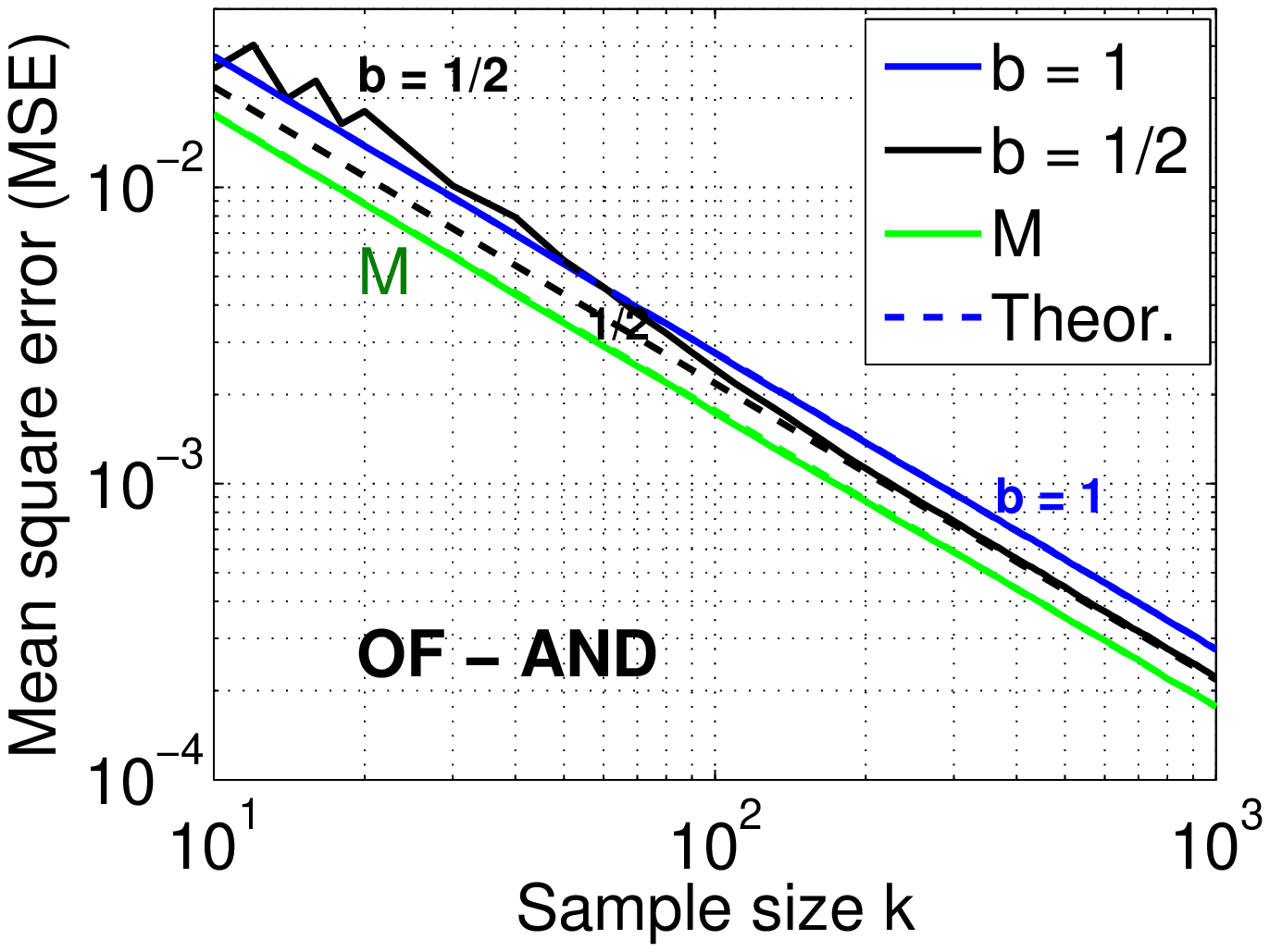}}\vspace{-0.2in}
\mbox{
\includegraphics[width = 1.8in]{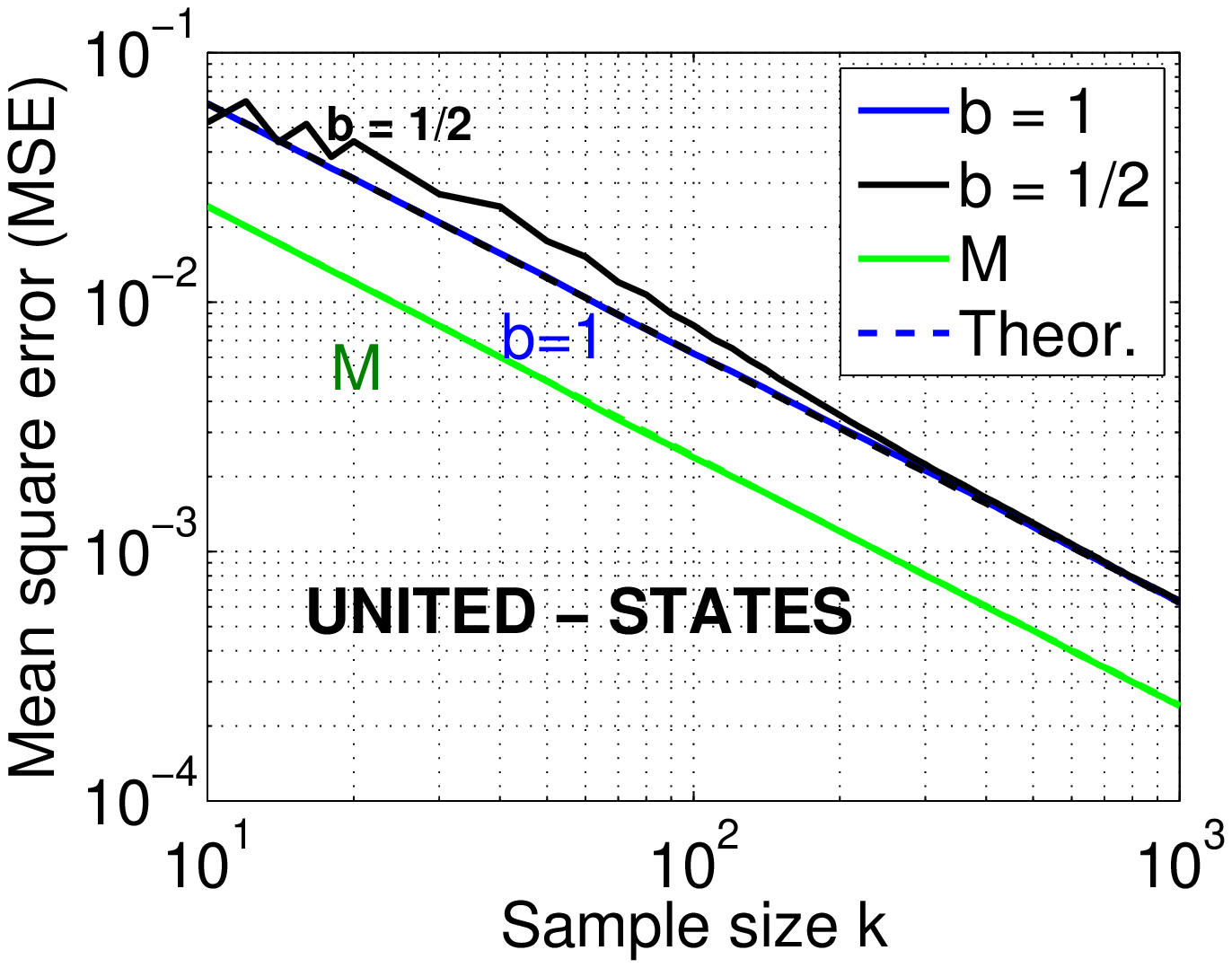}\vspace{-0.1in}
\includegraphics[width = 1.8in]{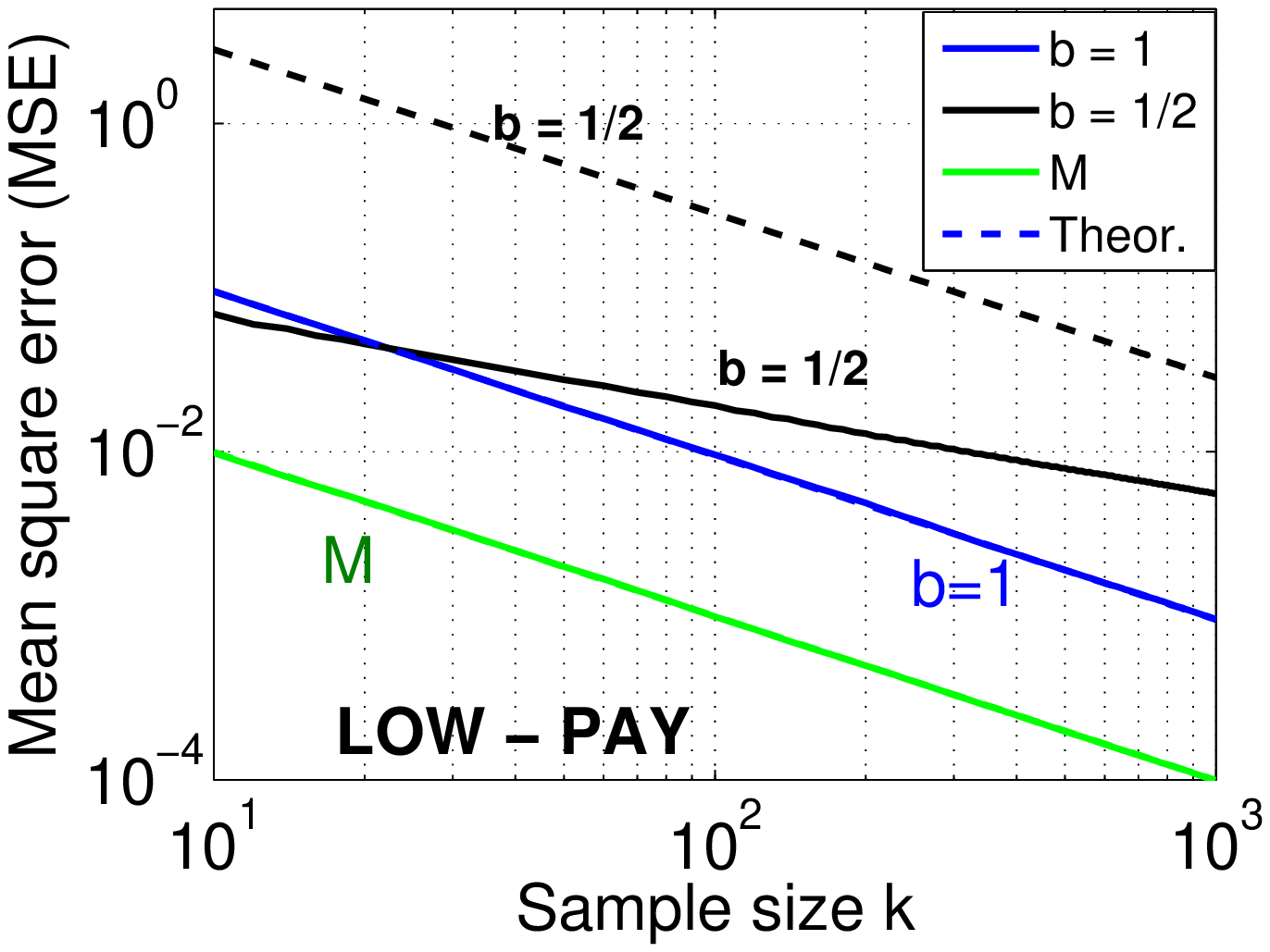}}\vspace{-0.25in}
\end{center}
\caption{\textbf{MSEs} for comparing $\hat{R}_{1/2}$ with $\hat{R}_{1}$ and $\hat{R}_M$. Due to the bias, the theoretical variances $\text{Var}\left(\hat{R}_{1/2}\right)$, i.e., (\ref{eqn_Var_1/2}), deviate from the empirical MSEs when the sample size $k$ is not large.  }\label{fig_R_1/2}
\end{figure}

Figure \ref{fig_R_1/2} plots the empirical MSEs for four word pairs in Experiment 1, for $\hat{R}_{1/2}$, $\hat{R}_{1}$, and $\hat{R}_M$:
\begin{itemize}
\item For the highly similar pair, ``KONG-HONG,'' $\hat{R}_{1/2}$ exhibits superb performance compared to $\hat{R}_1$.
\item For the fairly similar pair, ``OF-AND,''  $\hat{R}_{1/2}$ is still considerably better.
\item For ``UNITED-STATES,'' whose $R=0.591$, $\hat{R}_{1/2}$ performs similarly to $\hat{R}_{1}$.
\item For ``LOW-PAY,'' whose $R=0.112$ only, the theoretical variance of $\hat{R}_{1/2}$ is very large. However, owing to the variance-bias trade-off, the empirical performance of $\hat{R}_{1/2}$ is not too bad.
\end{itemize}

In a summary, while the idea of combining two bits is interesting, it is mostly useful  in applications which only care about pairs of very high similarities.
%

\section{Conclusion}

The {\em minwise hashing} technique has been  widely used as a standard approach in information retrieval, for efficiently computing set similarity in massive data sets (e.g., duplicate detection). Prior studies commonly used  $64$ or 40 bits to store each hashed value,

In this study, we propose the theoretical framework of \textbf{\em $b$-bit minwise hashing}, by only storing the lowest $b$ bits of each hashed value. We theoretically prove that, when the similarity is reasonably high (e.g., resemblance $\geq0.5$), using $b=1$ bit per hashed value can, even in the worse case, gain a space improvement by $10.7\sim16$-fold, compared to storing each hashed value using 32 bits. The improvement would be even more significant (e.g., at least $21.3\sim32$-fold) if the original hashed values are stored using 64 bits.

We also discussed the idea of combining 2  bits from different hashed values, to further enhance the improvement, when the target similarity is very high.

Our proposed method is simple and requires only minimal modification to the original minwise hashing algorithm. We expect our method will be adopted in practice.




\appendix
\section{Proof of Theorem 1}\label{app_proof_basic}

Consider two sets, $S_1$ and $S_2$,
\begin{align}\notag
&S_1, S_2 \subseteq \Omega = \{0,1, 2, ..., D-1\},\\\notag
&f_1=|S_1|,\ f_2 = |S_2|, \ a = |S_1\cap S_2|
\end{align}

We apply a random permutation $\pi$ on $S_1$ and $S_2$:
\begin{align}\notag
\pi: \Omega \longrightarrow \Omega.
\end{align}

Define the minimum values under  $\pi$ to be $z_1$ and $z_2$:
\begin{align}\notag
z_1 = \min\left(\pi\left(S_1\right)\right), \hspace{0.2in} z_2 = \min\left(\pi\left(S_2\right)\right).
\end{align}

Define $e_{1,i} = i$th lowest bit of $z_1$, and $e_{2,i} = i$th lowest bit of $z_2$.  The task is to derive the analytical expression for
 \begin{align}\notag
 \mathbf{Pr}\left(\prod_{i=1}^b 1\{e_{1,i}=e_{2,i}\} =1\right),
 \end{align}
which can be decomposed to be
 \begin{align}\notag
& \mathbf{Pr}\left(\prod_{i=1}^b 1\{e_{1,i}=e_{2,i}\} =1, \ z_1 = z_2\right)\\\notag
 +&  \mathbf{Pr}\left(\prod_{i=1}^b 1\{e_{1,i}=e_{2,i}\} =1, \ z_1 \neq z_2\right)\\\notag
 =&\mathbf{Pr}\left(z_1 = z_2\right)+  \mathbf{Pr}\left(\prod_{i=1}^b 1\{e_{1,i}=e_{2,i}\} =1, \ z_1 \neq z_2\right)\\\notag
 =&R+  \mathbf{Pr}\left(\prod_{i=1}^b 1\{e_{1,i}=e_{2,i}\} =1, \ z_1 \neq z_2\right).
 \end{align}
where $R = \frac{|S_1\cap S_2|}{|S_1\cup S_2|} =\mathbf{Pr}\left(z_1 = z_2\right) $ is the resemblance.

When $b=1$, the task boils down to estimating
\begin{align}\notag
&\mathbf{Pr}\left(e_{1,1}=e_{2,1}, \ z_1 \neq z_2\right) \\\notag
=&\sum_{i=0,2,4, ...}\left\{\sum_{j\neq i, j=0, 2, 4, ...}\mathbf{Pr}\left(z_1 = i, z_2 = j\right)\right\}\\\notag
+&\sum_{i=1,3,5, ...}\left\{\sum_{j\neq i, j=1, 3, 5, ...}\mathbf{Pr}\left(z_1 = i, z_2 = j\right)\right\}.
\end{align}

Therefore, we need the following basic probability formula:
\begin{align}\notag
\mathbf{Pr}\left(z_1 = i,\ z_2 = j, \ i\neq j \right).
\end{align}
We will first start with
\begin{align}\notag
\mathbf{Pr}\left(z_1 = i,\ z_2 = j, \ i<j \right) = \frac{P_1+P_2}{P_3}
\end{align}
where
\begin{align}\notag
&P_3 = \binom{D}{a}\binom{D-a}{f_1-a}\binom{D-f_1}{f_2-a},\\\notag
&P_1= \binom{D-j-1}{a}\binom{D-j-1-a}{f_2-a-1}\binom{D-i-1-f_2}{f_1-a-1},\\\notag
&P_2= \binom{D-j-1}{a-1}\binom{D-j-a}{f_2-a}\binom{D-i-1-f_2}{f_1-a-1}.
\end{align}

The expressions for $P_1$, $P_2$, and $P_3$ can be understood by  the experiment of randomly throwing $f_1+f_2-a$ balls into $D$ locations, labeled $0, 1, 2, ..., D-1$.  Those $f_1+f_2-a$ balls belong to three disjoint sets: $S_1 - S_1\cap S_2$, $S_2 - S_1\cap S_2$, and $S_1\cap S_2$. Without any restriction, the total number of combinations should be $P_3$.

To understand $P_1$ and $P_3$, we need to consider two cases:
\begin{enumerate}
\item {\em The $j$th element is not in $S_1\cap S_2$ $\Longrightarrow P_1$.}
\item {\em The $j$th element is in $S_1 \cap S_2$ $\Longrightarrow P_2$.}\\
\end{enumerate}

The next task is to simplify the expression for the probability $\mathbf{Pr}\left(z_1 = i,\ z_2 = j, \ i<j \right)$. After conducing expansions and cancelations, we obtain
\begin{align}\notag
&\mathbf{Pr}\left(z_1 = i,\ z_2 = j, \ i<j \right) = \frac{P_1+P_2}{P_3}\\\notag
=&\frac{\left(\frac{1}{a}+\frac{1}{f_2-a}\right)\frac{(D-j-1)!(D-i-1-f_2)!}{(a-1)!(f_1-a-1)(f_2-a-1)!(D-j-f_2)!(D-i-f_1-f_2+a)!} }{\frac{D!}{a!(f_1-a)!(f_2-a)!(D-f_1-f_2+a)!}}\\\notag
=&\frac{f_2(f_1-a)(D-j-1)!(D-f_2-i-1)!(D-f_1-f_2+a)!}{D!(D-f_2-j)!(D-f_1-f_2+a-i)!}\\\notag
= &\frac{f_2(f_1-a)\prod_{t=0}^{j-i-2}(D-f_2-i-1-t)\prod_{t=0}^{i-1}(D-f_1-f_2+a-t}{\prod_{t=0}^{j} (D-t)}\\\notag
=&\frac{f_2}{D}\frac{f_1-a}{D-1}\prod_{t=0}^{j-i-2}\frac{D-f_2-i-1-t}{D-2-t}\prod_{t=0}^{i-1}\frac{D-f_1-f_2+a-t}{D+i-j-1-t}
\end{align}
For convenience, we introduce the following notation:
\begin{align}\notag
r_1 = \frac{f_1}{D}, \hspace{0.2in} r_2 = \frac{f_2}{D}, \hspace{0.2in} s = \frac{a}{D}.
\end{align}
Also, we assume $D$ is large (which is always satisfied in practice). Thus, we can obtain a reasonable approximation:
\begin{align}\notag
&\mathbf{Pr}\left(z_1 = i,\ z_2 = j, \ i<j \right) \\\notag
= &r_2(r_1-s)\left[1-r_2\right]^{j-i-1}\left[1-(r_1+r_2-s)\right]^i
\end{align}
Similarly, we obtain,  for large $D$,
\begin{align}\notag
&\mathbf{Pr}\left(z_1 = i,\ z_2 = j, \ i>j \right) \\\notag
= &r_1(r_2-s)\left[1-r_1\right]^{i-j-1}\left[1-(r_1+r_2-s)\right]^j
\end{align}

Now we have the tool to calculate the probability
\begin{align}\notag
&\mathbf{Pr}\left(e_{1,1}=e_{2,1}, \ z_1 \neq z_2\right) \\\notag
=&\sum_{i=0,2,4, ...}\left\{\sum_{j\neq i, j=0, 2, 4, ...}\mathbf{Pr}\left(z_1 = i, z_2 = j\right)\right\}\\\notag
+&\sum_{i=1,3,5, ...}\left\{\sum_{j\neq i, j=1, 3, 5, ...}\mathbf{Pr}\left(z_1 = i, z_2 = j\right)\right\}
\end{align}
For example, (again, assuming $D$ is  large)
\begin{align}\notag
&\mathbf{Pr}\left(z_1 =0, z_2 = 2, 4, 6, ...\right)\\\notag
=&r_2(r_1-s)\left([1-r_2]+[1-r_2]^3+[1-r_2]^5+...\right)\\\notag
=&r_2(r_1-s)\frac{1-r_2}{1-[1-r_2]^2}
\end{align}
\begin{align}\notag
&\mathbf{Pr}\left(z_1 =1, z_2 = 3, 5, 7, ...\right)\\\notag
=&r_2(r_1-s)[1-(r_1+r_2-s)]\left([1-r_2]+[1-r_2]^3+[1-r_2]^5+...\right)\\\notag
=&r_2(r_1-s)[1-(r_1+r_2-s)]\frac{1-r_2}{1-[1-r_2]^2}.
\end{align}
Therefore,
\begin{align}\notag
&\sum_{i=0,2,4, ...}\left\{\sum_{i<j, j=0, 2, 4, ...}\mathbf{Pr}\left(z_1 = i, z_2 = j\right)\right\}\\\notag
+&\sum_{i=1,3,5, ...}\left\{\sum_{i<j, j=1, 3, 5, ...}\mathbf{Pr}\left(z_1 = i, z_2 = j\right)\right\}\\\notag
=&r_2(r_1-s)\frac{1-r_2}{1-[1-r_2]^2}\times \\\notag
&\left(1+[1-(r_1+r_2-s)]+[1-(r_1+r_2-s)]^2+...\right)\\\notag
=&r_2(r_1-s)\frac{1-r_2}{1-[1-r_2]^2}\frac{1}{r_1+r_2-s}.
\end{align}
By symmetry, we know
\begin{align}\notag
&\sum_{j=0,2,4, ...}\left\{\sum_{i>j, i=0, 2, 4, ...}\mathbf{Pr}\left(z_1 = i, z_2 = j\right)\right\}\\\notag
+&\sum_{j=1,3,5, ...}\left\{\sum_{i>j, i=1, 3, 5, ...}\mathbf{Pr}\left(z_1 = i, z_2 = j\right)\right\}\\\notag
=&r_1(r_2-s)\frac{1-r_1}{1-[1-r_1]^2}\frac{1}{r_1+r_2-s}.
\end{align}
Combining the probabilities, we obtain
\begin{align}\notag
&\mathbf{Pr}\left(e_{1,1}=e_{2,1}, \ z_1 \neq z_2\right) \\\notag
=&\frac{r_2(1-r_2)}{1-[1-r_2]^2}\frac{r_1-s}{r_1+r_2-s}+\frac{r_1(1-r_1)}{1-[1-r_1]^2}\frac{r_2-s}{r_1+r_2-s}\\\notag
=&A_{1,1} \frac{r_2-s}{r_1+r_2-s} + A_{2,1} \frac{r_1-s}{r_1+r_2-s},
\end{align}
where
\begin{align}\notag
&A_{1,b} = \frac{r_1\left[1-r_1\right]^{2^b-1}}{1-\left[1-r_1\right]^{2^b}}, \hspace{0.2in}
A_{2,b} = \frac{r_2\left[1-r_2\right]^{2^b-1}}{1-\left[1-r_2\right]^{2^b}}.
\end{align}

Therefore, we can obtain the desired probability, for $b=1$,
 \begin{align}\notag
 &\mathbf{Pr}\left(\prod_{i=1}^{b=1} 1\{e_{1,i}=e_{2,i}\} =1\right)\\\notag
=&R +  A_{1,1} \frac{r_2-s}{r_1+r_2-s} + A_{2,1} \frac{r_1-s}{r_1+r_2-s}\\\notag
=&R + A_{1,1} \frac{f_2-a}{f_1+f_2-a} + A_{2,1} \frac{f_1-a}{f_1+f_2-a}\\\notag
=&R + A_{1,1} \frac{f_2-\frac{R}{1+R}(f_1+f_2)}{f_1+f_2-\frac{R}{1+R}(f_1+f_2)}+ A_{2,1} \frac{f_1-a}{f_1+f_2-a}\\\notag
=&R + A_{1,1} \frac{f_2 - Rf_1}{f_1+f_2} + A_{2,1} \frac{f_1 - Rf_2}{f_1+f_2}\\\notag
=&C_{1,1} + (1-C_{2,1})R
\end{align}
where
\begin{align}\notag
&C_{1,b} = A_{1,b} \frac{r_2}{r_1+r_2} + A_{2,b}\frac{r_1}{r_1+r_2}\\\notag
&C_{2,b} = A_{1,b} \frac{r_1}{r_1+r_2} + A_{2,b}\frac{r_2}{r_1+r_2}.
 \end{align}
To this end, we have proved the main result for $b=1$.

The proof for the general case, i.e., $b=2, 3, ...$, follows a similar procedure:
 \begin{align}\notag
 &\mathbf{Pr}\left(\prod_{i=1}^{b} 1\{e_{1,i}=e_{2,i}\} =1\right)\\\notag
=&R +  A_{1,b} \frac{r_2-s}{r_1+r_2-s} + A_{2,b} \frac{r_1-s}{r_1+r_2-s}\\\notag
=&C_{1,b} + (1-C_{2,b})R.
\end{align}

The final task is to show some useful properties of $A_{1,b}$ (same for $A_{2,b}$).  The first derivative of $A_{1,b}$ with respect to $b$ is
\begin{align}\notag
\frac{\partial A_{1,b}}{\partial b} =&\frac{r_1[1-r_1]^{2^b-1}\log(1-r_1)\log2\left(1-[1-r_1]^{2^b}\right)} {\left(1-[1-r_1]^{2^b}\right)^2}\\\notag
&-\frac{-[1-r_1]^{2^b}\log(1-r_1)\log2\ r_1\left(1-[1-r_1]^{2^b-1}\right)} {\left(1-[1-r_1]^{2^b}\right)^2}\\\notag
\leq&0 \hspace{0.5in} (\text{Note that} \ \log(1-r_1)\leq 0)
\end{align}
Thus, $A_{1,b}$ is a monotonically decreasing function of $b$.

Also,
\begin{align}\notag
\lim_{r_1\rightarrow 0}A_{1,b} =\lim_{r_1\rightarrow 0}\frac{\left[1-r_1\right]^{2^b-1}-r_1\left(2^b-1\right)[1-r_1]^{2^b-2}}{2^b[1-r_1]^{2^b-1}}=  \frac{1}{2^b},
\end{align}
and
\begin{align}\notag
\frac{\partial A_{1,b}}{\partial r_1} =& \frac{\left[1-r_1\right]^{2^b-1}-r_1\left(2^b-1\right)[1-r_1]^{2^b-2}}{\left(1-[1-r_1]^{2^b}\right)}\\\notag
&\hspace{0in}-\frac{2^b[1-r_1]^{2^b-1}r_1\left[1-r_1\right]^{2^b-1}}{\left(1-[1-r_1]^{2^b}\right)^2}\\\notag
=& \frac{[1-r_1]^{2^b-2}}{\left(1-[1-r_1]^{2^b}\right)^2}
\left(1-2^br_1-[1-r_1]^{2^b}\right)\leq0.
\end{align}
Note that $(1-x)^c \geq 1-cx$, for $c\geq1$ and $x\leq 1$.

Therefore $A_{1,b}$ is a monotonically decreasing function of $r_1$. We complete the whole proof.

\end{document}